\documentclass[11pt]{article}
\pdfoutput=1
\usepackage{graphicx}
\usepackage{amsmath,amscd}
\usepackage{amsfonts}
\usepackage{amssymb}
\usepackage[usenames,dvipsnames]{color}
\usepackage{epsf}
\usepackage{feynmf}
\usepackage[small,nohug,heads=vee]{diagrams}
\diagramstyle[labelstyle=\scriptstyle]
\usepackage{epstopdf} 
\usepackage{jcappub}
\usepackage{slashed}
\usepackage{eucal}

\newcommand{\be}{\begin{equation}}         
\newcommand{\ee}{\end{equation}}
\newcommand{\ba}{\begin{eqnarray}}
\newcommand{\ea}{\end{eqnarray}}
\newcommand{\nn}{\nonumber}

\newcommand\lsim{\mathrel{\rlap{\lower4pt\hbox{\hskip1pt$\sim$}}
        \raise1pt\hbox{$<$}}}
\newcommand\gsim{\mathrel{\rlap{\lower4pt\hbox{\hskip1pt$\sim$}}
        \raise1pt\hbox{$>$}}}

\def\k{{\bf k}}
\def\x{{\bf x}}

\title{Non-Gaussian Mode Coupling and the Statistical Cosmological Principle}
\author[a]{Marilena LoVerde}
\author[b]{Elliot Nelson}
\author[b]{Sarah Shandera}
\affiliation[a]{Enrico Fermi Institute,  Kavli Institute for Cosmological Physics,  Department of Astronomy \& Astrophysics,  University of Chicago, Chicago, Illinois 60637, U.S.A}
\affiliation[b]{Institute for Gravitation and the Cosmos, The Pennsylvania State University, University Park PA 16802}
\emailAdd{marilena@uchicago.edu}
\emailAdd{eln121@psu.edu}
\emailAdd{shandera@gravity.psu.edu}
 
\abstract{
Local-type primordial non-Gaussianity couples statistics of the curvature perturbation $\zeta$ on vastly different physical scales. Because of this coupling, statistics (i.e. the polyspectra)  of $\zeta$ in our Hubble volume may not be representative of those in the larger universe -- that is, they may be biased. The bias depends on the local background value of $\zeta$, which includes contributions from all modes with wavelength $k\lsim H_0$ and is therefore enhanced if the entire post-inflationary patch is large compared with our Hubble volume. We study the bias to locally-measured statistics for general local-type non-Gaussianity. We consider three examples in detail: (i) the usual $f_{NL}$, $g_{NL}$ model, (ii) a strongly non-Gaussian model with $\zeta \sim \zeta_G^p$, and (iii) two-field non-Gaussian initial conditions. In each scenario one may generate statistics in a Hubble-size patch that are weakly Gaussian and consistent with observations despite the fact that the statistics in the larger, post-inflationary patch look very different.}

\begin{document}
\maketitle
\tableofcontents

\section{Introduction}
\label{sec:intro}

A primary goal of observational cosmology is to learn about the physics of inflation. A key observable is the statistical distribution of the primordial curvature perturbation $\zeta$. In single-field models of inflation the statistics of $\zeta$ are inherited from quantum fluctuations in the inflaton itself \cite{Guth:1980zm,Mukhanov:1981xt,Hawking:1982cz,Starobinsky:1982ee,Guth:1982ec,Bardeen:1983qw}. If multiple light fields are present during inflation, such as in the curvaton scenario \cite{Mollerach:1989hu,Linde:1996gt,Enqvist:2001zp,Lyth:2001nq,Lyth:2002my} or modulated reheating \cite{Dvali:2003em,Dvali:2003ar}, fields other than the inflaton may generate $\zeta$ and the relationship between the post-inflationary curvature and quantum fluctuations generated during inflation can be non-linear (see e.g. \cite{Bartolo:2004if}).  In these scenarios, the inflaton and curvaton field may obey Gaussian statistics while the observed curvature perturbation is non-linearly related to the Gaussian fluctuations, allowing for the phenomenological parameterization:
\be
\label{eq:zetafNLgNL}
\zeta(\x)=\zeta_{G}(\x)+\frac{3}{5}f_{NL}\left(\zeta^2_G(\x)-\langle \zeta_G^2\rangle\right)+\frac{9}{25}g_{NL}\left(\zeta^3_G(\x)-3\langle\zeta^2_G\rangle \zeta_G(\x)\right)+\dots 
\ee
where $\zeta_G$ is a Gaussian random field and $f_{NL}$, $g_{NL}$ are constants specified by the particular inflationary model \cite{Salopek:1990jq,Gangui:1993tt,Komatsu:2001rj,Okamoto:2002ik,Enqvist:2008gk}. The field $\zeta$ in Eq.~(\ref{eq:zetafNLgNL}) obeys non-Gaussian statistics and the level of non-Gaussianity can be characterized by the products $f_{NL}\sqrt{\langle\zeta_G^2\rangle}$,  $g_{NL}\langle\zeta_G^2\rangle$\dots

Anisotropies in the cosmic microwave background (CMB), along with other cosmological datasets, provide stringent constraints on the statistics of $\zeta$ within our Hubble volume. The variance of the primordial curvature perturbation is determined to be $\Delta_{\zeta}^2(k)= \Delta^2_{WMAP}(k/k_{pivot})^{n_s-1}$ where $\Delta_{WMAP}^2 = 2.464\pm 0.072 \times 10^{-9}$, $k_{pivot}\equiv 0.002$/Mpc and $n_s=0.9608\pm 0.0080$ at $68\%$ confidence \cite{Hinshaw:2012fq}. The non-Gaussian parameters in Eq.~(\ref{eq:zetafNLgNL}) are bounded $-10 < f_{NL} < 74$ and $-12.34 \times 10^{5} < g_{NL} < 15.58 \times 10^5$  at $95\%$ confidence \cite{Komatsu:2010fb,Fergusson:2010gn}\footnote{There are, of course, a number of theoretically-motivated forms of primordial non-Gaussianity that are not captured by Eq.~(\ref{eq:zetafNLgNL}) (see, for instance \cite{Cheung:2007st} and references therein). However, in this paper we restrict our attention to ``local" non-Gaussianity which can couple Fourier modes of vastly different wavelengths.}. In this paper we ask the following question: Suppose that inflation lasted sufficiently long that our Hubble volume is small compared to the entire volume generated during inflation. Are the statistics of the curvature perturbation observed in our Hubble volume necessarily representative of the statistics of the curvature perturbation in the rest of the universe? 

The cosmological principle might make this question seem unnecessary, but (as has been noted by a number of authors \cite{Fan:1995aq,Linde:2005yw,Gordon:2005ai,Boubekeur:2005fj,Giddings:2011zd,Byrnes:2011ri,Schmidt:2012ky,Tasinato:2012js,Nelson:2012sb,Nurmi:2013xv}) even if the universe is statistically homogeneous and isotropic, mode-coupling introduced by non-linear terms like those in Eq.~(\ref{eq:zetafNLgNL}) correlates the statistics of $\zeta$ on very different scales. The locally observed, smaller-than-Hubble-scale statistics depend on the unobservable, super-horizon modes of $\zeta_G$ (i.e. long-wavelength modes of the Gaussian field that are nearly constant across our Hubble volume). Changes to the local statistics depend on a small parameter, $\mathcal{O}(\zeta_{super-horizon}) \ll 1$, but can nevertheless be important. In particular, the background mode in our Hubble volume is roughly a sum over all modes with wavelength $\pi/k$ larger than $c/H_0$, so the variance of the total background fluctuations is larger than the locally observed variance in a single $k$ mode by an amount dependent on the number of super-horizon e-folds. This situation, in which long-wavelength modes of $\zeta$ bias the local statistics, is in stark contrast to the case where $\zeta$ is Gaussian and different Fourier modes are strictly uncorrelated. For a Gaussian field, the local power spectrum may be randomly different from the globally averaged one, but is not systematically biased. 

In this paper we study the variation in local statistics between Hubble patches due to the coupling between modes within a Hubble patch and those that are longer wavelength. Within the curvaton framework, Linde and Mukhanov \cite{Linde:2005yw} have pointed out that there are a wide range of possibilities for the local statistics (including the level of non-Gaussianity) in Hubble patches with different local backgrounds (see also \cite{Demozzi:2010aj}). More recently, the possibly biased nature of local statistics in non-Gaussian cosmologies due to long-wavelength modes was studied by Nelson and Shandera \cite{Nelson:2012sb} and Nurmi, Byrnes, and Tasinato \cite{Nurmi:2013xv} (which appeared while this paper was in preparation). Nelson and Shandera studied the scaling behavior and squeezed limits of $n$-point correlation functions computed by observers with access to a finite subvolume of the universe. They pointed out that some regions can be so biased that the local statistics appear only weakly non-Gaussian, even if the statistics in the larger universe are strongly non-Gaussian (a situation we consider in detail in \S \ref{sec:strongNG}). Nurmi, Byrnes, and Tasinato study possible relations between the local model parameters in Eq.~(\ref{eq:zetafNLgNL}) as measured by an observer in our Hubble volume. While these authors focus slightly more on the implications of a potential non-detection of primordial non-Gaussianity, there is significant overlap with our analysis in \S \ref{sec:weakNG}. In contrast to \cite{Linde:2005yw,Demozzi:2010aj}, who work within particular two-field inflationary constructions, this work as well as \cite{Nelson:2012sb,Nurmi:2013xv} considers the problem of mode coupling in the curvature perturbation from a completely statistical perspective -- i.e. we ignore the question of how the field $\zeta$ is generated. Our calculations are therefore more general than \cite{Linde:2005yw} in the sense that they do not rely on a particular early universe scenario. On the other hand, a within a microphysical model of inflation (such as the supercurvaton scenario \cite{Demozzi:2010aj})  there may be fixed relationships between observables that are neglected by our analysis but are nevertheless important for determining the final distribution of local statistics. 

In this paper we study the bias to the local curvature perturbation field $\zeta$ for non-Gaussian initial conditions that can be written as a non-linear function of a Gaussian field $\zeta_G$ that is local in configuration space. In \S \ref{sec:local} we introduce our notation and formulate the calculation of local statistics of $\zeta_{NG}$  in terms of a short-long wavelength split of the Gaussian field $\zeta_G$. In \S \ref{sec:weakNG} we study the mapping between the global and local values of non-Gaussian parameters in the case when the statistics in the larger universe are weakly non-Gaussian as in Eq.~(\ref{eq:zetafNLgNL}) (see also \cite{Nurmi:2013xv}).  In \S \ref{sec:strongNG} we present full calculations of a somewhat counterintuitive example first discussed in \cite{Nelson:2012sb} in which the local statistics of $\zeta$ can appear to be nearly Gaussian while the global ones are strongly non-Gaussian. In \S \ref{sec:twofieldNG} we study the relation between globally and locally determined statistics for two-field initial conditions. In \S \ref{sec:conclusions} we summarize our results and discuss some possible implications for interpreting potential measurements of primordial non-Gaussianity. Appendix \ref{sec:diagrams} contains a diagrammatic formulation of non-Gaussian statistics for a general local model, which is useful both computationally and conceptually for understanding the mapping between local and global statistics. For the brave at heart, calculations of global and local statistics for generic local-type non-Gaussian initial conditions are given in Appendices \ref{sec:global} and \ref{sec:mapping} and discussion of diagrammatic techniques in Fourier space is given in Appendix \ref{sec:Fourier}.

\section{Local Statistics in a Subvolume}
\label{sec:local}

\begin{figure}
\begin{center}
\includegraphics[width=0.8\textwidth]{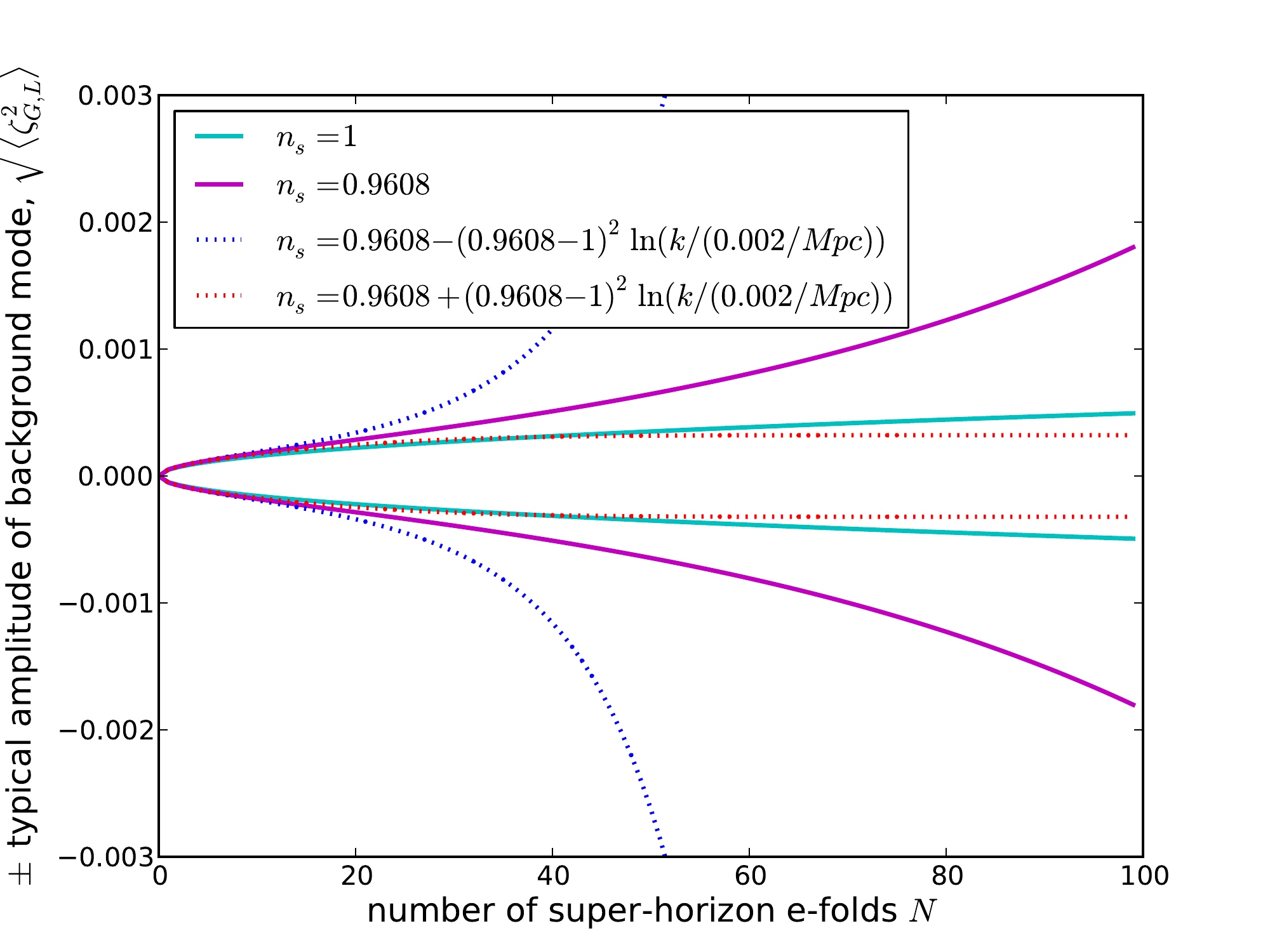} 
\caption{\label{Fig:Sigma_LW} Quantities in our Hubble volume are different from those in the larger universe by terms proportional to $\zeta_{G,L}$ -- the long wavelength fluctuations that appear to be constant within our Hubble volume. Plotted is $\langle\zeta_{G,L}^2\rangle^{1/2}$ assuming that $\Delta_{G}^2=2.464\times 10^{-9}\left({k}/{k_{piv.}}\right)^{n_s-1}$ for two constant values of $n_s$, along with two examples that include running. Note that $\langle\zeta_{G,L}^2\rangle^{1/2}$ is larger than the $10^{-5}$ amplitude in an individual Fourier mode because $\langle\zeta^2_{G,L}\rangle$  is roughly a sum over $\langle \zeta_{G}^2(\k)\rangle$ for $k < H_0$. Throughout this paper we assume constant $n_s=0.9608$ \cite{Hinshaw:2012fq} or $n_s=1$, but keep in mind that the results for $N \gsim 20$ are very sensitive to the (unknown) infrared behavior of $\Delta_G^2(k)$.}
\end{center}
\end{figure}
We are interested in mode-coupling for the non-linear curvature perturbation $\zeta$. Following \cite{Lyth:2004gb}, we let $\tilde{a}$ be the locally defined scale factor and factor $\tilde{a}$ into a spatially homogeneous piece and a perturbation,
\be
\tilde{a}=ae^{\zeta(\x)}
\ee
where $\zeta$ is a perturbation with volume average equal to zero over a volume $V_L$ so that,
\be
a(t)= e^{ \frac{1}{V_L}\int_{V_L} d^3\x\, \ln \tilde{a}} \quad{\rm and }\quad  \zeta(\x) = \ln \left(\frac{\tilde{a}}{a}\right)\,.
\ee
The locally observed perturbation with respect to some smaller volume $V_S$ will be
\ba
\label{eq:volzetadefs}
\zeta_S(\x) &=& \ln \tilde{a}-\frac{1}{V_S}\int_{V_S}d^3\x \, \ln \tilde{a}\\
&=&\underbrace{\ln \tilde{a}-\frac{1}{V_L}\int_{V_L}d^3\x \, \ln \tilde{a}}_{\zeta(\x)}-\underbrace{\left(\frac{1}{V_S}\int_{V_S}d^3\x \, \ln \tilde{a}-\frac{1}{V_L}\int_{V_L}d^3\x \, \ln \tilde{a}\right)}_{\equiv \zeta_L(\x)}
\ea
So that, 
\be
\zeta_S(\x)=\zeta(\x)-\zeta_L(\x)
\ee
i.e. a change in reference volume amounts to an additive shift in the perturbations (see also \cite{Giddings:2010nc,Giddings:2011zd,Senatore:2012nq}).

Suppose the non-Gaussian statistics in a volume $V_L$ can be written as a local, non-linear transformation of a Gaussian field as in Eq.~(\ref{eq:zetafNLgNL}), or more generally, $\zeta_{NG} = f(\zeta_G)-\langle f(\zeta_G)\rangle$.  The non-linearity of $f$ in the variable $\zeta_{G}$ couples Fourier modes of $\zeta_{NG}$ on different scales. Given a non-Gaussian curvature field  defined in a large volume $V_L$, how will the statistics appear to an observer with access to only a finite subvolume $V_S$? 

From Eq.~(\ref{eq:volzetadefs}), the local value of $\zeta$ in $V_S$ is
\be
\left.\zeta(\x)\right|_{S}=\zeta(\x)-\int \frac{d^3{\bf k}}{(2\pi)^3}e^{i\k\cdot\x}W_S(k)\zeta(\k)
\ee
where
\be
W_S(k) \equiv\frac{1}{V_S} \int_{V_S}d^3\x \,e^{-i\x\cdot\k}
\ee
giving
\be
\left.\zeta(\k)\right|_S=\zeta(\k)(1-W_S(k))
\ee
For simplicity we make the approximation that $W_S(k)$ is a top-hat function in Fourier space so that we can define 
\be
\zeta_{G,S}(\x) =  \int_{|{\bf k }|\ge k_*} \frac{d^3{\bf  k}}{(2\pi)^3}\zeta_G(\k)e^{i\k\cdot\x} \qquad {\rm and }\qquad \zeta_{G,L}(\x) =   \int_{|{\bf k}|< k_*} \frac{d^3{\bf  k}}{(2\pi)^3}\zeta_G(\k)e^{i\k\cdot\x} 
\ee
where $k* \sim 2\pi/V_S^{1/3}$. Now the local Gaussian curvature can be split into short and long-wavelength pieces,
\be
\zeta_G(\x)=\zeta_{G,S}(\x)+\zeta_{G,L}(\x)\
\ee
with auto-correlations
\be
\label{eq:zetaGs}
\langle \zeta_{G,S}(\x)\zeta_{G,S}(\x')\rangle  =  \int_{|{\bf k }|\ge k_*} \frac{d^3{\bf  k}}{(2\pi)^3}P_G(k)e^{i\k\cdot(\x-\x')}
\ee
\be
\label{eq:zetaGl}
\langle \zeta_{G,L}(\x)\zeta_{G,L}(\x')\rangle  =  \int_{|{\bf k }|< k_*} \frac{d^3{\bf  k}}{(2\pi)^3}P_G(k)e^{i\k\cdot(\x-\x')} 
\ee
and vanishing cross-correlation\footnote{The more realistic assumption of a top-hat window function in real space with radius $V_S^{1/3}$ generates $\sim \Delta^2_G(k\sim 2.5/V_S^{1/3})/4$ corrections to Eq.~(\ref{eq:zetaGls}), where $kV_S^{1/3}\sim 2.5$ is the peak of $W_S(k)(1-W_S(k))$ for a top-hat in real space. For $n_s=1$ this gives $\langle \zeta_{G,L}(\x)\zeta_{G,S}(\x')\rangle/\langle \zeta_{G,L}(\x)\zeta_{G,L}(\x')\rangle \sim 1/(4N)$, and $\langle \zeta_{G,L}(\x)\zeta_{G,S}(\x')\rangle/\langle \zeta_{G,S}(\x)\zeta_{G,S}(\x')\rangle \sim -1/(4\ln(|\x-\x'|H_0))$ for $|\x-\x'|/V_S^{1/3}\lsim 10$. }

\be
\label{eq:zetaGls}
\langle \zeta_{G,L}(\x)\zeta_{G,S}(\x')\rangle  = 0\,.
\ee
We've defined the power spectrum of the Gaussian field as
\be
\langle \zeta_G(\k)\zeta_G(\k')\rangle = (2\pi)^3P_G(k)\delta_{D}(\k+\k')\,.
\ee
In what follows we assume, 
\be
\label{eq:lwvariance}
\langle \zeta_{G,S}^2\rangle_{S}\approx \langle \zeta_{G,S}^2\rangle\qquad  {\rm while}\qquad  \langle \zeta_{G,L}^n\rangle_{S}\approx \zeta_{G,L}^n
\ee
where $\langle \rangle_{S}$ indicates averages over the volume $V_S$ and $\langle \rangle$ indicates an average over the entire volume $V_L$. That is, we assume that the locally measured, small-scale Gaussian power spectrum is representative of the globally defined one and that the variation in long-wavelength modes ($k_L<k*$) across the volume $V_S$ is negligible.

\begin{figure}[ht!]
\begin{center}
\begin{tabular}{|c|p{10cm}|}
\multicolumn{2}{c} {\bf Dictionary of Frequently Used Symbols} \\
\hline
Quantity & Definition  \\
\hline 
$\zeta_{NG}$, $P_{NG}(k)$ & The non-Gaussian curvature perturbation and its power spectrum\\
\hline
$\zeta_{G}$, $P_G(k)$ & A Gaussian random field used to generate $\zeta_{NG}$ and its power spectrum\\
\hline
$\Delta_X^2(k)$ & The variance of fluctuations of the field $X$, $\Delta_X^2(k)\equiv k^3P_X(k)/(2\pi^2)$\\
\hline
$V_L$ & The volume over which $\zeta_{NG}$ is defined, e.g. the entire post-inflationary patch\\
\hline
$V_S$, $W_S$ & A subvolume of $V_L$ and the corresponding window function, $W_S(k)=\int_{V_S}d^3\x\, e^{-i\x\cdot\k}$. For most of this paper we take $V_S$ to be our Hubble volume \\
\hline
$\zeta_{G,S}$, $\zeta_{G,L}$ & The short and long wavelength components of $\zeta_G$. See Eq.~(\ref{eq:zetaGs}),  Eq.~(\ref{eq:zetaGl}) \\
\hline
$N$ & The number of super-horizon e-folds, $\frac{1}{3}\ln(V_L/V_S)$ \\
\hline
$\Delta^2_{WMAP}$ &    The amplitude of scalar-perturbations at $k_{piv}=0.002/Mpc$ as measured by WMAP \cite{Hinshaw:2012fq}  \\
\hline
$n_s$&    The scalar spectral index $n_s-1=d\ln\Delta^2/d\ln k$ \\
\hline
$\Omega_k$, $H_0$&    The spatial curvature and Hubble scale today \\
\hline
$f_{NL}$, $g_{NL}$, $\tau_{NL}$, $h_{NL}$ & Non-Gaussian parameters given in Eq.~(\ref{eq:zetaNGfull}) (or as defined by the squeezed limits of the bispectrum and trispectrum in Eq.~(\ref{eq:fNLdef}), Eq.~(\ref{eq:gNLdef}) and Eq~(\ref{eq:tauNLdef}))\\
\hline 
$\left. X \right|_S$ &    The value of the quantity $X$ measured in $V_S$  \\
    \hline
    $\chi_G$, $P_\chi$ & In \S \ref{sec:weakNG} and \S \ref{sec:strongNG} we rewrite $\left.\zeta_{NG}\right|_S$ in terms of $\chi_G(\x)\propto \zeta_{G}(\x)$\\
    \hline
       $\sigma_G$, $\phi_G$, $\xi$ & Gaussian random fields we use to define $\zeta_{NG}$ in the two-field example in \S \ref{sec:twofieldNG}, and the ratio of their power spectra, $\xi^2\equiv P_\phi/P_\sigma$ \\
    \hline
\end{tabular}
\end{center}
\end{figure}

In this limit an observer in the small volume $V_S$ is unable to distinguish between $\zeta_{G,L}$ and the background. However, the nonlinear coupling of $\zeta_{G,L}$ to short wavelength modes $\zeta_{G,S}$ will cause the local, small-scale non-Gaussian statistics of $\zeta_{NG}$ to differ from the global ones {\em in a way that depends on the local value of $\zeta_{G,L}$} -- that is, the local statistics are {\em biased} by $\zeta_{G,L}$. 

The typical size of the bias is characterized by the variance of long-wavelength fluctuations, 
\be
\label{eq:varL}
\langle \zeta_{G,L}^2\rangle= \int_{\Lambda}^{k*} \frac{dk}{k}\Delta^2_{G}(k) 
\ee
where $\Delta_G^2(k)=\frac{4\pi}{(2\pi)^3}k^3 P_G(k)$ and $\Lambda \sim 2\pi/V_L^{1/3}$, the infrared cutoff corresponding to the larger volume where the perturbations are set up. In the example calculations and plots we take $V_S$ to be our Hubble volume $ \sim H_0^{-3}$, but the expressions in the paper are completely general. The general expressions may be relevant for making comparisons between theory and particular observables measured in a volume smaller than our Hubble volume.

Letting the power spectrum for $\zeta_G$ be a power law, $d\ln\Delta_G^2/d\ln k\equiv {n_s-1}$ with $n_s-1=const.$ gives closed-form expressions for the variance of long-wavelength modes
\ba
\label{eq:LWVarns1}
\left.\langle \zeta_{G,L}^2\rangle \right|_{n_s=1} & = & \Delta_G^2 N \\
\label{eq:LWVar}
\left.\langle \zeta_{G,L}^2\rangle \right|_{n_s\neq 1} &=& \Delta_G^2(H_0)\frac{\left(1-e^{-(n_s-1)N}\right)}{n_s-1}
\ea
where $N\equiv \ln (H_0/\Lambda)$ is the number of super-horizon e-folds from the start of inflation to the time when the comoving scale of the observable universe crossed into the horizon. In Figure \ref{Fig:Sigma_LW} we plot $\pm\langle \zeta_{G,L}^2\rangle^{1/2}$, the typical amplitude of the unobservable background mode, as a function of the number of super-horizon e-folds $N$. The power spectrum of $\zeta_G$ is of course unknown for $k \lsim H_0$, but as a starting point we consider constant $n_s$ as in Eq.~(\ref{eq:LWVarns1})-Eq.~(\ref{eq:LWVar}). As can be seen from Figure \ref{Fig:Sigma_LW}, even the modest red-tilt that is currently favored ($n_s=0.9608$) dramatically increases the typical amplitude of super-horizon fluctuations relative to that for a flat spectrum $n_s=1$. This difference becomes significant for $N\gtrsim -(n_s-1)^{-1}\simeq25$ e-folds -- precisely when  $\mathcal{O}((n_s-1)^2)$ contributions to the running are expected to change $n_s-1$ by order unity \cite{Liddle:2000cg}. The specific shapes of $\langle\zeta_{G,L}^2\rangle^{1/2}$ plotted in Figure \ref{Fig:Sigma_LW} should therefore be interpreted with caution, particularly for $N\gg 1/(n_s-1)$.  For reference, we also plot examples of $\langle \zeta_{G,L}^2\rangle$ with running spectral indices given by $n_s(k)=n_s(k_{piv})\pm (n_s(k_{piv})-1)^2\ln(k/k_{piv})$.

Long wavelength modes of $\zeta_{NG}$ will also contribute to the mean spatial curvature measured within our Hubble volume\footnote{Here, we are using the scalar curvature on spatial hypersurfaces $R^{(3)} =-4 \nabla^2\zeta(\x)$, and taking $\left.\Omega_k \right|_S =- \frac{1}{6} H_0\int_{H_0^{-3}} d^3 \x\, R^{(3)}(\x)$, however see \cite{Knox:2005hx,Waterhouse:2008vb,Vardanyan:2009ft,Erickcek:2008jp,Guth:2012ww,Kleban:2012ph} for more detailed discussions of constraints on $\zeta(\k)$ contributions to $\Omega_k$ as measured in our Hubble volume. }
\be
\left.\Omega_k\right|_s = \frac{-2}{3H_0^2}\int \frac{d^3\k}{(2\pi)^3} k^2\zeta_{NG}(\k) W_S(k)\,.
\ee
In a given subvolume $V_S$, knowing the value of the background mode $\zeta_{G,L}$, is insufficient to specify $\left.\Omega_k\right|_s $. Nevertheless, we can estimate the typical amplitude of $\left.\Omega_k\right|_s $ in the scenarios we consider
\be
\langle \left.\Omega_k\right|_s^2\rangle = \frac{4}{9H_0^4}\int \frac{dk}{k} k^4 \Delta_\zeta^2(k) |W_S(k)|^2= \frac{4\Delta_\zeta^2(H_0)}{9(n_s+3)}\left(1-e^{-(n_s+3)N}\right)\approx\frac{4\Delta_\zeta^2(H_0)}{9(n_s+3)}
\ee
where in the final $\approx$ we have assumed that the power spectrum is not too red (e.g. for $n_s+3 \gsim 1$). So, the dominant contributions to $ \left.\Omega_k\right|_s$ come from modes with $k \sim H_0$ and, in contrast to Eq.~(\ref{eq:LWVarns1})-Eq.~(\ref{eq:LWVar}), there is no enhancement from $N\gg1$. We therefore ignore constraints on $\zeta_{G,L}$ coming from constraints in $\Omega_k$ because only the first few modes outside the horizon lead to spatial curvature and in fact we are working in the limit that $\zeta_{G,L}$ is independent of $\x$ within our Hubble volume anyway (e.g. Eq.~(\ref{eq:lwvariance})). However, it would be interesting to revisit these constraints and their implication for the bias of local statistics, particularly if local-type non-Gaussianity is detected. For a further discussion of the physical effects of long wavelength modes on local observables see, for instance \cite{Giddings:2011zd,Senatore:2012nq}.

\section{Example I: Weakly Non-Gaussian Initial Conditions}
\label{sec:weakNG}

\begin{figure}
\begin{center}
\includegraphics[width=0.8\textwidth]{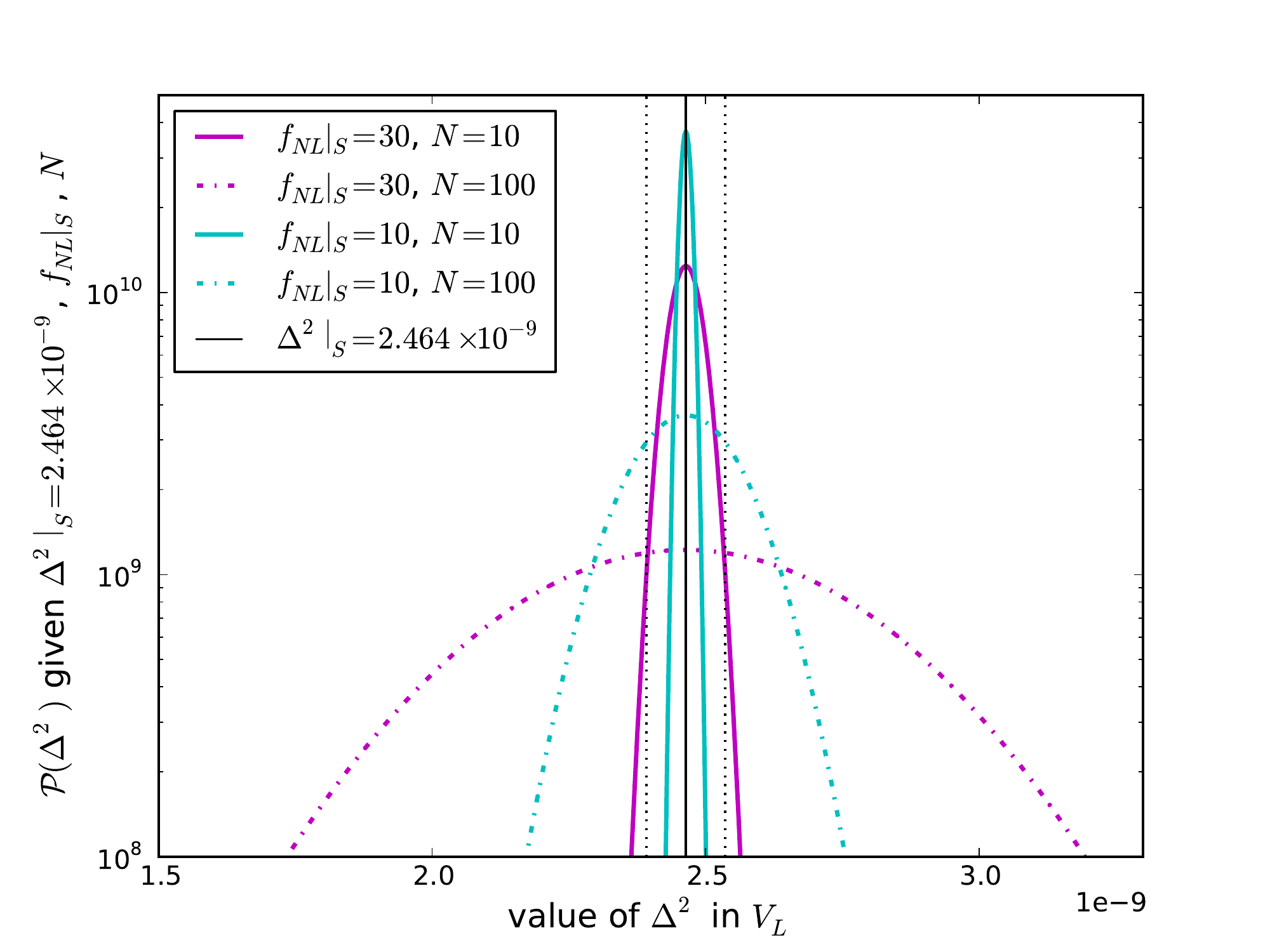} 
\caption{\label{Fig:Pzeta2} The observed amplitude of scalar fluctuations in our Hubble volume systematically differs from the average value in the entire universe by a fractional amount $12/5f_{NL}\zeta_{G,L}$ where $\zeta_{G,L}$ is the (unobservable) background mode in our Hubble patch. Plotted is an estimate of the probability distribution for the true value of $\Delta^2$, given the locally observed value $\Delta_{WMAP}^2=2. 464\times 10^{-9}$ for two values of $\left.f_{NL}\right|_S$ and two values of $N$, the number of super-horizon e-folds. An observer in our Hubble volume cannot measure $N$, and therefore is unable to determine which probability distribution correctly describes our universe.  The vertical dashed lines show the $68\%$ confidence interval ($\pm 0.072\times 10^{-9}$) on $\Delta^2_{WMAP}$ from the eCMB+BAO+H0 dataset \cite{Hinshaw:2012fq}.}
\end{center}
\end{figure}

\begin{figure}
\begin{center}
$\begin{array}{cc}
\includegraphics[width=0.5\textwidth]{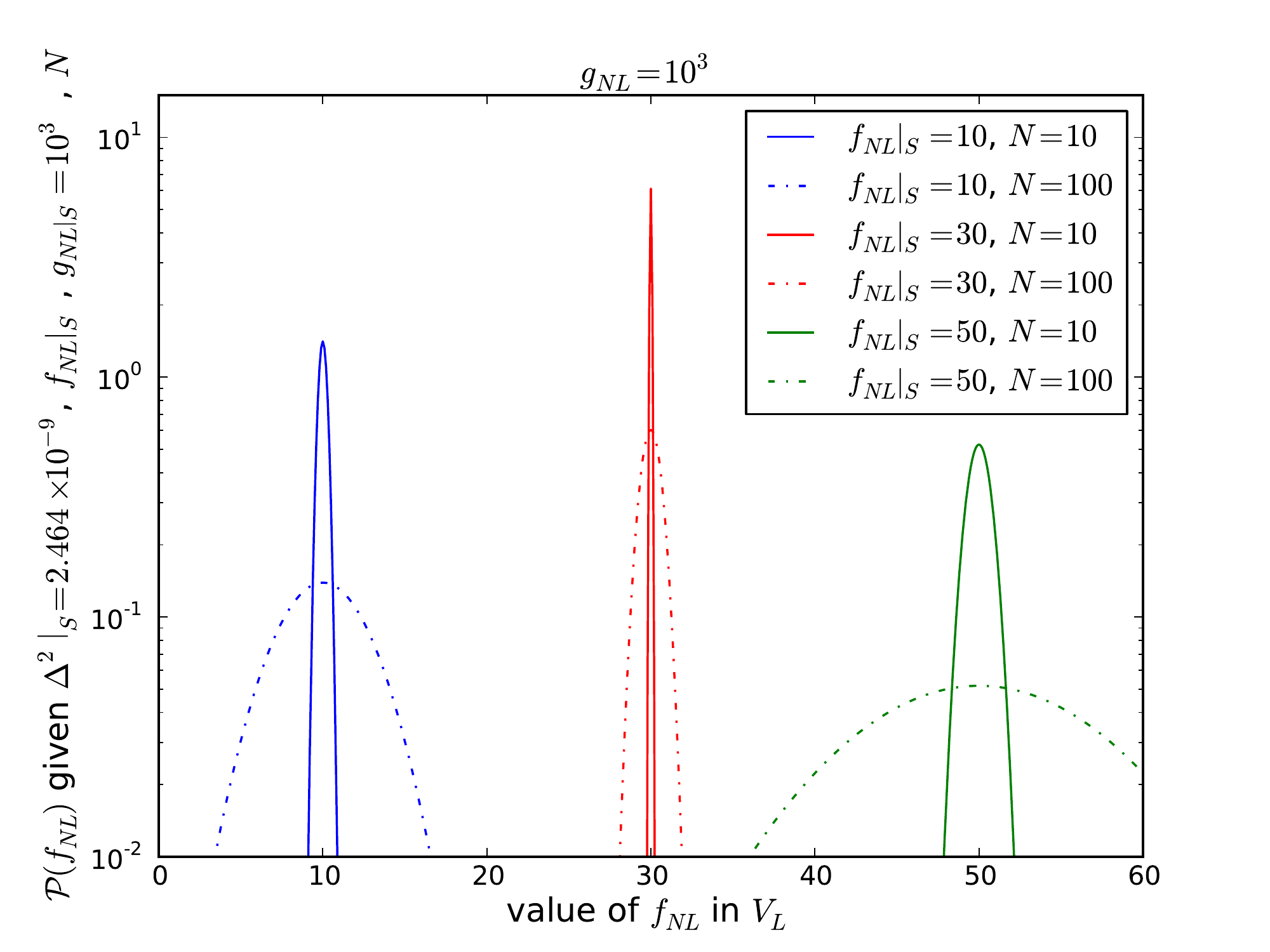}  &\includegraphics[width=0.5\textwidth]{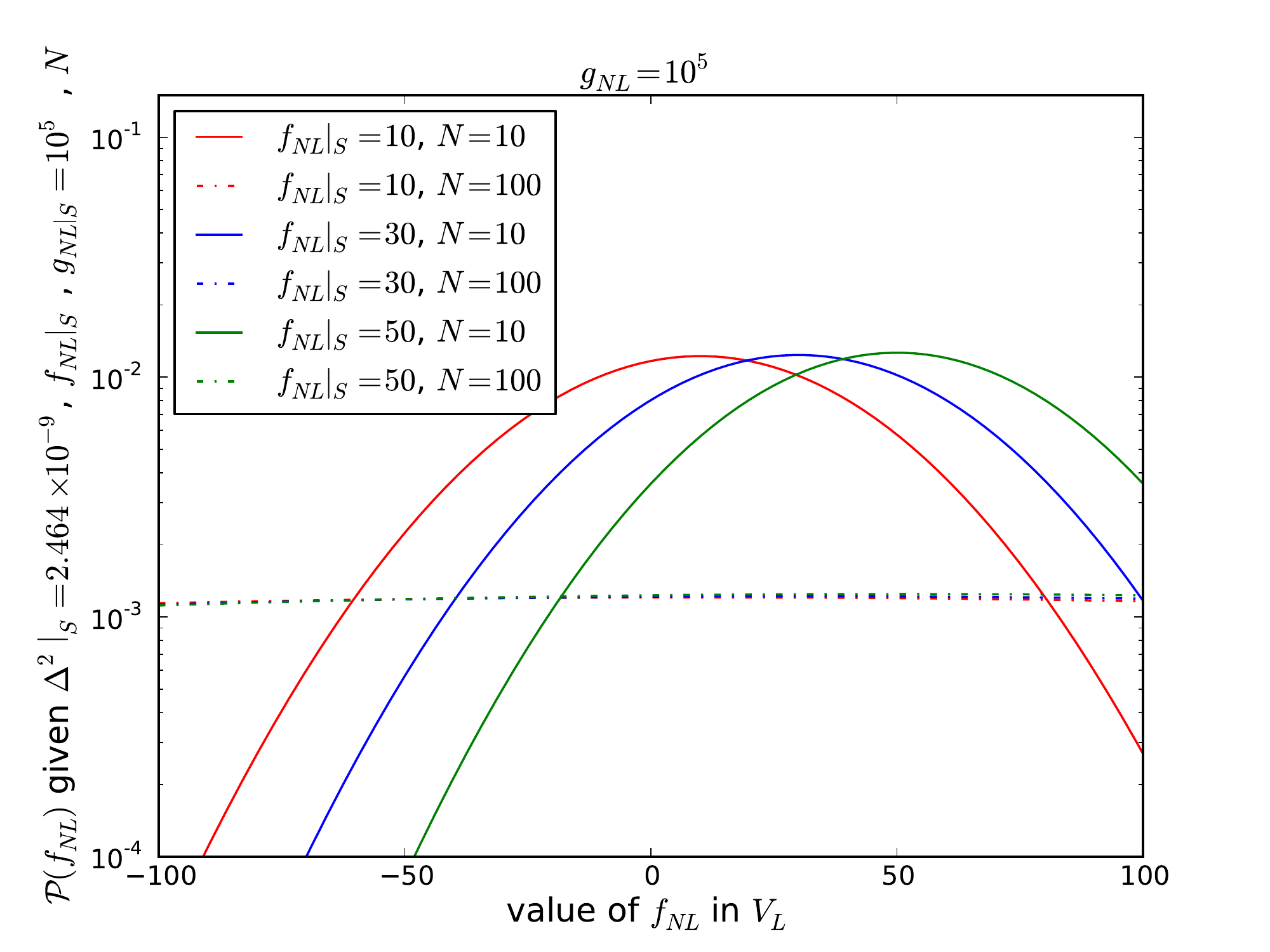} 
\end{array}$
\caption{\label{Fig:probfNL} The observed amplitude of $f_{NL}$ in our Hubble volume, $\left.f_{NL}\right|_S$, differs from the average value in entire universe depending on $\zeta_{G,L}$, $f_{NL}$ and $g_{NL}$. Plotted is an estimate of the probability distribution for the true value of $f_{NL}$ given $\Delta_{WMAP}^2=2.464\times 10^{-9}$ and fixed $g_{NL}=10^3$ (left panel) or  $g_{NL}=10^5$ (right panel). For $f_{NL}=30$ and $g_{NL}=1000$ there is a partial cancellation between the $g_{NL}$ and $f_{NL}^2$ terms in Eq.~(\ref{eq:weaklocalfNL}) causing the distribution to be narrower than in the $f_{NL}=10,50$ cases. The width of the distributions depends on the RMS value of $\zeta_{G,L}$. We assume $n_s=const.=0.9608$, which in comparison to $n_s=1$, gives a difference in $\langle \zeta_{G,L}^2\rangle^{1/2}$ of $\sim 10\%$ at $N=1$ and a factor of $\sim 3.5$ by $N=100$.}\end{center}
\end{figure}

In this section we imagine that the statistics in the larger volume $V_L$ can be described by the usual local ansatz\footnote{Here we subtract the $3g_{NL}\langle\zeta_G^2\rangle\zeta_{G}(\x)$ so that the power spectrum is unaffected at linear order in $g_{NL}$ and subtract $6h_{NL}\langle \zeta_G^2\rangle\zeta_G^2(\x)$ so that the bispectrum is unchanged at linear order in $h_{NL}$. This helps to isolate how each coefficient changes the statistics of $\zeta$, particularly in the case where a lower-order coefficient is vanishing (e.g. $f_{NL}=0$, but $h_{NL}\neq 0$). See Appendix \ref{sec:global} for general expressions relating the coefficients in Eq.~(\ref{eq:zetaNGfull}) and series coefficients in a general local map between $\zeta_{G}(\x)$ and $\zeta_{NG}(\x)$}
\ba
\label{eq:zetaNGfull}
\zeta_{NG}(\x)&=&\zeta_G(\x)+\frac{3}{5}f_{NL}\left(\zeta^2_G(\x)-\langle \zeta_G^2\rangle\right)+\frac{9}{25}g_{NL}\left(\zeta_G^3(\x)-3\langle\zeta_G^2\rangle\zeta_G(\x)\right)\\
&&+\,\frac{27}{125}h_{NL}\left(\zeta_G^4-6\langle\zeta_G^2\rangle\zeta_G^2(\x)+3\langle\zeta_G^2\rangle^2\right)\nn
\ea
where the non-zero coefficients satisfy 
\be
\label{eq:weakNGscaling}
1\gg f_{NL}\sqrt{\langle \zeta_G^2\rangle} \gg g_{NL}\langle \zeta_G^2\rangle \gg  h_{NL}(\langle \zeta_G^2\rangle)^{3/2}\;.
\ee
The equation above is the definition of weak non-Gaussianity for this model. Single-source non-Gaussian models with coefficients with this scaling will generate non-Gaussian polyspectra that scale as $\langle \zeta_{NG}^n\rangle_c \sim (\Delta_{NG}^2)^{n-1}$, where $\Delta_{NG}$ is the observed variance (for further discussion see Appendix \ref{sec:global}). We can then apply the condition in Eq.~(\ref{eq:weakNGscaling}) to require that the power spectrum of $\zeta_{NG}$ on CMB scales agrees with the power spectrum of $\zeta_G$ to some accuracy, that is we could require that the $\mathcal{O}(f_{NL}^2)$ terms are not important. Note that depending on the shape of the power spectrum, this requirement may be much stronger condition than requiring that  $f_{NL}\sqrt{\Delta^2_{WMAP}(k)}$, $g_{NL}\Delta_{WMAP}^2(k)$, $h_{NL}\Delta_{WMAP}^{3/2}(k)\ll1$ on CMB scales. We have checked that the examples plotted in Figures \ref{Fig:Pzeta2} and \ref{Fig:probfNL} satisfy  Eq.~(\ref{eq:weakNGscaling}) for the assumed $n_s=const.$ power spectra.  

In the larger volume, $V_L$ the field $\zeta_{NG}(\x)$ given in Eq.~(\ref{eq:zetaNGfull}) has power spectrum
\be
P_{NG}(k)=P_G(k)\left(1+\mathcal{O}(\langle\zeta_G^2\rangle)\right)\,,
\ee
and the bispectrum and trispectrum are characterized by the coefficients $f_{NL}$, $g_{NL}$ and $\tau_{NL}=(\frac{6}{5}f_{NL})^2$ up to corrections $\mathcal{O}(\langle\zeta_G^2\rangle)$. 

An observer in a finite region $V_S$ with background field value $\zeta_{G,L}$ will see 
local statistics described by
\ba
\left.\zeta_{NG}\right|_{S} & = & \chi_G(\x)+\frac{3}{5}\left.f_{NL}\right|_{S}\left(\chi_G^2(\x)-\langle\chi_G^2\rangle\right)+\frac{9}{25}\left.g_{NL}\right|_{S}\left(\chi_G^3-3\langle\chi^2\rangle\chi_G(\x)\right)+\dots
\ea
where we've defined
\be
\chi_G(\x)=\left(1+\frac{6}{5}f_{NL}\zeta_{G,L}+\mathcal{O}(\zeta_{G,L}^2)\right)\zeta_{G,S}
\ee
which we require to give the locally observed power spectrum 
\be
\langle \chi_G(\k)\chi_G(\k')\rangle = (2\pi)^3\delta_D(\k+\k') P_{\chi}(k) \qquad {\rm where}\qquad P_{\chi} (k)= \frac{2\pi^2\Delta^2_{{\rm {\tiny WMAP}}}(k)}{k^3}\,.
\ee
The local power spectrum $P_\chi$ is related to the globally defined one $P_{NG}(k)$ through
\be
\label{eq:weaklocalP}
\left.P_{NG}(k)\right|_S\equiv P_\chi(k)=\left(1+\frac{12}{5}f_{NL}\zeta_{G,L}+\mathcal{O}(\zeta_{G,L}^2)\right)P_G(k)
\ee
and the locally observed non-Gaussian parameters are 
\ba
\label{eq:weaklocalfNL}
\left.f_{NL}\right|_{S} &=& f_{NL}+\frac{9}{5}g_{NL}\zeta_{G,L}-\frac{12}{5}f_{NL}^2\zeta_{G,L}+\mathcal{O}(\zeta_{G,L}^2) \\
\label{eq:weaklocalgNL}
\left.g_{NL}\right|_{S} &=& g_{NL}+\frac{12}{5}h_{NL}\zeta_{G,L}-\frac{18}{5}f_{NL}g_{NL}\zeta_{G,L}+\mathcal{O}(\zeta_{G,L}^2)\,.
\ea
Eq.~(\ref{eq:weaklocalP})-Eq.~(\ref{eq:weaklocalgNL}) show that the connected $n+1$-point functions of $\zeta_{NG}$ adjust the $n$-point functions of $\left.\zeta_{NG}\right|_S$ by terms $\mathcal{O}(\zeta_{G,L})$ and cause the locally observed statistics to differ from the global ones. For a strictly $f_{NL}$ model (i.e. $g_{NL}$, $h_{NL}$,$\dots =0$) with $f_{NL}>0$, a positive background fluctuation $\zeta_{G,L}$ boosts the local power relative to the local bispectrum,  the net effect is to make the local statistics appear more Gaussian than they are in the larger volume $V_L$ (i.e. $\left. f_{NL}\right|_{S} < f_{NL}$ and $\left. f_{NL}\right|_{S}\sqrt{\Delta^2_{S}}< f_{NL}\sqrt{\Delta^2}$). Negative $f_{NL}$ or background fluctuations will, of course, have the opposite effect. On the other hand if $g_{NL}\sim f_{NL}^2$, then leakage from the trispectrum into the bispectrum can compensate and the local $f_{NL}$ value can be representative of the globally defined one. The cancellation between $\frac{9}{25}g_{NL}$ and $\frac{12}{5}f_{NL}^2$ in Eq.~(\ref{eq:weaklocalfNL}) is precisely what happens in the curvaton model when the curvaton dominates the energy density of the universe at the time of decay \cite{Sasaki:2006kq}. However, the {\em level} of non-Gaussianity as quantified by $\left.f_{NL}\sqrt{\Delta^2}\right|_S$ is still adjusted.

If $f_{NL}\neq 0$, the measured value of the scalar power spectrum in our Hubble volume $\Delta^2_{WMAP}$ differs from the average value in the larger universe $V_L$ by an unknown amount $\frac{12}{5}f_{NL}\zeta_{G,L}$ -- unknown because we don't know the values of $\zeta_{G,L}$ or $f_{NL}$. In Figure \ref{Fig:Pzeta2} we plot an estimate of the probability distribution for $\Delta^2$ in $V_L$ for fixed values of $f_{NL}$, assuming the observed value is $\Delta^2_{WMAP}$, and that $\zeta_{G,L}$ is drawn from a power law spectrum as in Eq.~({\ref{eq:LWVar}). Similarly, the local $f_{NL}$ and $g_{NL}$ values in $V_L$ are related to the observed ones by amounts dependent on $\zeta_{G,L}$. In Figure \ref{Fig:probfNL} we plot estimates for the distribution of $f_{NL}$  values in $V_L$ assuming the locally observed power spectrum, and several possible values of $\left.f_{NL}\right|_S$, $\left.g_{NL}\right|_S$. The probability distributions plotted in Figure \ref{Fig:Pzeta2} and Figure \ref{Fig:probfNL} are estimates of the probability distributions in that: (i) we don't allow all the observed parameters ($\left.\Delta^2_G\right|_S$, $\left.f_{NL}\right|_S$, $\left.g_{NL}\right|_S$) to vary simultaneously and (ii) we neglect terms $\mathcal{O}(\zeta_{G,L}^2)$ in relating values of parameters $f_{NL}$, $g_{NL}$ measured in our Hubble volume to those in the larger universe $V_L$. A more realistic, but more involved calculation would be to calculate the posterior probability distribution of ($\Delta^2_G$, $f_{NL}$, $g_{NL}$) given the observed values $\left.\Delta^2_G\right|_S$, $\left.f_{NL}\right|_S$, $\left.g_{NL}\right|_S$ along with their observational uncertainties, and the fact that $\zeta_{G,L}$ is Gaussian distributed. We are merely interested in illustrating the range of possibilities and leave a thorough exploration of parameters for another study. We further emphasize that observationally, we don't have observational access to $N$ -- a parameter we have held fixed in Figure \ref{Fig:Pzeta2} and Figure \ref{Fig:probfNL}. The variation in the probability distributions for different $N$ values should therefore be interpreted as an additional observational uncertainty. 

\subsection{Implications for Model Builder}
When constructing a model of inflation, one typically specifies some set of fields relevant for inflation and the primordial fluctuations, as well as any interactions the fields may have. This guarantees the existence of an inflating solution and fluctuations and determines the possible shapes of the correlation functions. Adjustable parameters then allow the model builder to match the observed amplitude of fluctuations and to tune any non-Gaussianity to an amplitude consistent with observational constraints. The length of slow-roll inflation may or may not be an independently tunable microphysical parameter. How should the model builder decide if a given set of microphysical parameters gives rise to a significant number of Hubble volumes consistent with the one we see? In non-Gaussian models the necessarily statistical nature of making predictions from inflation for our Hubble volume becomes much more important, even for relatively short durations of inflation. 

To illustrate this point, consider a very simple (if unrealistic) model with only quadratic non-Gaussianity, $f_{NL}$, and a constant spectral index. Expressing the amplitude of fluctuations and of non-Gaussianity in terms of the parameters of the large volume theory and the subsample bias gives a sense of how the local statistics can differ from the global statistics: 
\ba
f_{\text{NL}}|_S&=&f_{\text{NL}}\left(1-\frac{12}{5}f_{\text{NL}}\sqrt{\langle\zeta_{G}^2\rangle} \,B\right)\;, \\\nonumber
\label{fP}
f_{\text{NL}}\Delta|_S&=&f_{\text{NL}}\Delta\left(1-\frac{6}{5}f_{\text{NL}}\sqrt{\langle\zeta_{G}^2\rangle}\,B\right). 
\ea
where $\langle \zeta_G^2\rangle$ is the correlation function at zero separation -- a constant\footnote{One might worry that we are scaling quantities by a loop factor $\langle \zeta^2_G\rangle$, which is  dependent on the power spectrum over the entire range of scales (and, without a cutoff is formally divergent for a scale invariant power spectrum). However, $\langle \zeta_G^2\rangle$ is merely a placeholder and the actual value cancels when calculating observed quantities -- our results do not depend on the unknown UV behavior of the power spectrum $P_{G}$.}. We've defined the bias, $B$, as in \cite{Nelson:2012sb} so that it increases as $N$ increases and for fixed $N$ is larger for rarer fluctuations:
\be
B \equiv\frac{\zeta_{G,L}}{\langle \zeta_{G}^2\rangle^{1/2}}\;.
\ee
In this example we restrict to weak non-Gaussianity, so by the condition in Eq.~(\ref{eq:weakNGscaling}), $f_{\text{NL}}\sqrt{\langle\zeta_{G}^2\rangle}\ll1$. Since the long wavelength modes are only a fraction of the total modes contributing to $\langle \zeta_{G}^2\rangle$, for a scale invariant spectrum the bias is also less than one in magnitude except for extremely rare fluctuations, $|B<1|$.

Notice that for non-Gaussian inflation models, matching parameters in the theory to agree exactly with our local observations makes sense only if the number of e-folds in the model is not too large. One way of visualizing this criteria is plotted in Figure \ref{Fig:Var}. If the number of e-folds in the theory is larger, the parameters should {\it not} be matched identically to what we observe on CMB scales. In that case sub-volumes that have statistics identical to the parent will be rare, and so our observed universe will not be the typical outcome of those models. Finally, we note that because both the amplitude of fluctuations and the value of $f_{NL}$ are changing in typical subsamples as we look on different scales, it is useful to plot the quantity that shows how non-Gaussian the subsamples are on average. The relative amplitude of non-Gaussianity in the subvolume to that in the large volume is shown in Figure \ref{Fig:Var}. Note that for positive $f_{\text{NL}}$, an overdensity $\zeta_{G,L}>0$ causes the non-Gaussianity to be smaller in the small volume. Similarly, an underdensity $\zeta_{G,L}<0$ causes the non-Gaussianity to be larger.

\begin{figure}
\begin{center}
$\begin{array}{cc}
\includegraphics[width=.5\textwidth]{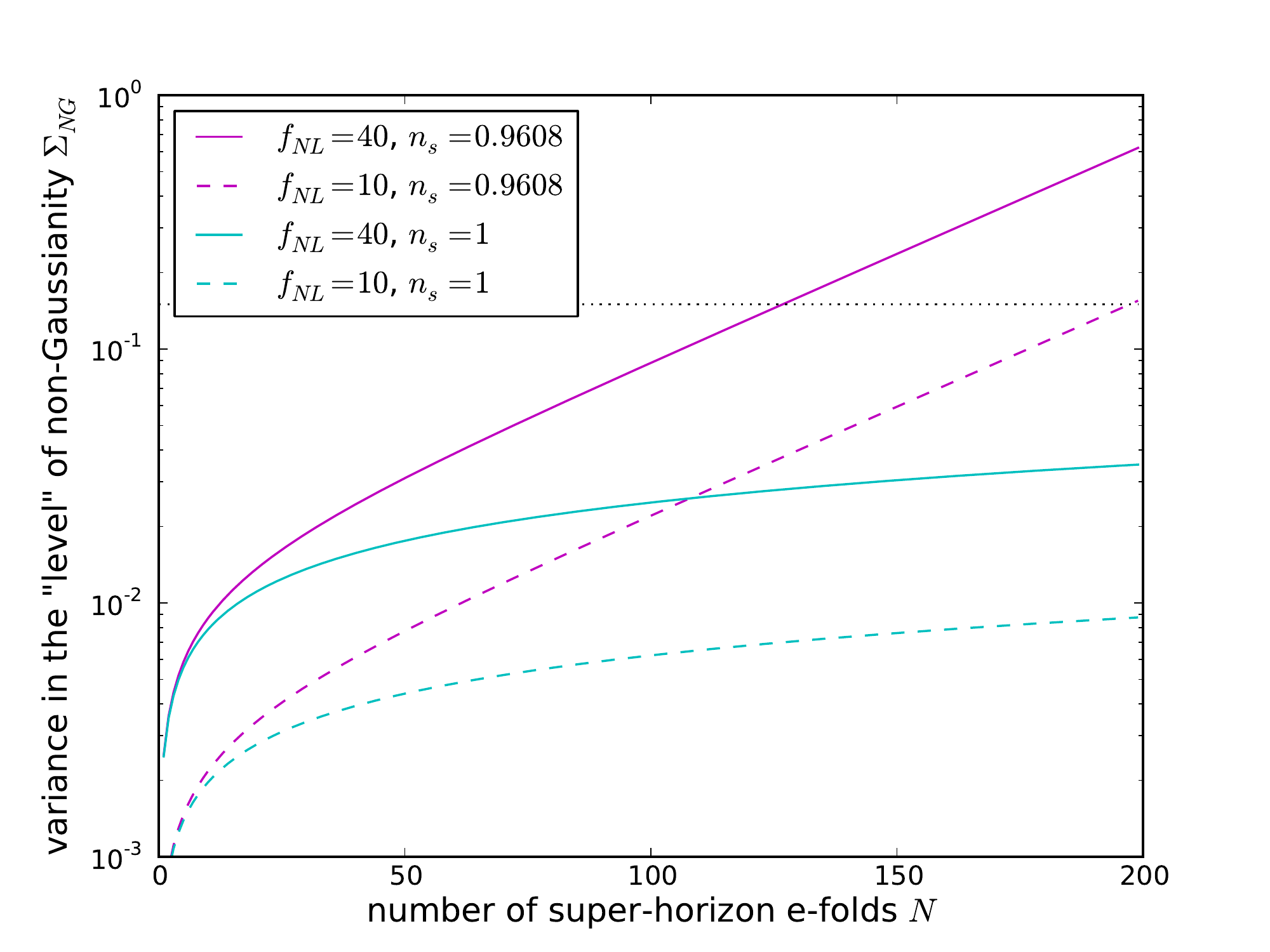} &\includegraphics[width=.5\textwidth]{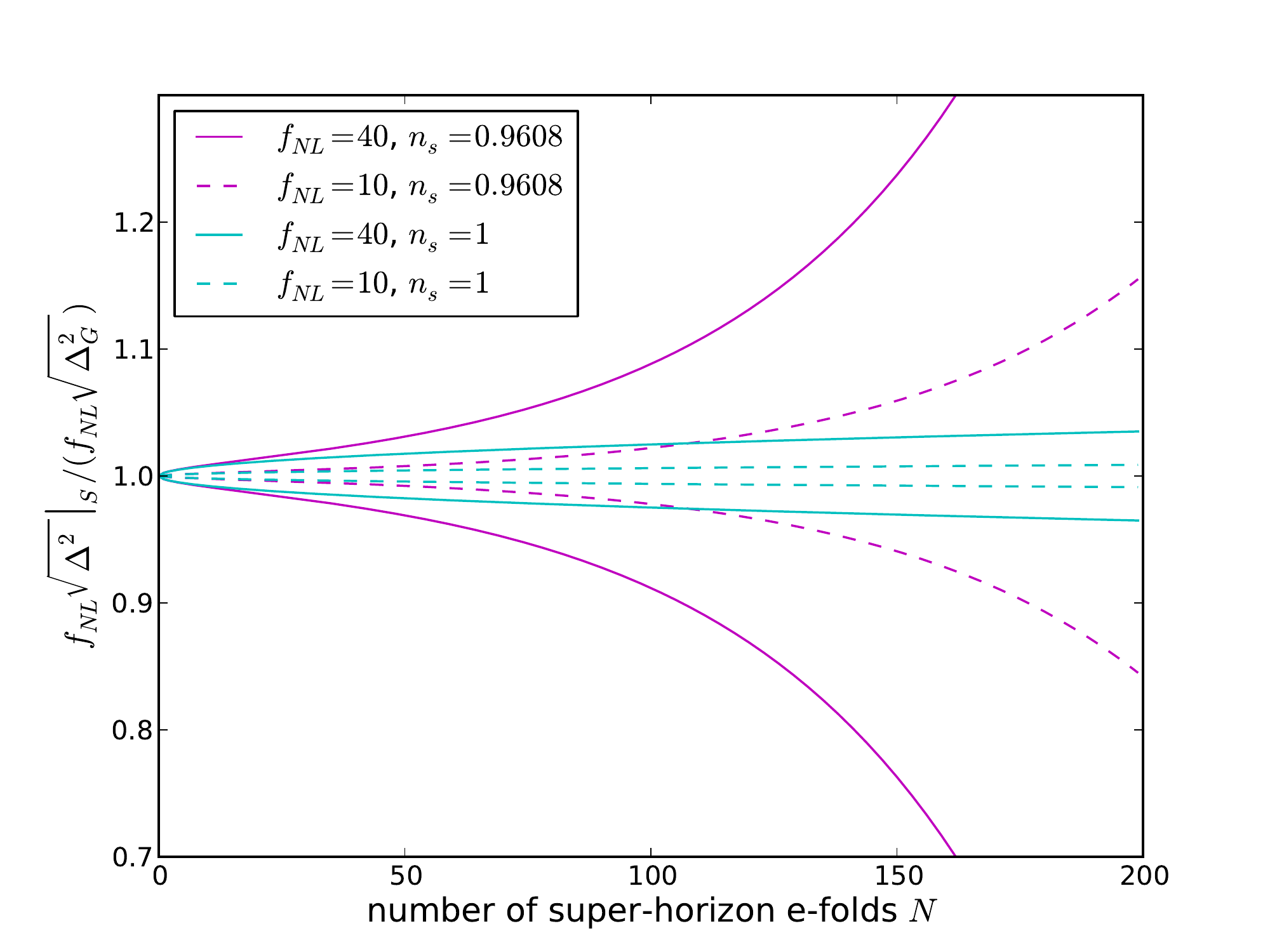}
\end{array}$
\caption{\label{Fig:Var}Left: The degree of variation in non-Gaussianity in subvolumes as quantified by the variance  $\Sigma_{NG}^2\equiv\big\langle\big(\frac{(f_{NL}\sqrt{\Delta^2_G})|_S}{f_{NL}\sqrt{\Delta^2_G}}\big)^2\big\rangle-1=\big(\frac{6}{5}f_{NL}\big)^2\langle\zeta_G^2\rangle\langle B^2\rangle=\big(\frac{6}{5}f_{NL}\big)^2\langle\zeta_{G,L}^2\rangle$. The horizontal, dotted line indicates $\Sigma_{NG}=0.15$, roughly when uncertainty in $f_{NL}$ due to super-horizon correlations becomes comparable to the expected error on $f_{NL}$ from Planck (we assume $\pm 5$), if $f_{NL}=40$ in $V_L$.  Right: The fractional change in the level of non-Gaussianity, $\frac{(f_{NL}\sqrt{\Delta^2})|_S}{f_{NL}\sqrt{\Delta^2}}$, vs. number of super-horizon e-folds $N$. The upper and lower curves correspond to $\zeta_{G,L}=\langle\zeta_{G,L}\rangle^{1/2}$ and $\zeta_{G,L}=-\langle\zeta_{G,L}\rangle^{1/2}$, respectively.  In these figures we've assumed $g_{NL}$, $h_{NL}$\dots $=0$. }
\end{center}
\end{figure}

\section{Example II: Strongly Non-Gaussian Initial Conditions}
\label{sec:strongNG}

\begin{figure}[ht!]
\begin{center}
$\begin{array}{cc}
\includegraphics[width=0.5\textwidth]{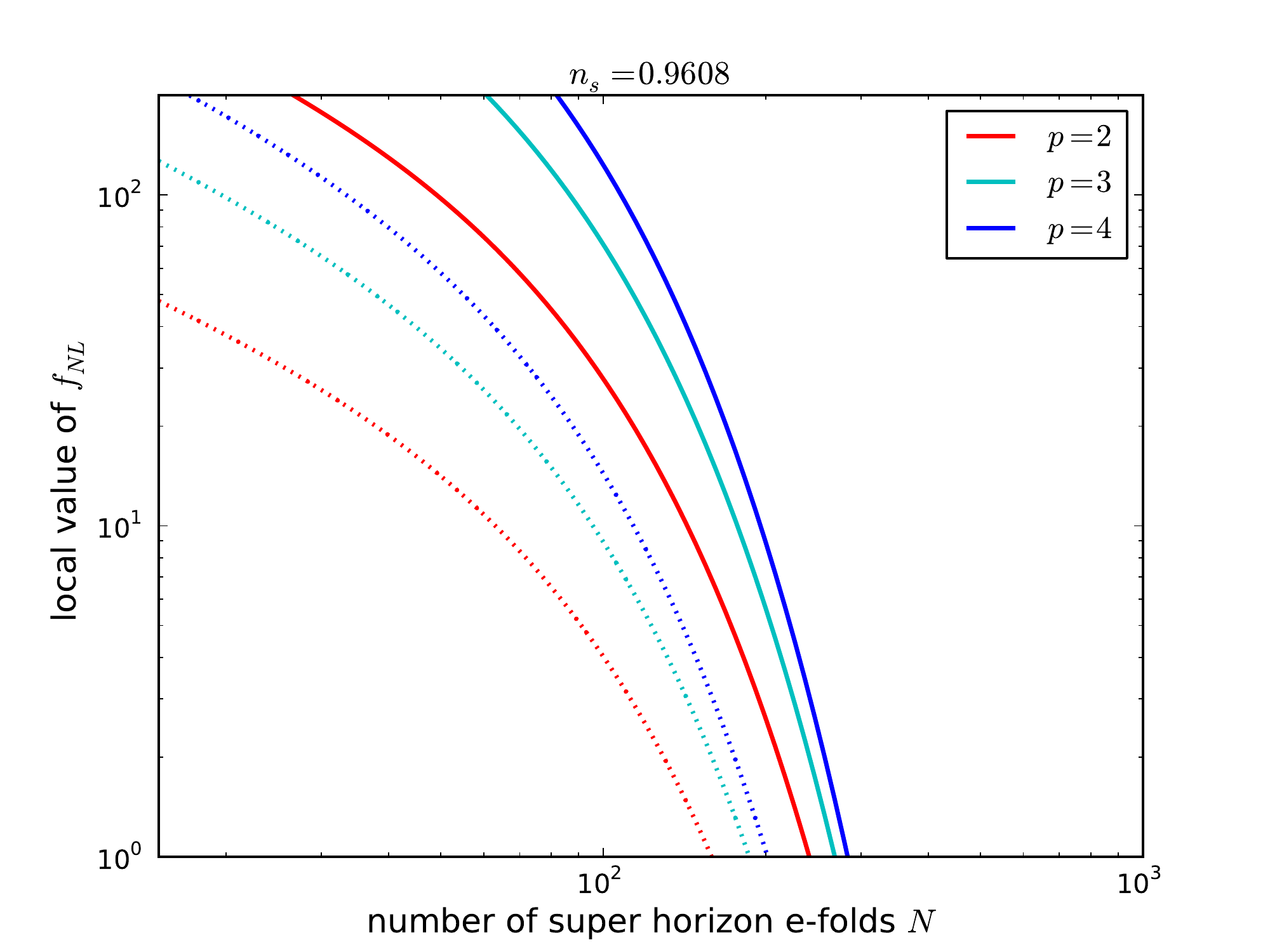} & \includegraphics[width=0.5\textwidth]{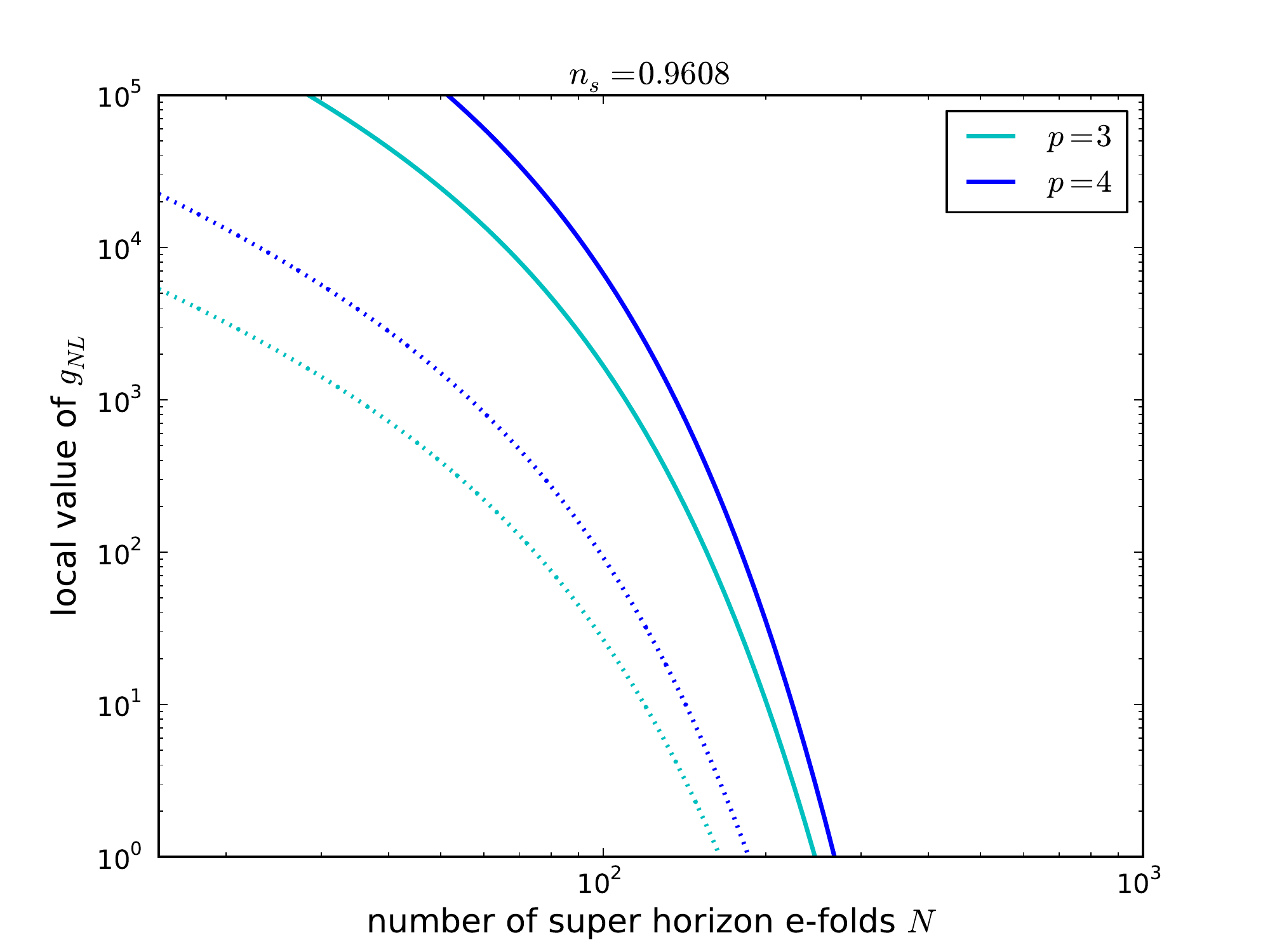}\\
{\bf (a)} &{\bf (b)} \\
\includegraphics[width=0.5\textwidth]{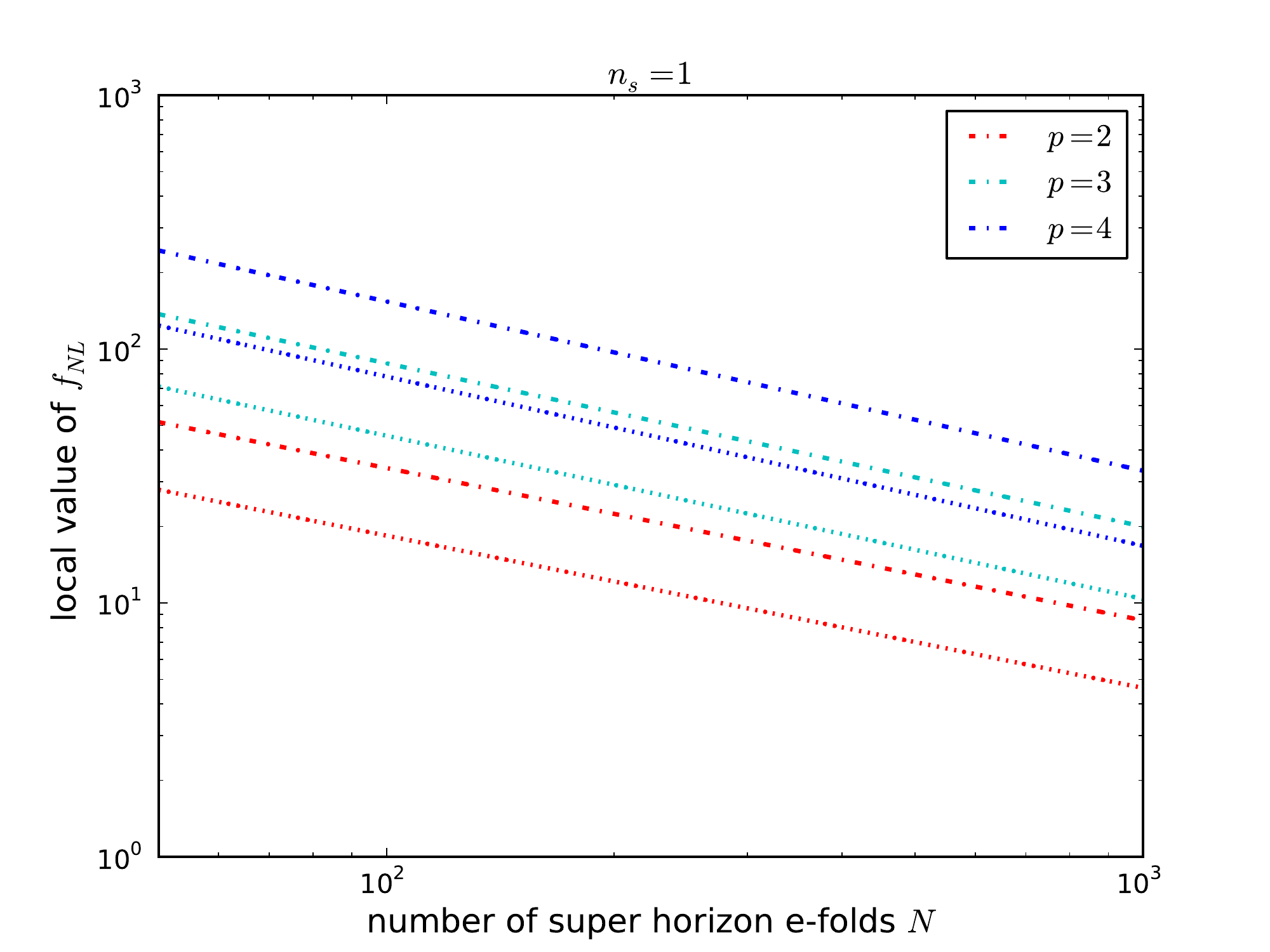} & \includegraphics[width=0.5\textwidth]{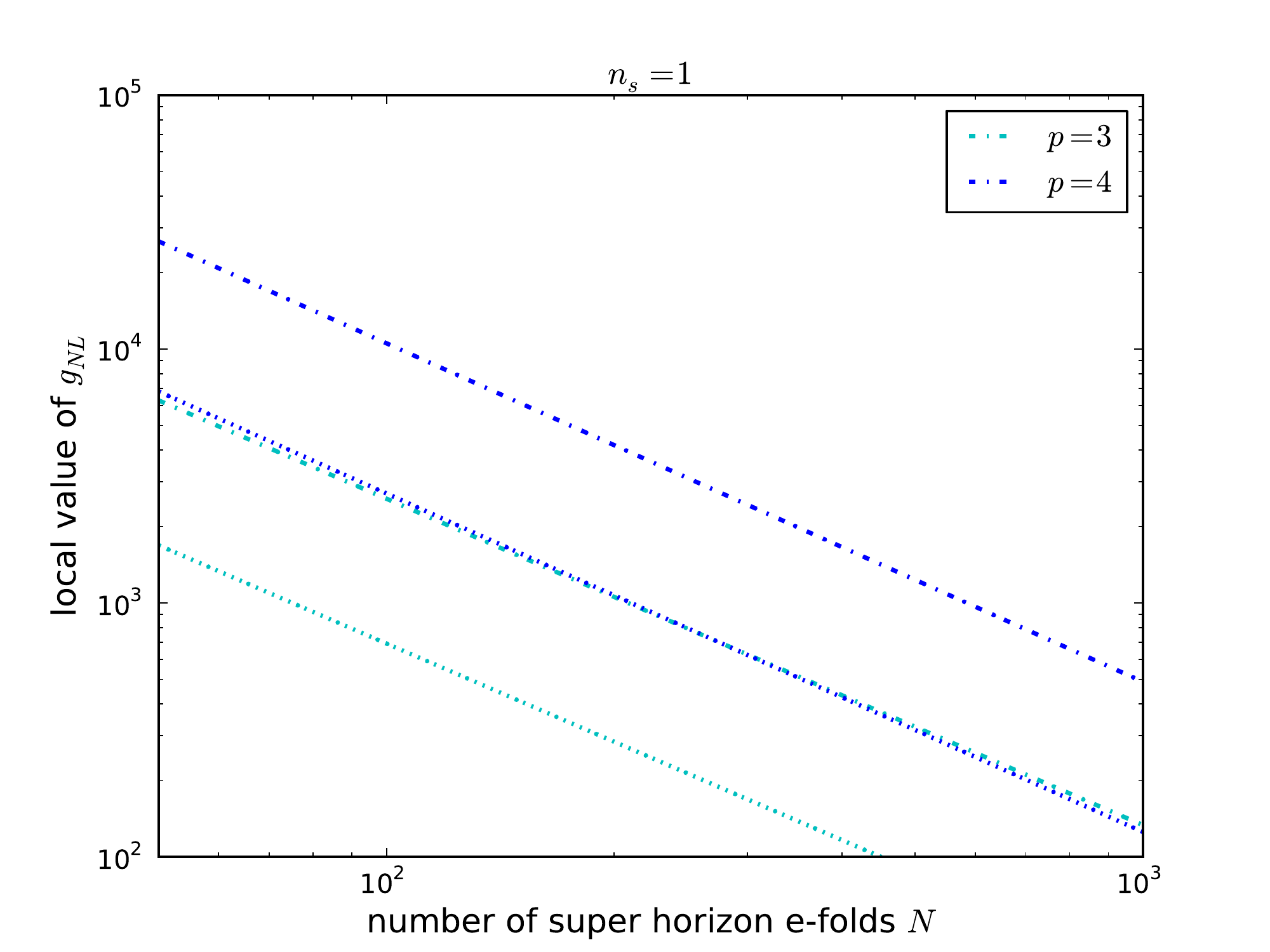}\\
{\bf (c)} &{\bf (d)} \\
\end{array}$ 
\caption{\label{Fig:zetap}  Here we suppose that the curvature perturbation in $V_L$ is given by $\zeta_{NG}(\x)=\zeta_G(\x)^p-\langle\zeta_G^p\rangle$, and ask whether the curvature perturbations in our Hubble patch could appear to be described by the weakly non-Gaussian series (e.g. Eq.~(\ref{eq:localfNLeasy})). We require a large local background fluctuation $\zeta_{G,L} \gg \sqrt{\langle\zeta_{G,S}^2\rangle}$ and consistency with the observed power spectrum $\Delta^2_{WMAP}$. Plotted are the corresponding values of $f_{NL}$ (left panels) and $g_{NL}$ (right panels) for different values of the power law index $p$ and $\zeta_{G,L}$: $\zeta_{G,L}=\sqrt{\langle\zeta_{G,L}^2\rangle}$ (solid lines), $\zeta_{G,L}=3\sqrt{\langle\zeta_{G,L}^2\rangle}$ (dashed lines) and $\zeta_{G,L}=5\sqrt{\langle\zeta_{G,L}^2\rangle}$ (dotted lines).The upper row uses $n_s=0.9608$ for $\Delta^2_G$, the lower row uses $n_s=1$. The bend in the plots for $n_s=0.9608$ occurs at $N(n_s-1)/p \sim 1$.}
\end{center}
\end{figure}

Suppose the non-Gaussian curvature perturbation, $\zeta_{NG}$, in the larger volume $V_L$ is given by
\be
\label{eq:stongNG}
\zeta_{NG}(\x)=\zeta_G^p(\x)-\langle\zeta_G^p\rangle
\ee
where $p$ is a positive integer $>1$. This field has statistics that are not accurately characterized by an expansion of the form Eq.~(\ref{eq:zetafNLgNL}), in particular the polyspectra have a different shape and scale dependence from the local shapes given in Eq.~(\ref{eq:fNLdef}), Eq.~(\ref{eq:gNLdef}), and Eq.~(\ref{eq:tauNLdef}). Nevertheless, in the squeezed limits that observationally define $f_{NL}$, $g_{NL}$, and $\tau_{NL}$ one finds
\be
f_{NL}\sim \frac{1}{\langle\zeta_G^2\rangle^{p/2}}\,,\quad g_{NL}\sim \frac{1}{\langle\zeta_G^2\rangle^p}\,,\quad \tau_{NL} \sim \frac{1}{\langle\zeta_G^2\rangle^p}\qquad {\rm for  }\,\, p\,\,{\rm even} 
\ee
\be
f_{NL}= 0 \,,\quad g_{NL}\sim \frac{1}{\langle\zeta_G^2\rangle^p}\,,\quad \tau_{NL} \sim \frac{1}{\langle\zeta_G^2\rangle^p} \qquad {\rm for  }\,\, p\,\,{\rm odd} \,.
\ee
In contrast to the weakly non-Gaussian case in \S \ref{sec:weakNG}, this field has $1\sim f_{NL}\sqrt{\Delta_{NG}^2}\sim g_{NL}\Delta_{NG}^2 \sim \tau_{NL}\Delta_{NG}^2$, where $\Delta^2_{NG}\sim \langle \zeta_G^{2p}\rangle$ is the observed variance. In general, the $f_{NL}$, $g_{NL}$ and $\tau_{NL}$ will also be scale-dependent functions of $k_s$, $k_l$, the long and short-wavelengths used to take the squeezed limits in Eq.~(\ref{eq:fNLdef}), Eq.~(\ref{eq:gNLdef}), and Eq.~(\ref{eq:tauNLdef}).  For a more thorough discussion of weak and strong local non-Gaussianity, see Appendix \ref{sec:global}.

Consider the local statistics of $\zeta_{NG}$ in a subvolume of size $V_S$. The local non-Gaussian curvature can be written in terms of short and long wavelength modes of $\zeta_G$ as in Eq.~(\ref{eq:zetaGs}), Eq.~(\ref{eq:zetaGl}) as
\be
\label{eq:zetapbinomialexpand}
\left.{\zeta}_{NG}(\x)\right|_S = p\zeta_{G,L}^{p-1}\zeta_{G,S}(\x)+\frac{p!}{2!(p-2)!}\zeta_{G,L}^{p-2}\left(\zeta_{G,S}^2(\x)-\langle \zeta_{G,S}^2\rangle\right)
+\frac{p!}{3!(p-3)!}\zeta_{G,L}^{p-3}\zeta_{G,S}^3+\dots
\ee
where we have suggestively ordered the series with the term linear in $\zeta_{G,S}$ first. Now, if we happen to be considering a small volume $V_S$ with a background fluctuation satisfying
\be
\label{eq:criterion}
\zeta_{G,L}\gg \sqrt{\langle \zeta_{G,S}^2\rangle}\,,
\ee
then to a local observer $\left.\zeta_{NG}\right|_S$ given in Eq.~(\ref{eq:zetapbinomialexpand}) appears to be a field described by a weakly non-Gaussian expansion of the form\footnote{Here we're assuming $p\lsim 10$, say, or small enough that the binomial coefficients $\frac{p!}{k!(p-k)!}$ don't spoil the smallness of the quantity $\frac{p!}{k!(p-k)!}\sqrt{\langle \zeta_{G,s}^2\rangle}/\zeta_{G,l}$ when $\sqrt{\langle \zeta_{G,s}^2\rangle}/\zeta_{G,l}\ll1$ as given in Eq.~(\ref{eq:zetapbinomialexpand}).}
\be
\label{eq:localfNLeasy}
\left.\zeta_{NG}(\x)\right|_S = \chi_{G}(\x)+\frac{3}{5}\left.f_{NL}\right|_S\left(\chi_G^2(\x)-\langle\chi_G^2\rangle\right)+\frac{9}{25}\left.g_{NL}\right|_S\left(\chi_G^3(\x)-3\chi_G(\x)\langle\chi_G^2\rangle\right)+\dots
\ee
where 
\be
\label{eq:chiG}
\chi_G(\x)=p \zeta_{G,L}^{p-1}\zeta_{G,S}(\x)+\mathcal{O}\left(\frac{\langle\zeta_{G,S}^2\rangle}{\zeta_{G,L}^2}\right)\,,
\ee
(the $\mathcal{O}\left(\langle\zeta_{G,S}^2\rangle/\zeta_{G,L}^2\right)$ is because we have subtracted $3\left. g_{NL}\right|_S\chi_G\langle\chi_G^2\rangle$ from the linear term in Eq.~(\ref{eq:localfNLeasy})) and
\be
\label{eq:fNLgNL}
\frac{3}{5}\left.f_{NL}\right|_S = \frac{p-1}{2 p\zeta_{G,L}^p}+\mathcal{O}\left(\frac{\langle\zeta_{G,S}^2\rangle}{\zeta_{G,L}^2}\right)\quad {\rm and}\qquad \frac{9}{25}\left.g_{NL}\right|_S = \frac{(p-1)(p-2)}{3! p^2 \zeta_{G,L}^{2p}}+\mathcal{O}\left(\frac{\langle\zeta_{G,S}^2\rangle}{\zeta_{G,L}^2}\right)\, .
\ee
Now, $\zeta_{G,L} \ll 1$ so the field in Eq.~(\ref{eq:localfNLeasy}) should have {\em large} local non-Gaussianity. However, it is possible for the local statistics to appear only weakly non-Gaussian, i.e.
\be
\left.f_{NL}\right|_S\sqrt{\langle \chi_G^2\rangle}, \left.g_{NL}\right|_S\langle \chi_G^2\rangle, \dots \ll1
\ee
on top of sufficiently large background fluctuations. Taking $V_S$ to be our Hubble volume and assuming that $\zeta_G$ has a power-law spectrum with constant spectral index as in Eq.~(\ref{eq:LWVar}) the criterion given in Eq.~(\ref{eq:criterion})  for observing weak non-Gaussianity can be written
\ba
\label{eq:zetaLreqns1}
\frac{\zeta_{G,L}}{\sqrt{\langle \zeta_{G,L}^2\rangle}} &\gg&\sqrt{\frac{N_{S}}{N}}\quad {\rm for}\quad n_s=1\\
\label{eq:zetaLreqns}
\frac{\zeta_{G,L}}{\sqrt{\langle \zeta_{G,L}^2\rangle}} &\gg& \sqrt{\frac{e^{(n_s-1)N_{S}}-1}{1-e^{-(n_s-1)N} }}\quad {\rm for}\quad n_s={\rm const.}\neq 1
\ea
where as, before $N$ is the number of super-horizon e-folds and we have introduced $N_{S}$,  the number of sub-horizon e-folds. For reference, $N_{S}\sim 60$ gives $\sqrt{\frac{e^{(n_s-1)N_{S}}-1}{1-e^{-(n_s-1)N} }}\sim 1$ for $N \sim 15$ and $\sqrt{\frac{e^{(n_s-1)N_{S}}-1}{1-e^{-(n_s-1)N} }}\sim 0.05$ by $N\sim 150$ when $n_s=.9608$. 

Equations (\ref{eq:zetaLreqns1}),  (\ref{eq:zetaLreqns})  show that even for strongly non-Gaussian statistics of $\zeta_{NG}$ in $V_L$, the statistics in $V_S$ appear weakly Gaussian on top of very rare background fluctuations ($\zeta_{G,L}/\langle \zeta_{G,L}^2\rangle^{1/2}\gsim 1$). But for very large $N$ the statistics in $V_S$ appear weakly Gaussian even for typical values of $\zeta_{G,L}$.  In regions where Eq.~(\ref{eq:criterion}) is satisfied, the possible values of $\left.f_{NL}\right|_S$ in subvolumes depends qualitatively on the sign of $p$: for even $p$, $\left.f_{NL}\right|_S>0$ in all subvolumes, whereas for odd $p$ the sign of $\zeta_{G,L}^p$ is significant and $\left.f_{NL}\right|_S$ can be negative. On the other hand, $\left.g_{NL}\right|_S>0$ for all values of $p$.

Now we ask what the restrictions on $p$, $\Delta_G^2$, and $\zeta_{G,L}$ are in order to generate a curvature perturbation as in Eq.~(\ref{eq:localfNLeasy}) that satisfies the observational constraints on the power spectrum, $f_{NL}$ and $g_{NL}$ in our Hubble volume. If we fix the ratio $\zeta_{G,L}/\sqrt{\langle \zeta_{G,L}^2\rangle}$ (which is a measure of the rarity of our Hubble patch), the index $p$, and the observed level of fluctuations $\Delta_\chi^2=\Delta_{WMAP}^2$, Eq.~(\ref{eq:chiG}) and Eq.~(\ref{eq:fNLgNL}) allows us to solve for the variance of fluctuations in the (unobservable) background field $\zeta_G$ along with the observed values of $f_{NL}$ and $g_{NL}$ as a function of $N$. The results are plotted in Figure \ref{Fig:zetap}. We see that current constraints on the observed level of non-Gaussianity are indeed compatible with a scenario in which our Hubble patch is a biased subsample of a larger universe with strongly non-Gaussian initial curvature perturbations $\zeta_{NG}(\x)=\zeta_G^p(\x)-\langle\zeta_G^p\rangle$. 

\section{Example III: Two-field Initial Conditions}
\label{sec:twofieldNG}

In this section we consider initial conditions inspired by a version of the curvaton model \cite{Mollerach:1989hu,Linde:1996gt,Enqvist:2001zp,Lyth:2001nq,Lyth:2002my,Sasaki:2006kq} in which perturbations from both the inflaton $\phi$ and the curvaton $\sigma$ are responsible for generating $\zeta$ (see e.g. \cite{Byrnes:2008zy,Tseliakhovich:2010kf,Shandera:2010ei,Smith:2010gx}).  In this ``inflaton-curvaton" scenario,  the curvature perturbation in the larger volume is given by
\be
\label{eq:phisigma}
\zeta_{NG}(\x)=\phi_G(\x)+\sigma_G(\x)+\frac{3}{5}\tilde{f}_{NL}(\sigma_G^2(\x)-\langle\sigma_G^2\rangle)\,
\ee
We make the simplifying assumption that $\phi_G(\x)$ and $\sigma_G(\x)$ are statistically independent  (i.e. $\langle \phi(\k)\sigma(\k')\rangle =0$), Gaussian random fields with proportional power spectra
\be
\label{eq:xidef}
\xi^2\equiv \frac{P_\phi}{P_\sigma} \quad {\rm so\,\, that}\quad P_{NG}(k) = P_\sigma(k)\left(1+\xi^2+\frac{18}{25}\tilde{f}_{NL}^2I_\sigma(k)\right)
\ee
where $I_\sigma(k)\sim \Delta_\sigma^2$ is defined in Eq.~(\ref{eq:I1def}) and for simplicity we assume $\xi$ is a constant (however, see e.g. \cite{Byrnes:2009pe,Shandera:2010ei, Bramante:2013}). 

\subsection{Case 1: $\sigma$ Is Weakly Non-Gaussian}
\label{ssec:weaktwofieldNG}
First, we make the usual assumption that the curvaton contributions to the curvature perturbation are only weakly non-Gaussian. That is, we assume that $\tilde{f}_{NL}\sqrt{\Delta_\sigma^2} \ll 1$. The non-Gaussian parameters that characterize the bispectrum and trispectrum of $\zeta$ are 
\be
\label{eq:fNLtauNLweaktwofield}
f_{NL} = \frac{\tilde{f}_{NL}}{(1+\xi^2)^2} \,,\quad \tau_{NL} = \frac{\left(\frac{6}{5}\tilde{f}_{NL}\right)^2}{(1+\xi^2)^3}
\ee
and $g_{NL}=0$. In a subvolume, $V_S$ a local observer will see statistics described by
\ba
\label{eq:localPweaktwofield}
\left.P_{NG}\right|_{S} &=& P_{NG}\left(1+\frac{12}{5}\frac{\tilde{f}_{NL}}{1+\xi^2}\sigma_{G,L}\right)\\
\left.f_{NL}\right|_{S} &=& \frac{\tilde{f}_{NL}}{(1+\xi^2)^2}\left(1+\frac{12}{5}\frac{\xi^2-1}{\xi^2+1}\tilde{f}_{NL}\sigma_{G,L}\right)\\
\left.g_{NL}\right|_{S}&=&0\\
\label{eq:localtauweaktwofield}
\left.\tau_{NL}\right|_{S}&=&\frac{\left(\frac{6}{5}\tilde{f}_{NL}\right)^2}{(1+\xi^2)^3}\left(1+\frac{12}{5}\frac{\xi^2-2}{\xi^2+1}\tilde{f}_{NL}\sigma_{G,L}\right)\,.
\ea
In contrast to the case in \S \ref{sec:weakNG}, the local statistics are now modulated by long-wavelength modes of $\sigma_{G}$ {\em only}, as opposed to fluctuations in the total curvature fluctuation $\zeta_{G,L}=\phi_{G,L}+\sigma_{G,L}$. To compare with \S \ref{sec:weakNG}, we rewrite Eq.~(\ref{eq:localPweaktwofield}) - Eq.~(\ref{eq:localtauweaktwofield}) in terms of $f_{NL}$, $\tau_{NL}$ in $V_L$,
\ba
\left.P_{NG}\right|_{S} &=& P_{NG}\left(1+\frac{12}{5}f_{NL}\left(1+\xi^2\right)\sigma_{G,L}\right)\\
\left.f_{NL}\right|_{S} &=&f_{NL}\left(1+\frac{12}{5}\left(\frac{\left(\frac{5}{6}\right)^2\tau_{NL}-2f_{NL}^2}{f_{NL}}\right)\left(1+\xi^2\right)\sigma_{G,L}\right)
\ea
Now, the amount by which the power spectrum and $f_{NL}$ vary from place to place is the same as in Eq.~(\ref{eq:weaklocalP}) and Eq.~(\ref{eq:weaklocalfNL}) with $\zeta_{G,L}\rightarrow (1+\xi^2)\sigma_{G,L}$. The typical size of the modulation of statistics in $V_S$ is $\sqrt{\langle (1+\xi^2)^2\sigma_{G,L}^2\rangle} = \sqrt{(1+\xi^2)\langle \zeta_{G,L}^2\rangle}$. So for fixed $f_{NL}$, the typical modulation in the power spectrum is larger relative to the case where a single field, $\zeta_G$, generates density perturbations. That is, a one-sigma fluctuation in $\sigma_{G,L}$ generates a larger change than a one-sigma fluctuation in $\zeta_{G,L}$, larger by a factor $\sqrt{1+\xi^2}$. The $\sigma$ field itself must be more non-Gaussian to maintain a fixed $f_{NL}$ in the curvature as the power from $\sigma$ decreases ($\xi$ increases). In Figure \ref{Fig:probstauNL} we plot estimates for the distributions of $\Delta^2_\zeta$ and $f_{NL}$ in the total volume, given several values of $\xi^2$. As in \S \ref{sec:weakNG} these are estimates of the probability distributions in that we (i) neglect terms $\mathcal{O}(\zeta_{G,L}^2)$ in relating values of parameters measured in our Hubble volume to those in the larger universe $V_L$ and (ii) we don't allow the observed parameters ($\left.\Delta^2_G\right|_S$, $\left.f_{NL}\right|_S$) to all vary simultaneously. We have again fixed the number of e-folds for illustrative purposes even though this is also an unobservable quantity.

Note, that while the observed amplitude of the three-point and four-point functions are characterized by $f_{NL}\sqrt{\Delta_\zeta^2}$, $\tau_{NL}\Delta_\zeta^2$, when one considers the entire series of correlation functions neither product alone quantifies the level of non-Gaussianity in the field $\zeta_{NG}$. The single quantity that controls the level of non-Gaussianity is $\frac{\tilde{f}_{NL}\Delta_{\zeta}}{1+\xi^2}$ (rather than $f_{NL}\Delta_{\zeta}=\frac{\tilde{f}_{NL}\Delta_{\zeta}}{(1+\xi^2)^2}$). When this quantity is small, the series of cumulants is ordered and the amplitude of each consecutive cumulant is smaller by this factor. Each cumulant also has an extra factor of $1/(1+\xi^2)$ which does not affect their relative importance. In terms of the observed non-Gaussian parameters given in Eq.~(\ref{eq:fNLtauNLweaktwofield}), this criterion for weak non-Gaussianity, $\frac{\tilde{f}_{NL}\Delta_{\zeta}}{1+\xi^2}\ll 1$, is equivalent to requiring the kurtosis to be much smaller than the skewness: $ \tau_{NL}\Delta^2_\zeta\ll f_{NL}\sqrt{\Delta_\zeta^2}$.  As in the single field case, we can ask how the total amplitude of non-Gaussianity differs in biased subvolumes:
\be
\left.\frac{\tilde{f}_{NL}\sqrt{\Delta^2_{\zeta}}}{1+\xi^2}\right|_S=\frac{\tilde{f}_{NL}\sqrt{\Delta^2_{\zeta}}}{1+\xi^2}\left[1-\frac{6}{5}\frac{\tilde{f}_{NL}\langle\zeta_G^2\rangle^{1/2}}{1+\xi^2}B\right]
\ee
where the bias here is defined as
\be
B=\frac{\sigma_{G,L}}{\langle\zeta_G^2\rangle^{1/2}}\;.
\ee
The relationship between the amplitude of non-Gaussianity in $V_S$ and $V_L$ has the same structure as in the single field case, but the bias will generally be smaller (assuming the same total amplitude of fluctuations, $\sqrt{\Delta^2_\zeta}$) since the fluctuating field contributes only part of the total power. Notice that when $\xi=0$ this reduces to the single field expression, Eq.(\ref{fP}).

\begin{figure}
\begin{center}
$\begin{array}{cc}
\includegraphics[width=0.5\textwidth]{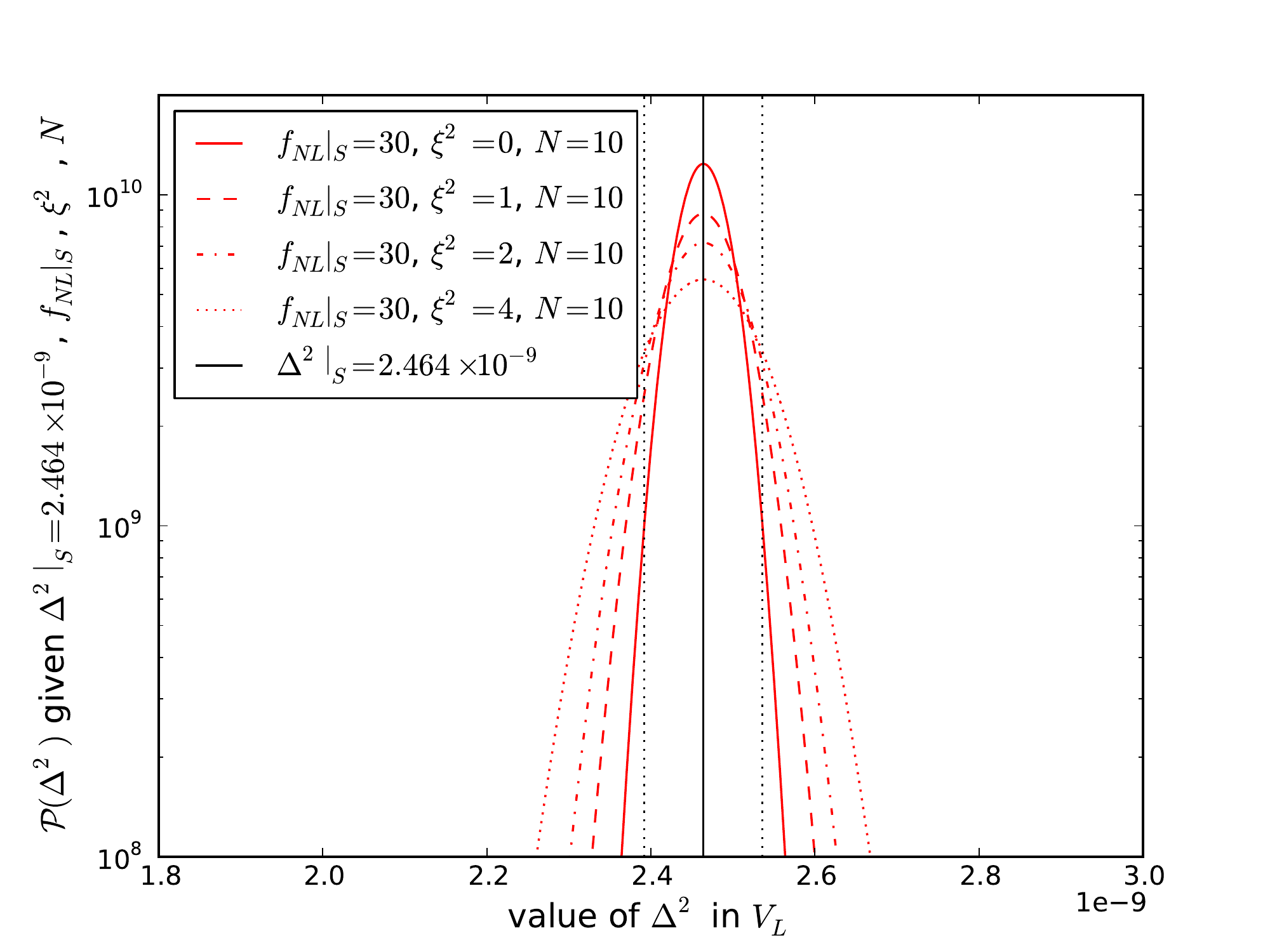}  &\includegraphics[width=0.5\textwidth]{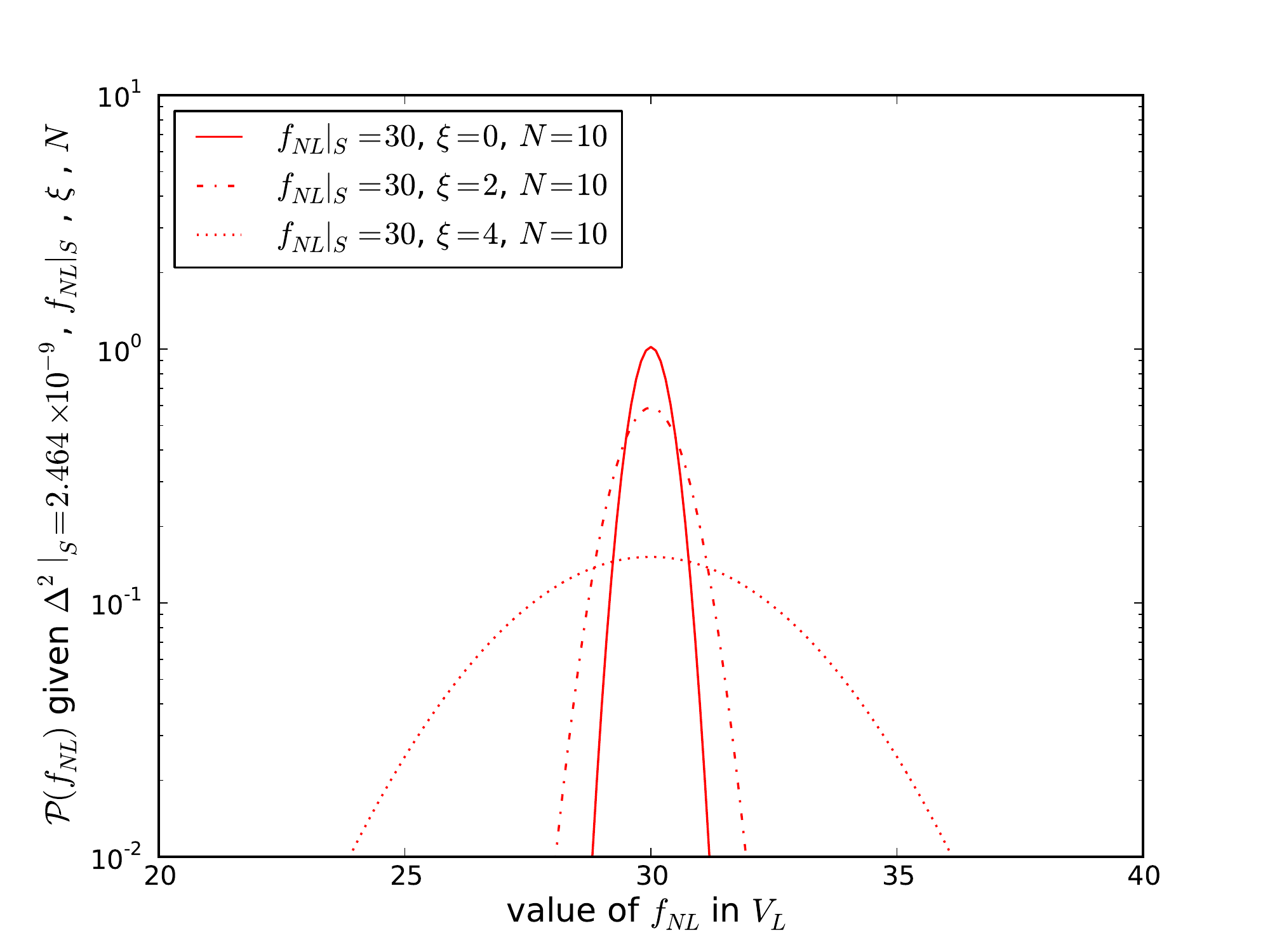} 
\end{array}$
\caption{\label{Fig:probstauNL} In the two-field ``weak non-Gaussianity" case the amplitude of scalar perturbations, $\left.\Delta^2\right|_S$ and of $f_{NL}$ in our Hubble volume, $\left.f_{NL}\right|_S$, differ from the average value due to the background value of $\sigma_{G,L}$, while the total curvature is set by $\sigma_{G,L}+\phi_{G,L}$. For fixed $f_{NL}$, changes to the local statistics are typically larger by a factor of $\sqrt{1+\xi^2}$ relative to the case in \S\ref{sec:weakNG}.  Plotted are estimates of the probability distributions for $\Delta^2$ (left panel) and $f_{NL}$ (right panel) in $V_L$, given the observed values in $V_S$ for different values of $\xi^2\equiv P_\phi/P_\sigma$. Note that for $\xi^2=1$, $\left.f_{NL}\right|_S =f_{NL}+\mathcal{O}(\sigma_{G,L}^2)$. }\end{center}
\end{figure}

\subsection{Case 2: $\sigma$ Is Strongly Non-Gaussian}
\label{ssec:strongtwofieldNG}
Now we assume the perturbations coming from the curvaton are strongly non-Gaussian $\tilde{f}_{NL}\sqrt{\Delta_\sigma^2}\sim 1$, but a subdominant contribution to the total curvature ($\xi^2\gg\tilde{f}_{NL}^2\Delta_\sigma^2\sim 1$). To understand the dependencies, it's helpful to define the $\mathcal{O}(\tilde{f}_{NL}^2)$ fractional change in the globally defined power spectrum in $V_L$
\be
\label{eq:epsilon}
\tilde{\epsilon}(k)\equiv 2\frac{\left(\frac{3}{5}\tilde{f}_{NL}\right)^2I_\sigma(k)}{\xi^2} 
\ee
By assumption $\tilde{\epsilon}\sim 1/\xi^2$ and again $I_\sigma\sim \Delta^2_\sigma$ as defined in Eq.~(\ref{eq:I1def}). In the larger box $V_L$ the power spectrum is
\be
P_{NG}(k)=\xi^2P_\sigma(k)\left(1+\tilde{\epsilon}(k)\right)\,.
\ee
Taking the squeezed and squashed limits in Eq.~(\ref{eq:fNLdef}), Eq.~(\ref{eq:gNLdef}), Eq.~(\ref{eq:tauNLdef}) gives scale-dependent non-Gaussian parameters:
\be
f_{NL}(k_l)=\frac{\tilde{f}_{NL}}{\xi^2}\tilde{\epsilon}(k_l)\,,\quad g_{NL}=0\,,\quad \tau_{NL}(k_l)= \left(\frac{6}{5}f_{NL}(k_l)\right)^2\frac{1}{\tilde{\epsilon}} 
\ee
where $k_l$ is the magnitude of the long-wavelength mode used to calculate the squeezed and squashed limits. The scale dependence of $f_{NL}$, $\tau_{NL}$ is given by the function $I_\sigma(k)$ in Eq.~(\ref{eq:I1def}). For  $k\gg \Lambda$,  where $\Lambda$ is the infrared cutoff in $\Delta^2(k)$, the scale dependence of $I_\sigma(k)$ is generally weak: for $n_s=1$,  $I_\sigma (k)\sim 2\Delta_\sigma^2\ln(k/\Lambda)$. 

In this example, the field $\zeta$ is weakly Gaussian with hierarchical cumulants in that $1 \gg f_{NL}\sqrt{\langle\zeta_G^2\rangle}\gg \tau_{NL}\langle\zeta_G^2\rangle$. However, in contrast to the weakly non-Gaussian, single-source case in \S\ref{sec:weakNG} where the cumulants scale as $\langle\zeta_{NG}^{n+1}\rangle_c/\langle\zeta_{NG}^n\rangle_c\sim \langle \zeta_{NG}^2\rangle$, the hierarchy of cumulants in this example scales as 
\be
\frac{\langle\zeta_{NG}^{n+1}\rangle_c}{\langle \zeta_{NG}^n\rangle_c}\sim (\tilde{f}_{NL}\sqrt{\langle\sigma_G^2\rangle} )\frac{\langle\zeta_{NG}^2\rangle^{1/2}}{\xi}\sim \frac{\langle\zeta_{NG}^2\rangle^{1/2}}{\xi}\,.
\ee
We have assumed that $\xi^2>>1$, but depending on the relative magnitudes of $\xi$ and $\langle \zeta_{NG}^2\rangle$,  the higher-order cumulants may be more important relative to the lower order ones than in the examples considered in \S \ref{ssec:weaktwofieldNG} and \S\ref{sec:weakNG}. 

In a subvolume $V_S$, an observer will see a local power spectrum
\be
\left.P_{NG}(k)\right|_{S} = \xi^2P_\sigma(k)\left(1+\tilde{\epsilon}_s(k)\left(1+\frac{2\sigma_{G,L}^2}{I_{\sigma_s}(k)}\right)\right)
\ee
and non-Gaussian parameters 
\be
\left.f_{NL}\right|_S= \frac{\tilde{f}_{NL}}{\xi^2}\tilde{\epsilon}_s\left(1+\frac{2\sigma_{G,L}^2}{I_{\sigma_s}(k_l)}\right)\,,\quad \left.g_{NL}\right|_S=0\,,\quad \tau_{NL} = \frac{\left(\frac{6}{5}\left.f_{NL}\right|_S\right)^2}{\tilde{\epsilon}_s\left(1+\frac{2\sigma_{G,L}^2}{I_{\sigma_s}(k)}\right)}
\ee
where $I_{\sigma_s}$ is Eq.~(\ref{eq:I1def}) with $P(k)\rightarrow P_\sigma(k)|1-W_S(k)|^2$ and $\tilde{\epsilon}_s$ is Eq.~(\ref{eq:epsilon}) with $I_\sigma \rightarrow I_{\sigma_s}$. In this case the difference between the local and global statistics is more complicated: since $I_\sigma\neq I_{\sigma_s}$ the local statistics in $V_S$ differ from those in $V_L$ even if $\sigma_{G,L}=0$. Averaging over $\sigma_{G,L}^2$ will recover the parameters in $V_L$.\footnote{The average of the small-volume polyspectra over the long wavelength modes must recover the large-volume polyspectra. However, since the parameters $f_{NL}$, $g_{NL}$ etc. are ratios of quantities dependent on the random variable $\zeta_{G,L}$, when terms non-linear in $\zeta_{G,L}$ are important the relationship between $\left.f_{NL}\right|_S$ and $f_{NL}$, say, is generally more complicated. In this example the non-Gaussian polyspectra are dependent on $\sigma$, but power spectrum is dominated by the Gaussian field $\phi$, and therefore averaging over $\sigma_{G,L}$ doesn't change the denominator in the ratios used to define the non-Gaussian parameters, and the expressions for $f_{NL}$ and $\tau_{NL}$ are easily recovered from $\left.f_{NL}\right|_S$ and $\left.\tau_{NL}\right|_S$ by averaging over $\sigma_{G,L}^2$. } To see this, note that $P_\sigma I_\sigma \sim I_{\sigma_s}P_{\sigma_s}+2P_{\sigma_s}\langle{\sigma_L}^2\rangle+I_{\sigma_L}P_{\sigma_L}$ so that $I_\sigma(k)\rightarrow I_{\sigma_s}(k)+2\langle{\sigma_L}^2\rangle$ for scales $k \gsim V_S^{1/3}$, so that $\langle\tilde{\epsilon}_s(1+2\sigma_{G,L}^2/I_{\sigma_s})\rangle\rightarrow \tilde{\epsilon}$.

\begin{figure}
\begin{center}
$\begin{array}{cc}
\includegraphics[width=0.5\textwidth]{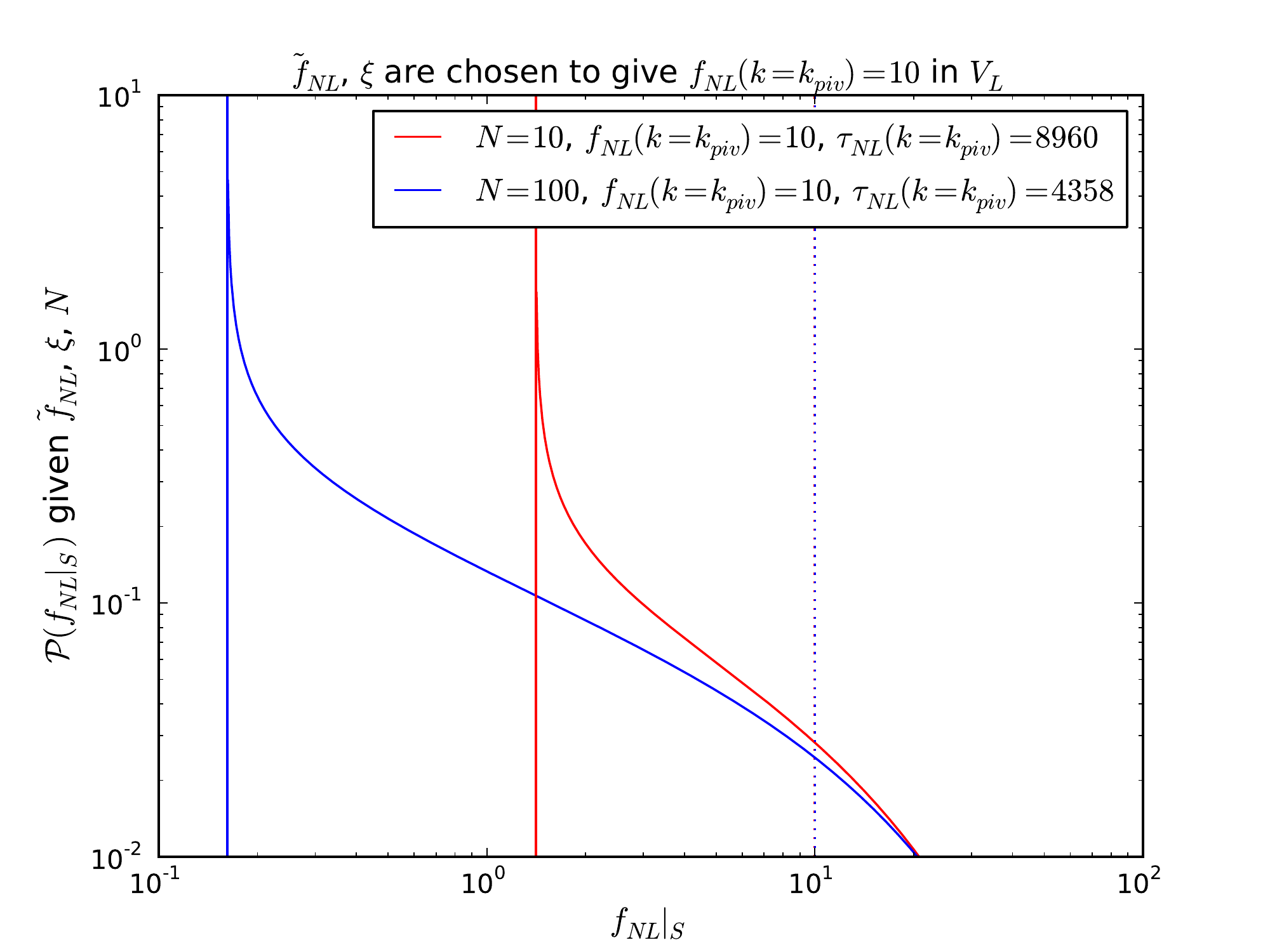}  &\includegraphics[width=0.5\textwidth]{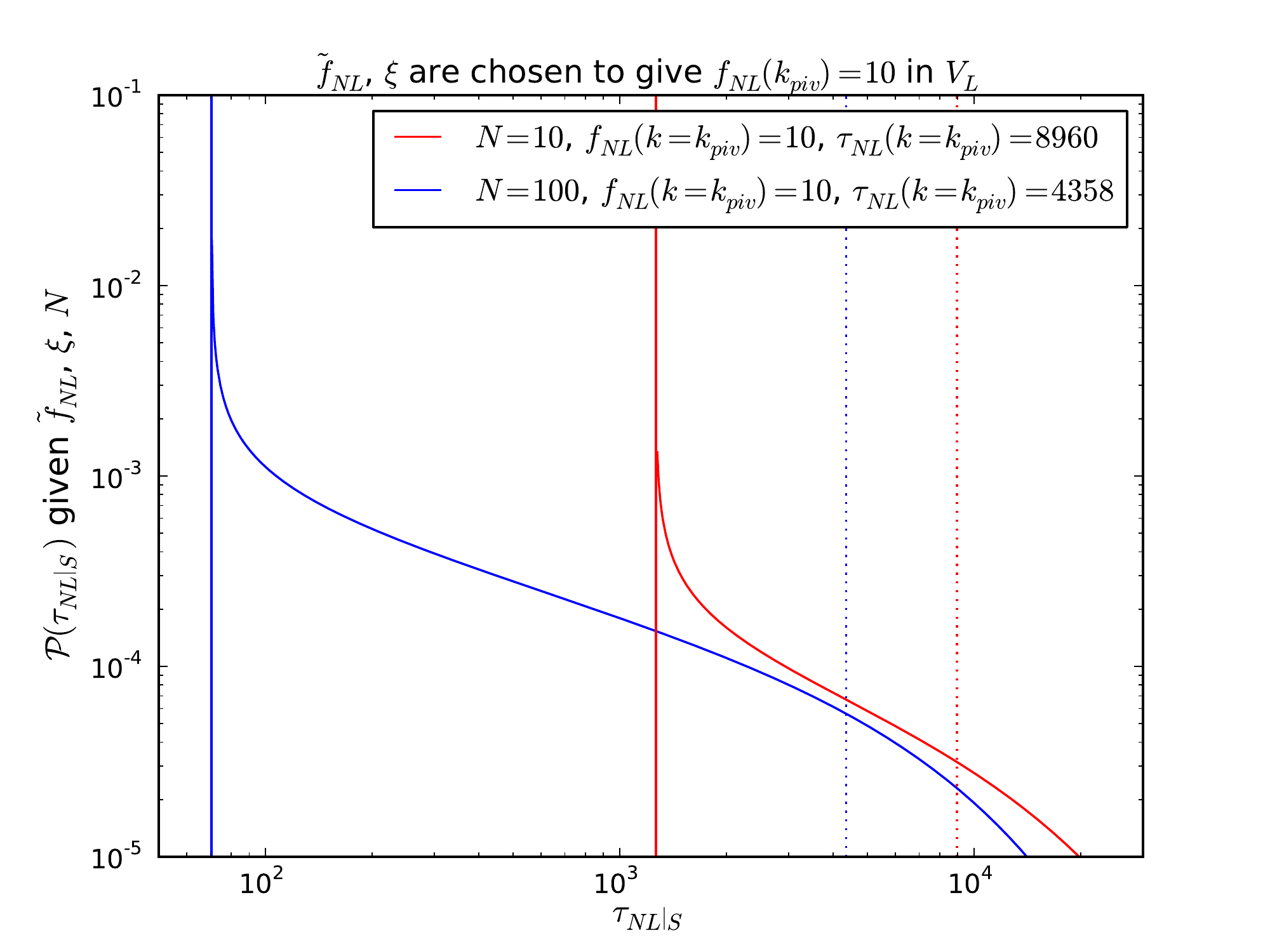} 
\end{array}$
\caption{\label{Fig:probstrongtwofield} In the two-field ``strong non-Gaussianity" case $\sigma_{G,L}^2$ modulates the local statistics leading to skewed probability distributions for the observed non-Gaussian parameters. The solid curves are the distributions for the observed values of $\left.f_{NL}\right|_S$ (left panel) and $\left.\tau_{NL}\right|_S$ (right panel) for different values of $N$ -- the number of super-horizon e-folds (in this plot we set $n_s=1$). In each case we've fixed $(\frac{3}{5}\tilde{f}_{NL})^2\Delta_\sigma^2=1$ and chosen the values of $\tilde{f}_{NL}$ and $\xi$ (see Eq.~(\ref{eq:phisigma}) - Eq.~(\ref{eq:xidef})) to produce $f_{NL} (k=0.002/Mpc)=10$ when averaged throughout $V_L$ (however, the corresponding $\tau_{NL}$ values are different). The values of $f_{NL}$ and $\tau_{NL}$ when averaged over $V_L$ are indicated by the vertical dotted lines. }\end{center}
\end{figure}

To study the statistics in subvolumes $V_S$,  consider the following example: fix $(\frac{3}{5}\tilde{f}_{NL})^2\Delta_\sigma^2 =1$, then we find expressions for the locally measured non-Gaussian parameters in terms of $\xi$ and the observed power spectrum $\Delta^2_{WMAP}$, 
\be
\label{eq:strongtwofieldexample}
\tilde{\epsilon}\rightarrow \frac{2}{\xi^2}\frac{I_\sigma(k)}{\Delta^2_\sigma}\,, \quad f_{NL}\rightarrow \pm \frac{10}{3\sqrt{\Delta^2_{WMAP}}}\frac{1}{\xi^3}\frac{I_\sigma(k)}{\Delta^2_\sigma}, \quad \tau_{NL}\, \rightarrow \left(\frac{6}{5}f_{NL}\right)^2\frac{\xi^2}{2}\frac{\Delta^2_\sigma}{I_\sigma(k)}\,.
\ee
For simplicity we'll assume that $n_s=1$ so that we can use the analytic expression for $I_\sigma(k)$ given in Eq.~(\ref{eq:I1scaleinv}). Eq.~(\ref{eq:strongtwofieldexample}) allows us to choose values of the ratio of inflaton to curvaton power that give particular values of $f_{NL}$, $\tau_{NL}$. We now choose $\xi$ such that $f_{NL}(k_{piv}) = 10$ when averaged over the entire volume $V_L$.  However, since the local value of $f_{NL}$ depends on $\sigma_{G,L}^2$, observers in a finite volume can easily measure $f_{NL}(k_{piv})\neq 10$.  In  Figure \ref{Fig:probstrongtwofield}, we plot the probability distribution of possible observed values of $\left.f_{NL}\right|_S$ and $\left.\tau_{NL}\right|_S$. We see that unlike the cases considered in \S \ref{sec:weakNG}, \S \ref{sec:strongNG}, and \S\ref{ssec:strongtwofieldNG} the probability distributions are extremely skewed and there is a large offset between the median and modes of the distribution of values of $\left.f_{NL}\right|_S$ and $\left.\tau_{NL}\right|_S$. 

\section{Conclusions}
\label{sec:conclusions}

Local type non-Gaussianity couples the small-scale statistics measured by an observer restricted to a small volume $V_S$ to the unobservable, long wavelength modes $\zeta_{G,L}$ that are nearly constant across $V_S$. In this paper we have systematically calculated the relationship between local and global statistical quantities (the power spectrum, bispectrum, and trispectrum) in models with local-type primordial non-Gaussianity. We demonstrate through explicit calculation, that broad classes of statistical distributions for the curvature perturbation in the larger universe $V_L$ are consistent with nearly Gaussian statistics  observed in our Hubble volume. This many-to-one nature of the mapping between statistics in $V_L$ and $V_S$ is potentially a challenge for using statistics measured in our Hubble volume to infer the statistics in the entire universe. The framework outlined in \S \ref{sec:local} and Appendix \ref{sec:Npointsglobal} is general, but we study three examples in detail: the usual local ansatz in \S \ref{sec:weakNG}, an example with strongly non-Gaussian initial conditions coming from a single field in \S \ref{sec:strongNG}, and finally a two-field example in \S \ref{sec:twofieldNG}. 

For the weakly non-Gaussian statistics for $\zeta_{NG}$  in \S \ref{sec:weakNG} we find, in agreement with \cite{Nurmi:2013xv}, that values of the non-Gaussian parameters $f_{NL}$, $g_{NL}$ consistent with current constraints can cause the statistics measured within our Hubble volume to differ from those in the larger universe, even for a modest number of super horizon e-folds ($N\sim \mathcal{O}(10)$, say). This is illustrated in Figures \ref{Fig:Pzeta2} and \ref{Fig:probfNL}. These results are dependent on the unknown behavior of the curvature power spectrum on super-horizon scales (examples for different IR extrapolations of the power spectrum are plotted in Figure~\ref{Fig:Sigma_LW}). Figures \ref{Fig:Pzeta2} and \ref{Fig:probfNL} assume that the power law spectrum on super-horizon scales remains unchanged out to $k/H_0\sim e^{N}$, which may be false -- the true behavior of $\Delta^2(k)$ for $k<H_0$ could increase or decrease the typical amplitude of $\zeta_{G,L}$. Typical changes to the level of non-Gaussianity in Hubble-size subvolumes are plotted in Figure \ref{Fig:Var} for both $n_s=0.9608$ and $n_s=1$. In a universe with local non-Gaussianity, constraints on global statistics (and therefore inflationary parameters) from observations in our Hubble patch are {\em necessarily probabilistic} because the locally observed power spectrum, bispectrum, trispectrum are dependent upon the unknown value of the random variable $\zeta_{G,L}$. While this has been known for a long time in the context of slow-roll inflation \cite{Starobinsky:1986fx, Salopek:1990re, Lyth:2007jh,Salem:2012ve}, we have shown that for inflationary models with local non-Gaussianity -- either strong non-Gaussianity or merely observable levels of non-Gaussianity -- the probabilistic relationship between theory and observations is important.

In \S \ref{sec:strongNG} we consider strongly non-Gaussian statistics, $\zeta_{NG}(\x)\sim \zeta_G^p$ in the larger universe $V_L$ that can appear only weakly non-Gaussian on sufficiently large background fluctuations $\zeta_{G,L}$ \cite{Nelson:2012sb}. We determine the restrictions on $p$, $\Delta^2_{\zeta_G}$ and $\zeta_{G,L}$ to produce statistics consistent with observations in our Hubble volume. The main results are illustrated in Figure \ref{Fig:zetap}. We see that for sufficiently large $N$, typical subvolumes (e.g. $\zeta_{G,L}/\sqrt{\langle\zeta^2_{G,L}\rangle}\sim 1$, corresponding to $\sim 30\%$ of Hubble-sized patches in the universe) will have statistics consistent with constraints on parameters in the weakly non-Gaussian ansatz Eq.~(\ref{eq:zetafNLgNL}), even if the curvature perturbation in the rest of the universe is strongly non-Gaussian. At very large $N$, this can be true of the vast majority of subsamples (not only $1\sigma$ and higher fluctuations) depending on the infrared behavior of the power spectrum. In this sense weakly non-Gaussian statistics may be considered `natural' as discussed in \cite{Nelson:2012sb}. While \S \ref{sec:strongNG} focuses on initial conditions that are a single power law $\zeta^p$ the qualitative results should hold for more general forms of strongly non-Gaussian initial conditions and we provide a framework for these calculations in Appendices \ref{sec:diagrams} and \ref{sec:global}. 

Finally, in \S \ref{sec:twofieldNG} we consider an example in which the initial curvature perturbation is given by a sum of two uncorrelated fields, one Gaussian $\phi$ and one non-Gaussian with a quadratic coupling, $\sigma=\sigma_{G}(\x)+\tilde{f}_{NL}(\sigma_G^2(\x)-\langle\sigma_G^2\rangle)$. Initial conditions of this type are consistent with observations for a range of values of $\tilde{f}_{NL}$ and/or $\xi^2=P_\phi/P_\sigma$. The qualitative difference between this scenario and those in \S \ref{sec:weakNG} and \S \ref{sec:strongNG} is that the field that modulates the local statistics, $\sigma_{G,L}$, is only partially correlated with the total curvature perturbation $\zeta=\phi+\sigma$. There is therefore greater freedom in finding statistics in $V_L$ that map to weakly Gaussian statistics in $V_S$. For $\sigma$ only weakly non-Gaussian, the results are similar to those in \S \ref{sec:weakNG} but, for fixed $f_{NL}$ and $\Delta_\zeta^2$, the typical size of the modulation in local statistics is enhanced by a factor $\sqrt{1+P_\phi/P_\sigma}$ (see Figure \ref{Fig:probstauNL}). On the other hand, if $\tilde{f}_{NL}\langle \sigma_G^2\rangle^{1/2}\sim 1$, the results are qualitatively different: the probability distribution for observed values of $f_{NL}$, $\tau_{NL}$ is highly skewed  (see Figure \ref{Fig:probstrongtwofield}). 

The calculations in this paper are completely statistical: we do not attempt to give a dynamical model that generates the examples of statistics in $V_L$ we have considered, nor do we attempt to understand how differences between local and global statistics alter inferences about particular inflationary scenarios. The parametric forms of initial conditions we have considered in \S \ref{sec:weakNG}, \S \ref{sec:strongNG}, and \S \ref{sec:twofieldNG} are simplified examples of initial conditions that can arise in the curvaton, or modulated reheating scenarios, but we have assumed that the parameters ($\Delta^2_{\zeta_G}$ and the coefficients of the non-linear terms, for instance) can be freely adjusted to tune the statistics in $V_S$, which is not necessarily the case. For a thorough analysis of the range of possibilities of local statistics in a Hubble-size patches throughout the universe within the curvaton framework, see \cite{Linde:2005yw,Demozzi:2010aj}. 

Throughout this paper we have made the simplifying assumption that the background mode $\zeta_{G,L}$ is precisely constant across our Hubble volume (see Eq.~(\ref{eq:lwvariance})). In reality slight variations in $\zeta_{G,L}$ from modes with wavelengths not too much larger than $c/H_0$, and correlations between these variations and the small-scale statistics of $\zeta$ may be detectable \cite{Knox:2005hx,Waterhouse:2008vb,Vardanyan:2009ft,Erickcek:2008jp,Guth:2012ww,Kleban:2012ph,Bull:2013fga}. While the assumptions we have made in Eq. ~(\ref{eq:lwvariance}) should be sufficiently precise for sub-horizon scales $k\gg H_0 $, and contributions to $\zeta_{G,L}$ from $k\ll  H_0$, it would be interesting to explore the potentially observable corrections for $k\sim H_0$.  \\

\noindent {\bf Acknowledgements}\\
ML is grateful for discussions with and feedback from Peter Adshead, Wayne Hu, Andrei Linde, and Matias Zaldarriaga. E.N. and S.S. thank Joe Bramante and Jason Kumar for discussions on closely related work. ML is supported by U.S. Dept. of Energy contract DE-FG02-90ER-40560. EN and SS are supported by the Eberly Research Funds of The Pennsylvania State University. The Institute for Gravitation and the Cosmos is supported by the Eberly College of Science and the Office of the Senior Vice
President for Research at the Pennsylvania State University. S.S. is grateful to the Aspen Center for Physics and the NSF Grant \#1066293 for hospitality while some of the ideas for this paper were developed. 

\appendix

\section{Diagrammatic Representations of $n$-point Functions}
\label{sec:diagrams}
We want to calculate $n$-point correlation functions of the non-Gaussian field $\zeta_{NG}$ defined by 
\be
\zeta_{NG}(\x)=f(\zeta_G(\x))-\langle f(\zeta_G)\rangle\,.
\ee
The $n$-point functions of $\zeta_{NG}$ can be written entirely in terms of two-point functions of $\zeta_G$ and the derivatives of $f$. However, the expressions quickly get messy so it's helpful to use connected diagrams to keep track of the terms (see also \cite{Byrnes:2007tm,Tasinato:2012js}).  \\

\noindent {\bf Dictionary of Diagrams:}\\
In this paper a line segment connecting two points $1$ and $2$ represents the real-space correlation function between the Gaussian fields $\zeta_G$ at two spatial points $\x_1$ and $\x_2$. 
\be
\langle\zeta_{G}(\x_1)\zeta_{G}(\x_2)\rangle \equiv \qquad 
\parbox{15mm}{
\begin{fmffile}{GaussP}
\begin{fmfgraph*}(20,15)
\fmfleft{v1}
\fmfright{v2}
\fmf{plain}{v1,v2}
\fmfdot{v1,v2}
\fmflabel{$1$}{v1}
\fmflabel{$2$}{v2}
\end{fmfgraph*}
\end{fmffile}
}\qquad\qquad  
\ee
while a double line segment indicates the square of the Gaussian correlation function
\be
\langle\zeta_{G}(\x_1)\zeta_{G}(\x_2)\rangle^2 \equiv \qquad
\parbox{15mm}{
\begin{fmffile}{DP4}
\begin{fmfgraph*}(20,15)
\fmfleft{v1}
\fmfright{v2}
\fmf{dbl_plain}{v1,v2}
\fmfdot{v1,v2}
\fmflabel{$1$}{v1}
\fmflabel{$2$}{v2}
\end{fmfgraph*}
\end{fmffile}
} 
\ee
and vertices with multiple line segments indicate products of correlation functions connected to different points
\be
\langle\zeta_G(\x_1)\zeta_G(\x_2)\rangle\langle\zeta_G(\x_1)\zeta_G(\x_3)\rangle \equiv\qquad
\parbox{15mm}{
\begin{fmffile}{bispectrumtree1}
\begin{fmfgraph*}(15,15)
\fmfleft{i}
\fmfright{o1,o2}
\fmf{phantom,tension=5}{i,v1}
\fmf{phantom,tension=5}{v2,o1}
\fmf{phantom,tension=5}{v3,o2}
\fmf{plain}{v1,v2}
\fmf{plain}{v1,v3}
\fmfdot{v1,v2,v3}
\fmflabel{$1$}{v1}
\fmflabel{$2$}{v2}
\fmflabel{$3$}{v3}
\end{fmfgraph*}
\end{fmffile}}
\ee
Circles represent $\zeta_G(\x)$ contracted with itself which is independent of $\x$. For instance, we can write
\ba
\langle \zeta_G(\x_1)\zeta_G(\x_2)\rangle \langle\zeta_G(\x_1)^2\rangle &=&\qquad \qquad  
\parbox{15mm}
{\begin{fmffile}{DP4loopleft}
\begin{fmfgraph*}(20,15)
\fmfleft{v1}
\fmfright{v2}
\fmf{plain}{v1,v2}
\fmf{plain}{v1,v1}
\fmfdot{v1,v2}
\fmflabel{$1$}{v1}
\fmflabel{$2$}{v2}
\end{fmfgraph*}
\end{fmffile}
}\\
&=&\qquad \qquad \parbox{15mm}{
\begin{fmffile}{GaussP}
\begin{fmfgraph*}(20,15)
\fmfleft{v1}
\fmfright{v2}
\fmf{plain}{v1,v2}
\fmfdot{v1,v2}
\fmflabel{$1$}{v1}
\fmflabel{$2$}{v2}
\end{fmfgraph*}
\end{fmffile}
} \qquad \qquad \times
\parbox{15mm}{
\begin{fmffile}{bubble}
\begin{fmfgraph*}(20,15)
\fmfleft{i}
\fmfright{o}
\fmf{phantom}{i,v1}
\fmf{phantom}{v2,o}
\fmf{plain,left,tension=.3}{v1,v2,v1}
\end{fmfgraph*}
\end{fmffile}
}\nn
\ea
so it doesn't matter which vertex a loop is connected to. \\

\noindent {\bf The Two-point Function:}\\
The two point correlation function of the non-Gaussian field $\zeta_{NG}$ is given by 
\ba
\langle\zeta_{NG}(\x_1)\zeta_{NG}(\x_2)\rangle &=&\qquad \parbox{15mm}{
\begin{fmffile}{GaussP}
\begin{fmfgraph*}(20,15)
\fmfleft{v1}
\fmfright{v2}
\fmf{plain}{v1,v2}
\fmfdot{v1,v2}
\fmflabel{$1$}{v1}
\fmflabel{$2$}{v2}
\end{fmfgraph*}
\end{fmffile}
}\qquad 
\qquad \left((f^{(1)})^2+f^{(1)}f^{(3)}
\parbox{15mm}{
\begin{fmffile}{bubble}
\begin{fmfgraph*}(20,15)
\fmfleft{i}
\fmfright{o}
\fmf{phantom}{i,v1}
\fmf{phantom}{v2,o}
\fmf{plain,left,tension=.3}{v1,v2,v1}
\end{fmfgraph*}
\end{fmffile}
}
\quad + \qquad \dots\right)\qquad \\
&+&\qquad \parbox{15mm}{
\begin{fmffile}{DP4}
\begin{fmfgraph*}(20,15)
\fmfleft{v1}
\fmfright{v2}
\fmf{dbl_plain}{v1,v2}
\fmfdot{v1,v2}
\fmflabel{$1$}{v1}
\fmflabel{$2$}{v2}
\end{fmfgraph*}
\end{fmffile}
}\qquad
\quad\left(
\frac{1}{2}(f^{(2)})^2
\quad+\dots\nn\right)\nn\\
&+&\qquad \dots \nn
\ea
where the $\dots$ in each parenthesis represent terms with higher-order loop contributions, and the $\dots$ in the last line indicate terms with additional shapes (but still $\mathcal{O}(\zeta_G^6)$). \\

\noindent {\bf The Three-point Function:}\\
The three point function of $\zeta_{NG}$ contains five types of terms at $\mathcal{O}(\zeta_G^6)$:
At this order the terms can be grouped into three distinct three-point function shapes:
\ba
\langle\zeta_{NG}(\x_1)\zeta_{NG}(\x_2)\zeta_{NG}(\x_3)\rangle_c &=&\\
&&\left( (f^{(1)})^2f^{(2)}+\left(\frac{1}{2}(f^{(1)})^2f^{(4)}+f^{(1)}f^{(2)}f^{(3)}\right)
\parbox{15mm}{
\begin{fmffile}{bubble}
\begin{fmfgraph*}(20,15)
\fmfleft{i}
\fmfright{o}
\fmf{phantom}{i,v1}
\fmf{phantom}{v2,o}
\fmf{plain,left,tension=.3}{v1,v2,v1}
\end{fmfgraph*}
\end{fmffile}
}
+\dots \right)\nn\\
&&
\times\left(\quad
\parbox{15mm}{
\begin{fmffile}{bispectrumtree1}
\begin{fmfgraph*}(15,15)
\fmfleft{i}
\fmfright{o1,o2}
\fmf{phantom,tension=5}{i,v1}
\fmf{phantom,tension=5}{v2,o1}
\fmf{phantom,tension=5}{v3,o2}
\fmf{plain}{v1,v2}
\fmf{plain}{v1,v3}
\fmfdot{v1,v2,v3}
\fmflabel{$1$}{v1}
\fmflabel{$2$}{v2}
\fmflabel{$3$}{v3}
\end{fmfgraph*}
\end{fmffile}}
\quad + \quad
\parbox{15mm}{
\begin{fmffile}{bispectrumtree2}
\begin{fmfgraph*}(15,15)
\fmfleft{i}
\fmfright{o1,o2}
\fmf{phantom,tension=5}{i,v1}
\fmf{phantom,tension=5}{v2,o1}
\fmf{phantom,tension=5}{v3,o2}
\fmf{plain}{v1,v2}
\fmf{plain}{v2,v3}
\fmfdot{v1,v2,v3}
\fmflabel{$1$}{v1}
\fmflabel{$2$}{v2}
\fmflabel{$3$}{v3}
\end{fmfgraph*}
\end{fmffile}}
\quad + \quad 
\parbox{15mm}{
\begin{fmffile}{bispectrumtree3}
\begin{fmfgraph*}(15,15)
\fmfleft{i}
\fmfright{o1,o2}
\fmf{phantom,tension=5}{i,v1}
\fmf{phantom,tension=5}{v2,o1}
\fmf{phantom,tension=5}{v3,o2}
\fmf{plain}{v1,v3}
\fmf{plain}{v2,v3}
\fmfdot{v1,v2,v3}
\fmflabel{$1$}{v1}
\fmflabel{$2$}{v2}
\fmflabel{$3$}{v3}
\end{fmfgraph*}
\end{fmffile}}
\quad
\right)\nn\\ \quad \nn \\ \quad \nn \\
&+&\left
(\frac{1}{2}f^{(1)}f^{(2)}f^{(3)}+\dots\right)\left(
\parbox{15mm}{
\begin{fmffile}{bispectrumbar1a}
\begin{fmfgraph*}(15,15)
\fmfleft{i}
\fmfright{o1,o2}
\fmf{phantom,tension=5}{i,v1}
\fmf{phantom,tension=5}{v2,o1}
\fmf{phantom,tension=5}{v3,o2}
\fmf{dbl_plain}{v1,v2}
\fmf{plain}{v1,v3}
\fmfdot{v1,v2,v3}
\fmflabel{$1$}{v1}
\fmflabel{$2$}{v2}
\fmflabel{$3$}{v3}
\end{fmfgraph*}
\end{fmffile}}
\quad + \quad 
\parbox{15mm}{
\begin{fmffile}{bispectrumbar1b}
\begin{fmfgraph*}(15,15)
\fmfleft{i}
\fmfright{o1,o2}
\fmf{phantom,tension=5}{i,v1}
\fmf{phantom,tension=5}{v2,o1}
\fmf{phantom,tension=5}{v3,o2}
\fmf{plain}{v1,v2}
\fmf{dbl_plain}{v1,v3}
\fmfdot{v1,v2,v3}
\fmflabel{$1$}{v1}
\fmflabel{$2$}{v2}
\fmflabel{$3$}{v3}
\end{fmfgraph*}
\end{fmffile}}
\quad +\quad
\parbox{15mm}{
\begin{fmffile}{bispectrumbar2a}
\begin{fmfgraph*}(15,15)
\fmfleft{i}
\fmfright{o1,o2}
\fmf{phantom,tension=5}{i,v1}
\fmf{phantom,tension=5}{v2,o1}
\fmf{phantom,tension=5}{v3,o2}
\fmf{dbl_plain}{v1,v2}
\fmf{plain}{v2,v3}
\fmfdot{v1,v2,v3}
\fmflabel{$1$}{v1}
\fmflabel{$2$}{v2}
\fmflabel{$3$}{v3}
\end{fmfgraph*}
\end{fmffile}}\right.\nn\\\quad \nn \\ \quad \nn \\
&&+\left.\quad
\parbox{15mm}{
\begin{fmffile}{bispectrumbar2b}
\begin{fmfgraph*}(15,15)
\fmfleft{i}
\fmfright{o1,o2}
\fmf{phantom,tension=5}{i,v1}
\fmf{phantom,tension=5}{v2,o1}
\fmf{phantom,tension=5}{v3,o2}
\fmf{plain}{v1,v2}
\fmf{dbl_plain}{v2,v3}
\fmfdot{v1,v2,v3}
\fmflabel{$1$}{v1}
\fmflabel{$2$}{v2}
\fmflabel{$3$}{v3}
\end{fmfgraph*}
\end{fmffile}}
\quad+\quad
\parbox{15mm}{
\begin{fmffile}{bispectrumbar3a}
\begin{fmfgraph*}(15,15)
\fmfleft{i}
\fmfright{o1,o2}
\fmf{phantom,tension=5}{i,v1}
\fmf{phantom,tension=5}{v2,o1}
\fmf{phantom,tension=5}{v3,o2}
\fmf{dbl_plain}{v1,v3}
\fmf{plain}{v2,v3}
\fmfdot{v1,v2,v3}
\fmflabel{$1$}{v1}
\fmflabel{$2$}{v2}
\fmflabel{$3$}{v3}
\end{fmfgraph*}
\end{fmffile}}
\quad + \quad 
\parbox{15mm}{
\begin{fmffile}{bispectrumbar3b}
\begin{fmfgraph*}(15,15)
\fmfleft{i}
\fmfright{o1,o2}
\fmf{phantom,tension=5}{i,v1}
\fmf{phantom,tension=5}{v2,o1}
\fmf{phantom,tension=5}{v3,o2}
\fmf{plain}{v1,v3}
\fmf{dbl_plain}{v2,v3}
\fmfdot{v1,v2,v3}
\fmflabel{$1$}{v1}
\fmflabel{$2$}{v2}
\fmflabel{$3$}{v3}
\end{fmfgraph*}
\end{fmffile}}\right)\nn\\\quad \nn \\ \quad \nn \\
&+&\left((f^{(2)})^3+\dots\right) \quad
\parbox{15mm}{
\begin{fmffile}{bispectrumall}
\begin{fmfgraph*}(15,15)
\fmfleft{i}
\fmfright{o1,o2}
\fmf{phantom,tension=5}{i,v1}
\fmf{phantom,tension=5}{v2,o1}
\fmf{phantom,tension=5}{v3,o2}
\fmf{plain}{v1,v2}
\fmf{plain}{v2,v3}
\fmf{plain}{v3,v1}
\fmfdot{v1,v2,v3}
\fmflabel{$1$}{v1}
\fmflabel{$2$}{v2}
\fmflabel{$3$}{v3}
\end{fmfgraph*}
\end{fmffile}}\nn\\ \quad \nn \\ \quad \nn \\
&+&\dots\nn
\ea
\\
\noindent {\bf The Connected Four-point Function:}\\
Finally, we compute the connected four-point function up to $\mathcal{O}(\zeta_G^8)$:

\ba
\langle\zeta_{NG}(\x_1)\zeta_{NG}(\x_2)\zeta_{NG}(\x_3)\zeta_{NG}(\x_4)\rangle_c &=&\left( (f^{(1)})^3f^{(3)} +\left(\frac{3}{2}(f^{(1)})^2(f^{(3)})^2+\frac{1}{2}(f^{(1)})^3f^{(5)}\right) \!\!\!\!\!\!\!\!\!
\parbox{15mm}{
\begin{fmffile}{bubble}
\begin{fmfgraph*}(20,15)
\fmfleft{i}
\fmfright{o}
\fmf{phantom}{i,v1}
\fmf{phantom}{v2,o}
\fmf{plain,left,tension=.3}{v1,v2,v1}
\end{fmfgraph*}
\end{fmffile}
}
\right)\nn\\
&&\times \left(
\quad
\parbox{15mm}{
\begin{fmffile}{f13f3}
\begin{fmfgraph*}(15,10)
\fmfleft{v1,v2}
\fmfright{v4,v3}
\fmf{plain,tension = 1}{v2,v1}
\fmf{plain,tension = 0.5 }{v2,v3}
\fmf{plain}{v2,v4}
\fmfdot{v1,v2,v3,v4}
\fmflabel{$1$}{v1}
\fmflabel{$2$}{v2}
\fmflabel{$3$}{v3}
\fmflabel{$4$}{v4}
\end{fmfgraph*}
\end{fmffile}}
\quad + \quad 3\, {\rm perm.}\right)\quad\\ \quad \nn \\ \quad \nn \\
&+&\left((f^{(1)})^2(f^{(2)})^2+\left(f^{(1)}(f^{(2)})^2f^{(3)}+ (f^{(1)})^2f^{(2)}f^{(4)}\right)
\parbox{15mm}{
\begin{fmffile}{bubble}
\begin{fmfgraph*}(20,15)
\fmfleft{i}
\fmfright{o}
\fmf{phantom}{i,v1}
\fmf{phantom}{v2,o}
\fmf{plain,left,tension=.3}{v1,v2,v1}
\end{fmfgraph*}
\end{fmffile}
}
\right) \nn\\
&&\times \left(\quad
\parbox{15mm}{
\begin{fmffile}{f12f22a}
\begin{fmfgraph*}(15,10)
\fmfleft{v1,v2}
\fmfright{v4,v3}
\fmf{plain}{v1,v2}
\fmf{plain}{v2,v3}
\fmf{plain}{v3,v4}
\fmfdot{v1,v2,v3,v4}
\fmflabel{$1$}{v1}
\fmflabel{$2$}{v2}
\fmflabel{$3$}{v3}
\fmflabel{$4$}{v4}
\end{fmfgraph*}
\end{fmffile}
}\quad +\quad
\parbox{15mm}{
\begin{fmffile}{f12f22b}
\begin{fmfgraph*}(15,10)
\fmfleft{v1,v2}
\fmfright{v4,v3}
\fmf{plain}{v1,v3}
\fmf{plain}{v2,v3}
\fmf{plain}{v2,v4}
\fmfdot{v1,v2,v3,v4}
\fmflabel{$1$}{v1}
\fmflabel{$2$}{v2}
\fmflabel{$3$}{v3}
\fmflabel{$4$}{v4}
\end{fmfgraph*}
\end{fmffile}
}\quad 
+\quad 5\, {\rm perm.} \right)\nn\\ \quad \nn \\ \quad \nn \\
&+&\quad (f^{(2)})^4\left(\quad
\parbox{15mm}{
\begin{fmffile}{f24}
\begin{fmfgraph*}(15,10)
\fmfleft{v1,v2}
\fmfright{v4,v3}
\fmf{plain}{v1,v2,v3,v4,v1}
\fmfdot{v1,v2,v3,v4}
\fmflabel{$1$}{v1}
\fmflabel{$2$}{v2}
\fmflabel{$3$}{v3}
\fmflabel{$4$}{v4}
\end{fmfgraph*}
\end{fmffile}
}+2\,{\rm perm.}\right)\nn\\ \quad \nn \\ \quad \nn \\
&+&
\frac{1}{2}f^{(1)}(f^{(2)})^2f^{(3)}\left(\quad
\parbox{15mm}{
\begin{fmffile}{f1f22f3aa}
\begin{fmfgraph*}(15,10)
\fmfleft{v1,v2}
\fmfright{v4,v3}
\fmf{plain}{v1,v2,v3}
\fmf{dbl_plain}{v3,v4}
\fmfdot{v1,v2,v3,v4}
\fmflabel{$1$}{v1}
\fmflabel{$2$}{v2}
\fmflabel{$3$}{v3}
\fmflabel{$4$}{v4}
\end{fmfgraph*}
\end{fmffile}
}\quad +\quad
\parbox{15mm}{
\begin{fmffile}{f1f22f3bb}
\begin{fmfgraph*}(15,10)
\fmfleft{v1,v2}
\fmfright{v4,v3}
\fmf{plain}{v1,v4}
\fmf{plain}{v3,v4}
\fmf{dbl_plain}{v2,v3}
\fmfdot{v1,v2,v3,v4}
\fmflabel{$1$}{v1}
\fmflabel{$2$}{v2}
\fmflabel{$3$}{v3}
\fmflabel{$4$}{v4}
\end{fmfgraph*}
\end{fmffile}
}+\quad 11\,{\rm perm.}\right)\nn\\ \quad \nn \\ \quad \nn\\ 
&+&\frac{1}{2}(f^{(1)})^2(f^{(3)})^2\left(
\quad 
\parbox{15mm}{
\begin{fmffile}{f12f32bb}
\begin{fmfgraph*}(15,10)
\fmfleft{v1,v2}
\fmfright{v4,v3}
\fmf{plain}{v2,v1}
\fmf{dbl_plain}{v2,v3}
\fmf{plain}{v3,v4}
\fmfdot{v1,v2,v3,v4}
\fmflabel{$1$}{v1}
\fmflabel{$2$}{v2}
\fmflabel{$3$}{v3}
\fmflabel{$4$}{v4}
\end{fmfgraph*}
\end{fmffile}}
\quad +\quad
\parbox{15mm}{
\begin{fmffile}{f12f32bc}
\begin{fmfgraph*}(15,10)
\fmfleft{v1,v2}
\fmfright{v4,v3}
\fmf{plain}{v2,v4}
\fmf{dbl_plain}{v2,v3}
\fmf{plain}{v1,v3}
\fmfdot{v1,v2,v3,v4}
\fmflabel{$1$}{v1}
\fmflabel{$2$}{v2}
\fmflabel{$3$}{v3}
\fmflabel{$4$}{v4}
\end{fmfgraph*}
\end{fmffile}}
\quad + 5\, {\rm perm.}\right)\nn\\ \quad \nn \\ \quad \nn \\ \quad \nn\\
& +& f^{(1)}(f^{(2)})^2f^{(3)}\left(
\quad
\parbox{15mm}{
\begin{fmffile}{f1f22f3c}
\begin{fmfgraph*}(15,10)
\fmfleft{v1,v2}
\fmfright{v4,v3}
\fmf{plain}{v1,v2,v3,v4,v2}
\fmfdot{v1,v2,v3,v4}
\fmflabel{$1$}{v1}
\fmflabel{$2$}{v2}
\fmflabel{$3$}{v3}
\fmflabel{$4$}{v4}
\end{fmfgraph*}
\end{fmffile}
}\quad + 11 \,{\rm  perm.}
 \right)\nn\\ \quad \nn \\ \quad \nn \\
&+&\frac{1}{2}
(f^{(1)})^2f^{(2)}f^{(4)}\left(
\quad
\parbox{15mm}{
\begin{fmffile}{f12f2f4}
\begin{fmfgraph*}(15,10)
\fmfleft{v1,v2}
\fmfright{v4,v3}
\fmf{plain}{v2,v1}
\fmf{plain}{v2,v3}
\fmf{dbl_plain}{v2,v4}
\fmflabel{$1$}{v1}
\fmflabel{$2$}{v2}
\fmflabel{$3$}{v3}
\fmflabel{$4$}{v4}
\fmfdot{v1,v2,v3,v4}
\end{fmfgraph*}
\end{fmffile}}\quad + 11\,{\rm perm.}\right)\nn
\ea
\\

\section{Statistics in $V_L$}
\label{sec:global}

\noindent{\bf Scaling of the Non-Gaussian Cumulants:}\\
Suppose that the statistics of the curvature perturbation can be written as a non-linear transformation of a Gaussian field
that is local in real space:
\be
\zeta_{NG}=f(\zeta_G(\x))-\langle f(\zeta_G(\x)\rangle \,.
\ee
We can calculate the statistics of $\zeta_{NG}$ in terms of $\langle\zeta_{G}^2\rangle \ll 1 $ and derivatives of $f$
\be
f^{(n)}\equiv\left. \frac{\partial ^{(n)} f}{\partial \zeta_G^n}\right|_{\zeta_G=0}\,.
\ee
For the moment, we'll ignore the shape dependence of the $n$-point functions of $\zeta_{NG}$ and just consider the scaling of the cumulants in terms of $\langle \zeta_G^2\rangle$ (we consider the shape dependence in \S \ref{sec:Npointsglobal}). In terms of $f$ and $\zeta_G$ we have,
 \ba
 \label{eq:globalcum1}
 \langle \zeta_{NG}^2\rangle& =& \langle \zeta_{G}^2\rangle\left\{(f^{(1)})^2+\frac{1}{2}\left((f^{(2)})^2+2f^{(1)}f^{(3)}\right)\langle \zeta_G^2\rangle\right.\\
 &&\left.+\, \frac{1}{12}\left(5(f^{(3)})^2+6f^{(2)}f^{(4)}+3f^{(1)}f^{(5)}\right)\langle \zeta_G^2\rangle^2+\dots\right\}\nn\\
 \langle \zeta_{NG}^3\rangle &=& \langle \zeta_G^2\rangle ^2\left\{3(f^{(1)})^2f^{(2)}+\left((f^{(2)})^3+6f^{(1)}f^{(2)}f^{(3)}+\frac{3}{2}(f^{(1)})^2f^{(4)}\right)\langle \zeta_G^2\rangle+\dots\right\}\\
 \langle \zeta_{NG}^4\rangle_c & =& \langle \zeta_G^2\rangle ^3 \left(4 (f^{(1)})^3f^{(3)}+12 (f^{(1)})^2(f^{(2)})^2+\left(3(f^{(2)})^4+36f^{(1)}(f^{(2)})^2f^{(3)}\right.\right.\\
 &&\left.\left.+12(f^{(1)})^2(f^{(3)})^2+\,18(f^{(1)})^2f^{(2)}f^{(4)}+2(f^{(1)})^3f^{(5)}\right)\langle\zeta_G^2\rangle+\dots \right)\nn\\
 \langle \zeta_{NG}^5\rangle_c&=& \langle \zeta_G^2\rangle ^4\left( 5(f^{(1)})^4f^{(4)}+60 (f^{(1)})^2(f^{(2)})^3+60(f^{(1)})^3f^{(2)}f^{(3)}+\dots\right)\\
 &\dots&\nn\\
 \label{eq:globalcumn}
 \langle  \zeta_{NG}^n\rangle_c &=& \langle\zeta_G^2\rangle^{n-1}\left(n(f^{(1)})^{n-1}f^{(n-1)}+n(n-1)(n-2)(f^{(1)})^{n-2}f^{(2)}f^{(n-2)}+\dots\right)\nn
  \ea
where the subscript $_c$ indicates the connected part and $\dots$ indicate terms higher-order in $\langle \zeta_G^2\rangle$. In Eq.~(\ref{eq:globalcum1})-Eq.(\ref{eq:globalcumn}) we've kept a number of subleading (in $\langle \zeta_G^2\rangle$) terms in order to help illustrate the following points:
\begin{itemize}
\item  For $f'\neq 0$ and $f^{(n)} /f^{(1)}\langle \zeta_G^2\rangle^{(n-1)/2} \ll  1$, cumulants scale as $\langle \zeta_{NG}^n\rangle_c \sim \langle\zeta_{G}^2\rangle^{n-1}\approx \langle\zeta_{NG}^2\rangle^{n-1}$. So, higher cumulants are suppressed by powers of the {\em observed} variance. We refer to this type of statistics as {\em weakly non-Gaussian}. 

\item If $f' =0$ the cumulants still scale with increasing powers of $\langle\zeta_G^2\rangle$, but the observed variance is $\langle \zeta_{NG}^2\rangle \sim \langle \zeta_{G}^2\rangle^p$ where $p$ is the order of the first nonzero derivative of $f$, so the relative scaling of each $\langle \zeta_{NG}^n\rangle_c$ is different\footnote{What we really mean by $f^{(n)}=0$ is $f^{(n)}\sim \mathcal{O}\left(\langle\zeta_{G,s}^2\rangle\right)$, our expansion parameter, so that the next order terms are comparable.}. In particular the suppression of higher cumulants can be weaker than in the $f'\neq 0$ case.\footnote{To be more precise, the scaling of cumulants with the non-Gaussian variance depends on which terms are non-vanishing in $f(x)=\sum_n\frac{1}{n!}f^{(n)}x^n$. If the only non-vanishing term has an odd power $n$, then the cumulants scale with powers of $\langle \zeta_{NG}^2\rangle$, but generically higher terms scale with $\sim\langle \zeta_{G}^2\rangle > \langle \zeta_{NG}^2\rangle$ and are less suppressed than in the $f' \neq 0$ case.} We refer to this type of statistics as {\em strongly non-Gaussian}.

\end{itemize}
Further note that for series coefficients $f^{(n+1)}\sqrt{\langle \zeta_G^2\rangle} \ll  f^{(n)}$ when $f^{(n)}$ is non-zero, non-Gaussianity will first be evident in either the skewness or kurtosis. That is, $\langle \zeta_{NG}^3\rangle \ge \langle \zeta_{NG}^n\rangle_c$ and/or $\langle \zeta_{NG}^4\rangle_c \ge \langle \zeta_{NG}^n\rangle_c$ for $n>4$.
\\

\noindent {\bf Shape and Scale-dependence of the $n$-point Functions:}\\
\label{sec:Npointsglobal}
We now consider the shape and scale-dependence of the power spectrum, bispectrum, and trispectrum of $\zeta_{NG}$ in terms of the Gaussian field $\zeta_G$. The 
statistics of $\zeta_G$ are completely specified by the two point function:
\be
\langle \zeta_G(\k)\zeta_G(\k')\rangle = (2\pi)^3 \delta_D(\k+\k') P_G(k)\quad {,}\quad \Delta_G^2 (k)= k^3P_G(k)/(2\pi^2)
\ee 
where $\delta_D$ is the Dirac delta function.  The power spectrum, bispectrum, and trispectrum of $\zeta_{NG}$ are defined through 
\ba
\langle \zeta_{NG}(\k_1)\zeta_{NG}(\k_2)\rangle &\equiv& (2\pi)^3 \delta_D(\k_1+\k_2) P_{NG}(k_1)\\
\langle \zeta_{NG}(\k_1)\zeta_{NG}(\k_2)\zeta_{NG}(\k_3)\rangle &\equiv &(2\pi)^3 \delta_D(\k_1+\k_2+\k_3)B(\k_1,\k_2,\k_3)\\
\langle \zeta_{NG}(\k_1)\zeta_{NG}(\k_2)\zeta_{NG}(\k_3)\zeta_{NG}(\k_4)\rangle_c&\equiv& (2\pi)^3\delta_D(\k_1+\k_2+\k_3+\k_4) T(\k_1,\k_2,\k_3,\k_4)
\ea
Fourier transforming the real-space diagrammatic expressions in Appendix \ref{sec:diagrams} we find that the non-Gaussian power spectrum is given by
\be
P_{NG}(k)=P_{G}(k)\left\{\left((f^{(1)})^2+f^{(1)}f^{(3)}\langle\zeta_G^2\rangle+\dots\right)+I_{\zeta_G}(k)\left(\frac{1}{2}(f^{(2)})^2+\dots\right)+\dots\right\}\, ,
\ee
where $\langle \zeta_G^2\rangle$ is the two-point function at zero separation (a constant) and we have defined,
\be
\label{eq:I1def}
I_{\zeta_G}(k)\equiv \frac{1}{P_G(k)}\int \frac{d^3\k'}{(2\pi)^3}P_G(k')P_G(|\k+\k'| )\,.
\ee
For a scale-invariant spectrum, $n_s=1$, this becomes
\be
\label{eq:I1scaleinv}
I_{\zeta_G}(k) = -\Delta^2_{G}\left\{\ln\frac{\Lambda^2}{k^2-\Lambda^2}+\frac{k^2}{k^2-\Lambda^2}\right\}\quad {\rm and }\quad \langle\zeta_G^2\rangle = \Delta_G^2\ln\left(\frac{k_{max}}{\Lambda}\right)
\ee
where $k_{max}$ is the UV cutoff of the power spectrum and $\Lambda\sim 2\pi/L$ is the IR cutoff. For $k < \sqrt{e k_{max}\Lambda }$, $\langle \zeta_G^2\rangle > I_{\zeta_G}(k)$, but for $k>\sqrt{e k_{max} \Lambda}$, $\langle \zeta_G^2\rangle < I_{\zeta_G}(k)$. $I_{\zeta_G}(k)$ is plotted in Figure \ref{Fig:I1} for $n_s=1$.

The bispectrum is given by 
\ba
B(\k_1,\k_2,\k_3)&=& \left(P_{G}(k_1)P_G(k_2) +2\,{\rm perm.}\right)\left((f^{(1)})^2f^{(2)}+\left(f^{(1)}f^{(2)}f^{(3)}+\frac{1}{2}(f^{(1)})^2f^{(4)}\right)\langle\zeta_G^2\rangle+\dots \right)\nn\\
&+&\frac{1}{3}\left(J(\k_1,\k_3)P_G(k_1)P_G(k_3)+2\, {\rm perm.})\right)\left((f^{(2)})^3+\dots\right)\nn\\
&+&\left(I(k_2)P_G(k_1)P_G(k_2)+5\,{\rm perm.}\right)\left(\frac{f^{(1)}f^{(2)}f^{(3)}}{2}+\dots \right)+\dots\nn
\ea
where $J(\k_1,\k_2) \sim \mathcal{O}(\Delta^2_G)$ is a function that depends on both the magnitudes of $\k_1$ and $\k_2$, and the angle between them
\be
\label{eq:J1def}
J(\k_1, \k_2) = \frac{1}{P_G(k_1)P_G(k_2)}\int \frac{d^3\k'}{(2\pi)^3}P_G(|\k_1-\k'|)P_G(|\k_2+\k'|)P_G(k')\, . 
\ee
Notice that in the squeezed limit $J(\k_l,\k_s) \rightarrow I(k_l)+\mathcal{O}(k_l^2/k_s^2)$. In another squeezed limit: 
$|\k_s|=|\k_s'|=|\k_s+\k_s'|$,  $J(\k_s,-\k_s'-\k_l)+J(\k_s',-\k_s-\k_l)+J(\k_s,\k_s+\k_s')\rightarrow 3\beta_{\zeta_G}(k_s)$ where, 
\be
\beta_{\zeta_G}(k_s) =  \frac{1}{4\pi}\Delta^2_G(k_s)\int d^3\x \,x^{n_s-4}(1+x^2-2\x\cdot\hat{\k}_s)^{\frac{n_s}{2}-2}(1+x^2+2\x\cdot\hat{\k}_s')^{\frac{n_s}{2}-2}
\ee
So, in the squeezed limit the angular dependence vanishes and the $J$ functions are just dependent on the magnitudes $k_s$, $k_l$. 

Finally, the trispectrum is given by 
\ba
T(\k_1,\k_2,\k_3,\k_4)&=&\left(P_G(k_1)P_G(k_2)P_G(k_3)+3\,{\rm perm.}\right)\\
&&\left((f^{(1)})^3f^{(3)}+\left(\frac{3}{2}(f^{(1)})^2(f^{(3)})^2+\frac{1}{2}(f^{(1)})^3f^{(5)}\right)\langle\zeta_G^2\rangle+\dots\right)\nn\\
&+&\left(P_G(k_1)P_G(k_2)\left(P_G(|\k_1+\k_3|)+P_G(|\k_1+\k_4|)\right)+5\,{\rm perm.}\right) \nn\\
&&\left((f^{(1)})^2(f^{(2)})^2+\left(f^{(1)}(f^{(2)})^2f^{(3)}+(f^{(1)})^2f^{(2)}f^{(4)}\right)\langle\zeta_G^2\rangle+\dots\right)\nn\\
&+&\left(K(\k_1,\k_2,\k_3)P_G(k_1)P_G(|\k_2+\k_3|)P_G(k_3)+2\,{\rm perm.}\right)\left((f^{(2)})^4+\dots\right)\nn\\
&+&\left((I_{\zeta_G}(k_1)+I_{\zeta_G}(k_2))P_G(k_1)P_G(k_2)(P_G(|\k_1+\k_3|)+ P_G(|\k_1+\k_4|))+5\,{\rm perm.}\right)\nn\\
&&\left(\frac{1}{2}f^{(1)}(f^{(2)})^2f^{(3)}+\dots\right)\nn\\
&+&\left(P_G(k_1)P_G(k_2)P_G(k_4)J(\k_2,\k_4)+11\,{\rm perm.}\right)\left(f^{(1)}(f^{(2)})^2f^{(3)}+\dots\right)\nn\\
&+&\left(P_G(k_1)P_G(k_2)\left(I_{\zeta_G}(|\k_1+\k_3|)P_G(|\k_1+\k_3|)+I_{\zeta_G}(|\k_2+\k_3|)P_G(|\k_2+\k_3|)\right)\right.\nn\\
&&\left.+5\,{\rm perm.}\right)\left(\frac{1}{2}(f^{(1)})^2(f^{(3)})^2+\dots\right)\nn\\
&+&\left(I_{\zeta_G}(k_1)P_G(k_1)P_G(k_2)P_G(k_3)+11\,{\rm perm.}\right)\left(\frac{1}{2}(f^{(1)})^2f^{(2)}f^{(4)}+\dots\right)\nn
\ea
where $K\sim \mathcal{O}(\Delta_G^2)$ and depends on the magnitudes of $\k_1,\k_2,\k_3$ and the relative angles between them 
\be
\label{eq:K1def}
K(\k_1,\k_2,\k_3) = \frac{\int\frac{d^3\k'}{(2\pi)^3}P_G(k')P_G(|\k_1+\k'|))P_G(|\k_1+\k_2+\k'|)P_G(|\k_3-\k'|) }{P_G(k_1)P_G(|\k_1+\k_2|)P_G(k_3) }\,.
\ee
In the squeezed limit needed to calculate $g_{NL}$ we find
\be
K(\k_s,\k_s',\k_l)\rightarrow I(k_l)
\ee
where in taking the limit we have fixed $|\k_s| = |\k_s'|=|\k_s+\k_s'|$. 
\begin{figure}
\begin{center}
$\begin{array}{cc}
\includegraphics[width=0.6\textwidth]{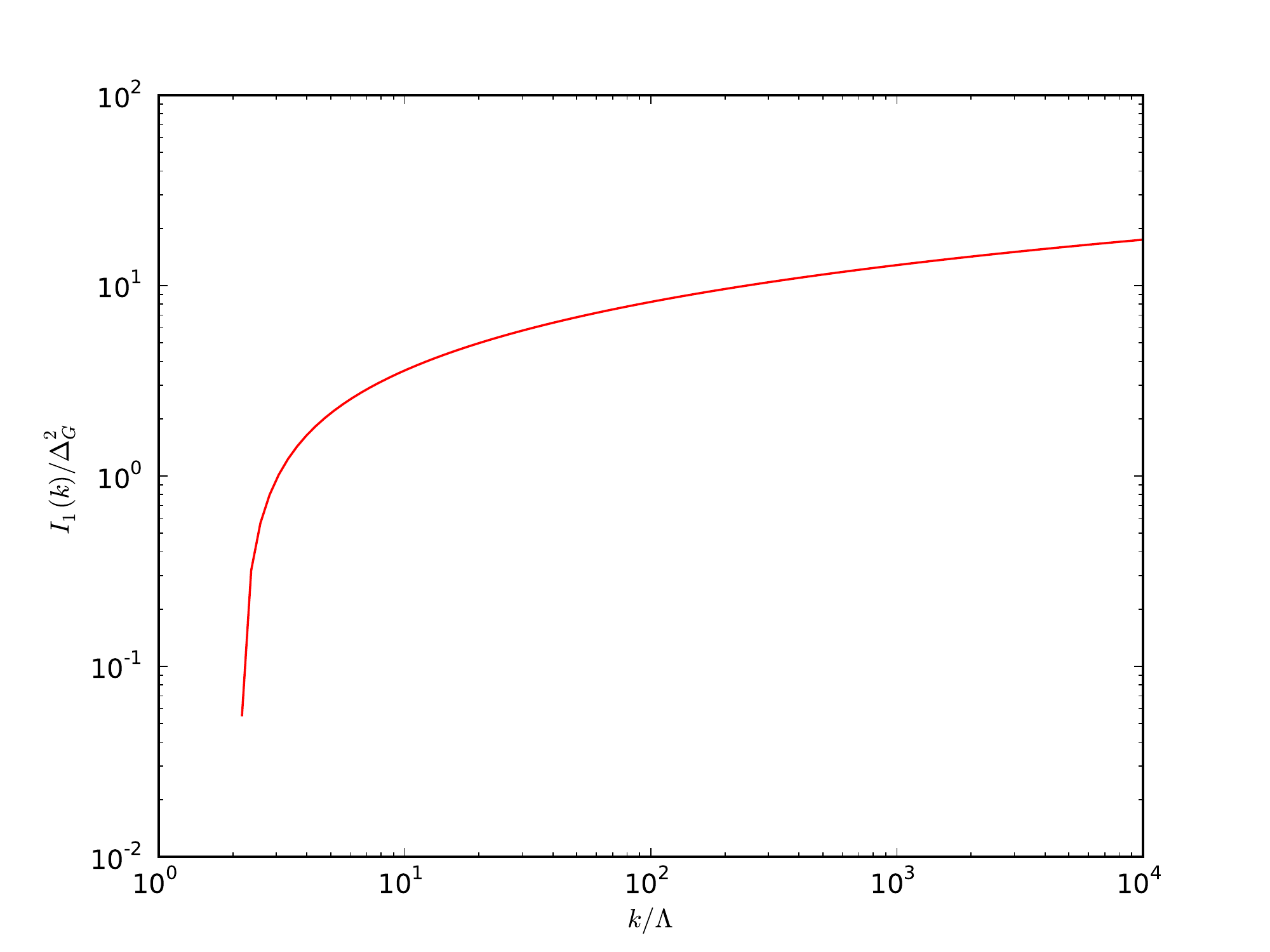} 
\end{array}$ 
\caption{\label{Fig:I1} Plotted is $I_{\zeta_G}(k)$, the higher-order correction to polyspectra of $\zeta_{NG}$ given in Eq.~(\ref{eq:I1def}). Shown here for a scale invariant spectrum $\Delta_G$ (e.g. Eq.~(\ref{eq:I1scaleinv})). }
\end{center}
\end{figure}

It is perhaps more useful to consider the usual local parameters (see, e.g. \cite{Smith:2011if}) : 
\ba
\label{eq:fNLdef}
\frac{3}{5}f_{NL} &\equiv&\frac{1}{4} \lim_{k_l\rightarrow 0} \frac{B(\k_l,\k_s,-\k_l-\k_s)}{P(k_s)P(k_l)}\\
\label{eq:tauNLdef}
\tau_{NL}&\equiv&\frac{1}{4}\lim_{\k_l\rightarrow 0}\frac{T(\k_s,-\k_s+\k_l,\k_s',-\k_s'-\k_l)}{P(k_s)P(k_s')P(k_l)}\\
\label{eq:gNLdef}
\frac{9}{25}g_{NL}&\equiv&\frac{1}{18}\lim_{k_l\rightarrow 0}\frac{T(\k_s,\k_s',\k_l,-\k_s-\k_s'-\k_l)}{P(k_s)P(k_s')P(k_l)} - \frac{1}{3}\tau_{NL}
\ea
where in the expression for $g_{NL}$ we fix $|\k_s|=|\k_s'|=|\k_s+\k_s'|$. Substituting the power spectrum, bispectrum and trispectrum into Eq.~(\ref{eq:fNLdef})-(\ref{eq:gNLdef})
gives
\be
\frac{3}{5}f_{NL}=\frac{f^{(2)}}{2(f^{(1)})^2}+\left(\frac{f^{(4)}}{4(f^{(1)})^2}-\frac{f^{(2)}f^{(3)}}{2(f^{(1)})^3}\right)\langle\zeta^2_G\rangle+\frac{f^{(2)}f^{(3)}}{4(f^{(1)})^3}I_{\zeta_G}(k_l)+\frac{1}{4}\left(\frac{f^{(2)}f^{(3)}}{(f^{(1)})^3}-\frac{(f^{(2)})^3}{(f^{(1)})^4}\right)I_{\zeta_G}(k_s)+\dots
\ee
\ba
\frac{9}{25}g_{NL} &=&\frac{f^{(3)}}{6(f^{(1)})^3}+\left(\frac{1}{12}\frac{f^{(5)}}{(f^{(1)})^3}-\frac{1}{4}\frac{(f^{(3)})^2}{(f^{(1)})^4}\right)\langle\zeta_G^2\rangle+\left(\frac{(f^{(3)})^2}{6(f^{(1)})^4}-\frac{(f^{(2)})^2f^{(3)}}{3(f^{(1)})^5}+\frac{f^{(2)}f^{(4)}}{6(f^{(1)})^4}\right)I_{\zeta_G}(k_s) \nn\\
&+&\left(\frac{1}{12}\frac{f^{(2)}f^{(4)}}{(f^{(1)})^4}-\frac{(f^{(3)})^2}{6(f^{(1)})^4}+\frac{1}{12}\frac{(f^{(2)})^2f^{(3)}}{(f^{(1)})^5}\right)I_{\zeta_G}(k_l)+\frac{1}{6}\frac{(f^{(2)})^2f^{(3)}}{(f^{(1)})^5}\beta(k_s)+\dots
\ea
and 
\ba
\tau_{NL}&=&\frac{(f^{(2)})^2}{(f^{(1)})^4}  +\left(\frac{f^{(2)}f^{(4)}}{(f^{(1)})^4}-2\frac{(f^{(2)})^2f^{(3)}}{(f^{(1)})^5}\right)\langle\zeta_G^2\rangle+\left(\frac{1}{2}\frac{(f^{(3)})^2}{(f^{(1)})^4}+\frac{(f^{(2)})^2f^{(3)}}
{(f^{(1)})^5}\right)I_{\zeta_G}(k_l)\\
&&+\left(\frac{(f^{(2)})^2f^{(3)}}{(f^{(1)})^5}-\frac{(f^{(2)})^4}{(f^{(1)})^6}\right)\frac{I_{\zeta_G}(k_s)+I_{\zeta_G}(k_s')}{2}+\dots\nn
\ea
The observed values of the non-Gaussian parameters $f_{NL}$, $g_{NL}$, $\tau_{NL}$ include the scale-independent loop contributions at $\mathcal{O}(\langle\zeta_G^2\rangle)$, rewriting the series $\zeta_{NG}(\x)=\sum_n \frac{1}{n!}f^{(n)}\zeta_G^n(\x)$ as $\zeta_{NG}(\x)=\sum_n h^{(n)}H_n(\zeta_G(\x))$, where $H_n$ are the Hermite polynomials (as we have done in Eq.~(\ref{eq:zetaNGfull})) cancels these lowest order loop terms. 

If instead we have $f^{(1)}=0$ and $f^{(2)}\neq 0$.For $f^{(1)}\neq 0$ then,
\be
\frac{3}{5}f_{NL}=\frac{1}{I_{\zeta_G}(k_s)f^{(2)}}+\dots \qquad \tau_{NL}=\frac{4}{(f^{(2)})^2I_{\zeta_G}(k_s)I_{\zeta_G}(k_s')}+\dots  \qquad g_{NL}= 0+\dots 
\ee

\section{Mapping Between Statistics in $V_L$ and Statistics in $V_S$}
\label{sec:mapping}
We would like to calculate the local statistics, that is correlation functions of $\zeta_{NG}|_S$. First, we rewrite the locally observed non-Gaussian field as
\ba
\left.\zeta_{NG}(\x)\right|_{S}&=& \left(f^{(1)}+f^{(2)}\zeta_{G,L}+\frac{f^{(3)}}{2}\left(\zeta_{G,L}^2-\langle\zeta_{G,L}^2\rangle\right)+\frac{f^{(4)}}{3!}\zeta_{G,L}^3+\dots\right)\zeta_{G,S}\\
&+&\frac{1}{2}\left(f^{(2)}+f^{(3)}\zeta_{G,L}+\frac{f^{(4)}}{2!}\left(\zeta_{G,L}^2-\langle\zeta_{G,L}^2\rangle\right)+\frac{f^{(5)}}{3!}\zeta_{G,L}^3+\dots\right)\left(\zeta_{G,S}^2-\langle\zeta_{G,S}^2\rangle\right)\nn\\
&+&\frac{1}{3!}\left(f^{(3)}+f^{(4)}\zeta_{G,L}+\frac{f^{(5)}}{2}\left(\zeta_{G,L}^2-\langle\zeta_{G,L}^2\rangle\right)+\dots\right)\zeta_{G,S}^3\nn\\
&+&\frac{1}{4!}\left(f^{(4)}+f^{(5)}\zeta_{G,L}+\frac{f^{(6)}}{2}\left(\zeta_{G,L}^2-\langle\zeta_{G,L}^2\rangle\right)+\dots \right)\left(\zeta_{G,S}^4-3\langle\zeta_{G,S}^2\rangle^2\right)\nn\\
&+&\dots\nn\\
&=& g^{(1)}\zeta_{G,S}+\frac{g^{(2)}}{2}\left(\zeta_{G,S}^2-\langle\zeta_{G,S}^2\rangle\right)+\frac{g^{(3)}}{3!}\zeta_{G,S}^3+\frac{g^{(4)}}{4!}\left(\zeta_{G,S}^4-\langle \zeta_{G,S}^4\rangle\right)+\dots
\ea
where $g\equiv f'(\zeta_{G,L})$. The coefficients $g^{(n)}$ are equal to $f^{(n)}$ up to corrections $\mathcal{O}(f^{(n+1)}\zeta_{G,L})$. So, the local statistics are similar to the global ones as long as the amplitude of the background mode obeys $\zeta_{G,L}^m< m! f^{(n)}/f^{(n+m)}$.

Under the approximation that the coefficients $g^{(n)}$ are constant across the volume $V_S$ we can use the expressions from \S \ref{sec:Npointsglobal} with $f\rightarrow g$ and $I_{\zeta_G}(k)\rightarrow I_{\zeta_{G,S}}(k)$ : 
\ba
\frac{3}{5}\left. f_{NL}\right|_{S}&=&\frac{g^{(2)}}{2(g^{(1)})^2}+\left(\frac{g^{(4)}}{4(g^{(1)})^2}-\frac{g^{(2)}g^{(3)}}{2(g^{(1)})^3}\right)\langle\zeta^2_{G,S}\rangle+\frac{g^{(2)}g^{(3)}}{4(g^{(1)})^3}I_{\zeta_{G,S}}(k_l)\\
&&+\,\frac{1}{4}\left(\frac{g^{(2)}g^{(3)}}{(g^{(1)})^3}-\frac{(g^{(2)})^3}{(g^{(1)})^4}\right)I_{\zeta_{G,S}}(k_s)+\dots\nn
\ea
\ba
\frac{9}{25}\left. g_{NL}\right|_{S} &=&\frac{g^{(3)}}{6(g^{(1)})^3}+\left(\frac{1}{12}\frac{g^{(5)}}{(g^{(1)})^3}-\frac{1}{4}\frac{(g^{(3)})^2}{(g^{(1)})^4}\right)\langle\zeta_{G,S}^2\rangle+\left(\frac{(g^{(3)})^2}{6(g^{(1)})^4}-\frac{(g^{(2)})^2g^{(3)}}{3(g^{(1)})^5}+\frac{g^{(2)}g^{(4)}}{6(g^{(1)})^4}\right)I_{\zeta_{G,S}}(k_s) \nn\\
&+&\left(\frac{1}{12}\frac{g^{(2)}g^{(4)}}{(g^{(1)})^4}-\frac{(g^{(3)})^2}{6(g^{(1)})^4}+\frac{1}{12}\frac{(g^{(2)})^2g^{(3)}}{(g^{(1)})^5}\right)I_{\zeta_{G,S}}(k_l)+\frac{1}{6}\frac{(g^{(2)})^2g^{(3)}}{(g^{(1)})^5}\beta_{\zeta_{G,S}}(k_s)+\dots\nn
\ea
and 
\ba
\left.\tau_{NL}\right|_{S}&=&\frac{(g^{(2)})^2}{(g^{(1)})^4}  +\left(\frac{g^{(2)}g^{(4)}}{(g^{(1)})^4}-2\frac{(g^{(2)})^2g^{(3)}}{(g^{(1)})^5}\right)\langle\zeta_{G,s}^2\rangle+\left(\frac{1}{2}\frac{(g^{(3)})^2}{(g^{(1)})^4}+\frac{(g^{(2)})^2g^{(3)}}
{(g^{(1)})^5}\right)I_{\zeta_{G,S}}(k_l)\\
&&+\left(\frac{(g^{(2)})^2g^{(3)}}{(g^{(1)})^5}-\frac{(g^{(2)})^4}{(g^{(1)})^6}\right)\frac{I_{\zeta_{G,S}}(k_s)+I_{\zeta_{G,S}}(k_s')}{2}+\dots\nn
\ea
where $k_s$, $k_l \ge 2\pi/V_S^{1/3}$ are the long and short wavelength modes used to measure $f_{NL}$, $g_{NL}$ $\tau_{NL}$ within $V_S$. 

So that
\ba
\frac{\left.f_{NL}\right|_{S}}{f_{NL}} &=&1+ \left(\frac{f^{(3)}}{f^{(2)}}-2\frac{f^{(2)}}{f^{(1)}}\right)\zeta_{G,L}+\left(3\left(\frac{f^{(2)}}{f^{(1)}}\right)^2-2\frac{f^{(3)}}{f^{(1)}}\right)\zeta_{G,L}^2\\
&&+\left(\frac{f^{(4)}}{2f^{(2)}}-\frac{f^{(3)}}{f^{(1)}}\right)\left(\zeta_{G,L}^2-\langle\zeta_{G,L}^2\rangle\right)+\left(\frac{f^{(3)}}{f^{(1)}}-\frac{f^{(4)}}{2f^{(2)}}\right)\left(\langle\zeta_G^2\rangle-\langle\zeta_{G,S}^2\rangle\right)\nn\\
&&+\frac{1}{2}\left(\frac{f^{(2)}}{f^{(1)}}\right)^2\left(I_{\zeta_{G}}(k_s)-I_{\zeta_{G,S}}(k_s)\right)-\frac{f^{(3)}}{2f^{(1)}}\left(I_{\zeta_G}(k_l)+I_{\zeta_G}(k_s)-I_{\zeta_{G,S}}(k_l)-I_{\zeta_{G,S}}(k_s)\right)\nn
\ea
and when $f^{(1)}=0$, $f^{(2)}\neq 0$, 
\be
\frac{\left.f_{NL}\right|_{S}}{f_{NL}} = \frac{8\zeta_{G,L}^2+2I_{\zeta_{G,S}}(k_l)}{f^{(2)}\left(I_{\zeta_{G,S}}+2\zeta_{G,L}^2\right)\left(I_{\zeta_{G,S}}+2\zeta_{G,L}^2\right)}+\dots
\ee

\section{Diagrammatic Rules for Fourier-Space Expressions}
\label{sec:Fourier}

The real-space diagrams shown in Appendix \ref{sec:diagrams} can also be calculated in Fourier space:
\be
\text{real-space expression of }(\x_1,\dots\x_n)=\int\prod_i^n\frac{d^3k_i}{(2\pi)^3}\left[\text{\textit{k}-space expression}\right]
e^{i\sum\mathbf{k}_i\cdot\mathbf{x}_i}.
\ee
Here we will show how momentum-space expressions such as those given above in Appendix \ref{sec:global} for the bispectrum and trispectrum can be quickly recovered from their corresponding diagrams.

For the local ansatz $\zeta_{NG}(\mathbf{x})=\sum_m\frac{1}{m!}f^{(m)}(\zeta^m_G(\mathbf{x})-\langle\zeta_G^m(\mathbf{x})\rangle)$, a particular $n$-point function is given by
\be
\langle\zeta_{\mathbf{k}_1}\zeta_{\mathbf{k}_2}\dots\zeta_{\mathbf{k}_n}\rangle
=\sum_{m_1}\sum_{m_2}\dots \sum_{m_n}\frac{f^{(m_1)}}{m_1!}\frac{f^{(m_2)}}{m_2!}\dots \frac{f^{(m_n)}}{m_n!}\langle(\zeta_G^{m_1})_{\mathbf{k}_1}(\zeta_G^{m_2})_{\mathbf{k}_2}\dots(\zeta_G^{m_n})_{\mathbf{k}_n}\rangle, \label{npt}
\ee
where the modes are given by the convolution integrals
\be
(\zeta_G^m)_{\mathbf{k}}=\int\prod_{i=1}^{m-1}\frac{d^3p_i}{(2\pi)^3}\zeta_G(\mathbf{p}_i)\zeta_G\Big(\mathbf{k}-\sum_{j=1}^{m-1}\mathbf{p}_j\Big).
\ee
A given term in \eqref{npt} is specified (up to permutations in the $\mathbf{k}_i$) by a set of numbers $V_m$, where $V_m$ is the number of times the $\zeta_G^m$ term appears. Thus, $\sum V_m=n$. Restricting to the connected part $\langle\zeta_{\mathbf{k}_1}\zeta_{\mathbf{k}_2}\dots\zeta_{\mathbf{k}_n}\rangle_c$ imposes the condition $\frac{1}{2}\sum m V_m-n+1\geq0$. This is not a sufficient condition for the contribution to be connected; the contractions must be made so that the corresponding diagram is connected, as discussed below. Note that $\sum m V_m$ is even for nonzero contributions.

There are $m-1$ momentum-space convolution integrals for each of the $(\zeta_G^m)_{\mathbf{k}_i}$ for a total of $\sum m V_m-n$ integrals. The factors of $\zeta_G$ in \eqref{npt} are contracted using Wick's theorem, giving $\frac{1}{2}\sum m V_m -1$ delta functions, not counting the final overall $\delta^3(\sum\mathbf{k}_i)$, so in the final expression there are $L\equiv\frac{1}{2}\sum m V_m-n+1$ integrals remaining. This is the number of loops that will appear in the diagram. If $L=0$ the graph is a tree graph. Tree graphs dominate contributions to the $n$-point functions for a weakly non-Gaussian series. If $L<0$ the diagram is disconnected.

The $(\zeta_G^m)_{\mathbf{k}_i}$ factors will be represented as $m$-point vertices, with $V_m$ of each type in the diagram, and a total of $n$ vertices. $V_1\equiv E$ denotes the number times the linear term contributes; these $1$-point vertices appear as external lines. Finally, each contraction between two factors of $\zeta_G$ yields a factor of the power spectrum and is represented by a line connecting two vertices, with a total of $P\equiv\frac{1}{2}\sum m V_m=n+L-1$ lines.

The rules for diagrams are as follows:
\begin{enumerate}
  \item Assign a momentum label $\mathbf{k}_i$ $(i=1,2...n)$ to each vertex, including external $1$-point vertices. Each $m$-point vertex is equivalent to a factor of $f^{(m)}$. (The $\frac{1}{m!}$ is cancelled by the $m!$ ways of contracting into the vertex.)
  \item Assign a momentum label to each line, with a direction. 
Lines contracted with $1$-point vertices share their momentum label. 
$L$ internal lines can be labelled with integrated momenta $\mathbf{p}_j$ $(j=1,2...L)$; these can be chosen arbitrarily among the lines forming loops. The remaining $P-E-L=(n-E)-1$ internal lines can be labelled with momenta $\mathbf{k}_I+\sum\mathbf{q}_k$, where $\mathbf{k}_I$ is the momentum of one of the vertices contracting with the line (either can be chosen), and $\mathbf{q}_k$ ($k=1,2...m-1$) denote the incoming momenta of the other lines being contracted into that vertex. This imposes momentum conservation at each vertex. These labels can be made by working into the diagram starting from the external lines. Each line is then equivalent to a factor of $P_G(q)$, where $\mathbf{q}$ is the momentum for that line.
  \item Integrate over the loop momenta by adding a factor $\int\frac{d^3p}{(2\pi)^3}$ for each loop. Note that loops at a single vertex contribute a factor $\langle\zeta_G^2\rangle$.
  \item Divide by the symmetry factor of the diagram. As in standard quantum field theory, the symmetry factor is determined by counting the number of ways of exchanging identical vertices or identical lines, as well as lines contracted at a single vertex.
  \item Sum over permutations of the $\mathbf{k}_j$ (momenta for the vertices). Sum over connected diagrams with $n$ vertices, to desired loop order, or level of approximation, and multiply by $(2\pi)^3\delta^3(\sum\mathbf{k}_i)$ to obtain the $n$-point function $\langle\zeta_{\mathbf{k}_1}\zeta_{\mathbf{k}_2}\dots\zeta_{\mathbf{k}_n}\rangle_c$.
\end{enumerate}

These diagrams are essentially equivalent to those considered in \cite{Byrnes:2007tm}, where the more general case of multiple fields contributing to the curvature perturbation was considered.

\bibliographystyle{ieeetr}
\bibliography{boxes}

\begin{thebibliography}{10}

\bibitem{Guth:1980zm}
A.~H. Guth, ``{The Inflationary Universe: A Possible Solution to the Horizon
  and Flatness Problems},'' {\em Phys. Rev.}, vol.~D23, pp.~347--356, 1981.

\bibitem{Mukhanov:1981xt}
V.~F. Mukhanov and G.~Chibisov, ``{Quantum Fluctuation and Nonsingular
  Universe. (In Russian)},'' {\em JETP Lett.}, vol.~33, pp.~532--535, 1981.

\bibitem{Hawking:1982cz}
S.~W. Hawking, ``{The Development of Irregularities in a Single Bubble
  Inflationary Universe},'' {\em Phys. Lett.}, vol.~B115, p.~295, 1982.

\bibitem{Starobinsky:1982ee}
A.~A. Starobinsky, ``{Dynamics of Phase Transition in the New Inflationary
  Universe Scenario and Generation of Perturbations},'' {\em Phys. Lett.},
  vol.~B117, pp.~175--178, 1982.

\bibitem{Guth:1982ec}
A.~H. Guth and S.~Y. Pi, ``{Fluctuations in the New Inflationary Universe},''
  {\em Phys. Rev. Lett.}, vol.~49, pp.~1110--1113, 1982.

\bibitem{Bardeen:1983qw}
J.~M. Bardeen, P.~J. Steinhardt, and M.~S. Turner, ``{Spontaneous Creation of
  Almost Scale - Free Density Perturbations in an Inflationary Universe},''
  {\em Phys. Rev.}, vol.~D28, p.~679, 1983.

\bibitem{Mollerach:1989hu}
S.~Mollerach, ``{ISOCURVATURE BARYON PERTURBATIONS AND INFLATION},'' {\em
  Phys.Rev.}, vol.~D42, pp.~313--325, 1990.

\bibitem{Linde:1996gt}
A.~D. Linde and V.~F. Mukhanov, ``{Nongaussian isocurvature perturbations from
  inflation},'' {\em Phys. Rev.}, vol.~D56, pp.~535--539, 1997.

\bibitem{Enqvist:2001zp}
K.~Enqvist and M.~S. Sloth, ``{Adiabatic CMB perturbations in pre - big bang
  string cosmology},'' {\em Nucl.Phys.}, vol.~B626, pp.~395--409, 2002.

\bibitem{Lyth:2001nq}
D.~H. Lyth and D.~Wands, ``{Generating the curvature perturbation without an
  inflaton},'' {\em Phys. Lett.}, vol.~B524, pp.~5--14, 2002.

\bibitem{Lyth:2002my}
D.~H. Lyth, C.~Ungarelli, and D.~Wands, ``{The primordial density perturbation
  in the curvaton scenario},'' {\em Phys. Rev.}, vol.~D67, p.~023503, 2003.

\bibitem{Dvali:2003em}
G.~Dvali, A.~Gruzinov, and M.~Zaldarriaga, ``{A new mechanism for generating
  density perturbations from inflation},'' {\em Phys.Rev.}, vol.~D69,
  p.~023505, 2004.

\bibitem{Dvali:2003ar}
G.~Dvali, A.~Gruzinov, and M.~Zaldarriaga, ``{Cosmological perturbations from
  inhomogeneous reheating, freezeout, and mass domination},'' {\em Phys.Rev.},
  vol.~D69, p.~083505, 2004.

\bibitem{Bartolo:2004if}
N.~Bartolo, E.~Komatsu, S.~Matarrese, and A.~Riotto, ``{Non-Gaussianity from
  inflation: Theory and observations},'' {\em Phys.Rept.}, vol.~402,
  pp.~103--266, 2004.

\bibitem{Salopek:1990jq}
D.~S. Salopek and J.~R. Bond, ``{Nonlinear evolution of long wavelength metric
  fluctuations in inflationary models},'' {\em Phys. Rev.}, vol.~D42,
  pp.~3936--3962, 1990.

\bibitem{Gangui:1993tt}
A.~Gangui, F.~Lucchin, S.~Matarrese, and S.~Mollerach, ``{The Three point
  correlation function of the cosmic microwave background in inflationary
  models},'' {\em Astrophys. J.}, vol.~430, pp.~447--457, 1994.

\bibitem{Komatsu:2001rj}
E.~Komatsu and D.~N. Spergel, ``{Acoustic signatures in the primary microwave
  background bispectrum},'' {\em Phys. Rev.}, vol.~D63, p.~063002, 2001.

\bibitem{Okamoto:2002ik}
T.~Okamoto and W.~Hu, ``{The Angular Trispectra of CMB Temperature and
  Polarization},'' {\em Phys. Rev.}, vol.~D66, p.~063008, 2002.

\bibitem{Enqvist:2008gk}
K.~Enqvist and T.~Takahashi, ``{Signatures of Non-Gaussianity in the Curvaton
  Model},'' {\em JCAP}, vol.~0809, p.~012, 2008.

\bibitem{Hinshaw:2012fq}
G.~Hinshaw, D.~Larson, E.~Komatsu, D.~Spergel, C.~Bennett, {\em et~al.},
  ``{Nine-Year Wilkinson Microwave Anisotropy Probe (WMAP) Observations:
  Cosmological Parameter Results},'' 2012.

\bibitem{Komatsu:2010fb}
E.~Komatsu {\em et~al.}, ``{Seven-Year Wilkinson Microwave Anisotropy Probe
  (WMAP) Observations: Cosmological Interpretation},'' {\em
  Astrophys.J.Suppl.}, vol.~192, p.~18, 2011.

\bibitem{Fergusson:2010gn}
J.~Fergusson, D.~Regan, and E.~Shellard, ``{Optimal Trispectrum Estimators and
  WMAP Constraints},'' 2010.

\bibitem{Cheung:2007st}
C.~Cheung, P.~Creminelli, A.~L. Fitzpatrick, J.~Kaplan, and L.~Senatore, ``{The
  Effective Field Theory of Inflation},'' {\em JHEP}, vol.~03, p.~014, 2008.

\bibitem{Fan:1995aq}
Z.-h. Fan and J.~M. Bardeen, ``{Distributions of Fourier modes of cosmological
  density fields},'' {\em Phys.Rev.}, vol.~D51, pp.~6714--6721, 1995.

\bibitem{Linde:2005yw}
A.~D. Linde and V.~Mukhanov, ``{The curvaton web},'' {\em JCAP}, vol.~0604,
  p.~009, 2006.

\bibitem{Gordon:2005ai}
C.~Gordon, W.~Hu, D.~Huterer, and T.~M. Crawford, ``{Spontaneous isotropy
  breaking: a mechanism for cmb multipole alignments},'' {\em Phys.Rev.},
  vol.~D72, p.~103002, 2005.

\bibitem{Boubekeur:2005fj}
L.~Boubekeur and D.~Lyth, ``{Detecting a small perturbation through its
  non-Gaussianity},'' {\em Phys.Rev.}, vol.~D73, p.~021301, 2006.

\bibitem{Giddings:2011zd}
S.~B. Giddings and M.~S. Sloth, ``{Cosmological observables, IR growth of
  fluctuations, and scale-dependent anisotropies},'' {\em Phys.Rev.}, vol.~D84,
  p.~063528, 2011.

\bibitem{Byrnes:2011ri}
C.~T. Byrnes, S.~Nurmi, G.~Tasinato, and D.~Wands, ``{Inhomogeneous
  non-Gaussianity},'' {\em JCAP}, vol.~1203, p.~012, 2012.

\bibitem{Schmidt:2012ky}
F.~Schmidt and L.~Hui, ``{CMB Power Asymmetry from Non-Gaussian Modulation},''
  {\em Phys. Rev. Lett. 110,}, vol.~011301, 2013.

\bibitem{Tasinato:2012js}
G.~Tasinato, C.~T. Byrnes, S.~Nurmi, and D.~Wands, ``{Loop corrections and a
  new test of inflation},'' 2012.

\bibitem{Nelson:2012sb}
E.~Nelson and S.~Shandera, ``{Statistical Naturalness and non-Gaussianity in a
  Finite Universe},'' 2012.

\bibitem{Nurmi:2013xv}
S.~Nurmi, C.~T. Byrnes, and G.~Tasinato, ``{A non-Gaussian landscape},'' 2013.

\bibitem{Demozzi:2010aj}
V.~Demozzi, A.~Linde, and V.~Mukhanov, ``{Supercurvaton},'' {\em JCAP},
  vol.~1104, p.~013, 2011.

\bibitem{Lyth:2004gb}
D.~H. Lyth, K.~A. Malik, and M.~Sasaki, ``{A General proof of the conservation
  of the curvature perturbation},'' {\em JCAP}, vol.~0505, p.~004, 2005.

\bibitem{Giddings:2010nc}
S.~B. Giddings and M.~S. Sloth, ``{Semiclassical relations and IR effects in de
  Sitter and slow-roll space-times},'' {\em JCAP}, vol.~1101, p.~023, 2011.

\bibitem{Senatore:2012nq}
L.~Senatore and M.~Zaldarriaga, ``{On Loops in Inflation II: IR Effects in
  Single Clock Inflation},'' {\em JHEP}, vol.~1301, p.~109, 2013.

\bibitem{Liddle:2000cg}
A.~R. Liddle and D.~Lyth, ``{Cosmological inflation and large scale
  structure},'' 2000.

\bibitem{Knox:2005hx}
L.~Knox, ``{On precision measurement of the mean curvature},'' {\em Phys.Rev.},
  vol.~D73, p.~023503, 2006.

\bibitem{Waterhouse:2008vb}
T.~Waterhouse and J.~Zibin, ``{The cosmic variance of Omega},'' 2008.

\bibitem{Vardanyan:2009ft}
M.~Vardanyan, R.~Trotta, and J.~Silk, ``{How flat can you get? A model
  comparison perspective on the curvature of the Universe},'' {\em
  Mon.Not.Roy.Astron.Soc.}, vol.~397, pp.~431--444, 2009.

\bibitem{Erickcek:2008jp}
A.~L. Erickcek, S.~M. Carroll, and M.~Kamionkowski, ``{Superhorizon
  Perturbations and the Cosmic Microwave Background},'' {\em Phys.Rev.},
  vol.~D78, p.~083012, 2008.

\bibitem{Guth:2012ww}
A.~H. Guth and Y.~Nomura, ``{What can the observation of nonzero curvature tell
  us?},'' {\em Phys.Rev.}, vol.~D86, p.~023534, 2012.

\bibitem{Kleban:2012ph}
M.~Kleban and M.~Schillo, ``{Spatial Curvature Falsifies Eternal Inflation},''
  {\em JCAP}, vol.~1206, p.~029, 2012.

\bibitem{Sasaki:2006kq}
M.~Sasaki, J.~Valiviita, and D.~Wands, ``{Non-gaussianity of the primordial
  perturbation in the curvaton model},'' {\em Phys. Rev.}, vol.~D74, p.~103003,
  2006.

\bibitem{Byrnes:2008zy}
C.~T. Byrnes, K.-Y. Choi, and L.~M. Hall, ``{Large non-Gaussianity from
  two-component hybrid inflation},'' {\em JCAP}, vol.~0902, p.~017, 2009.

\bibitem{Tseliakhovich:2010kf}
D.~Tseliakhovich, C.~Hirata, and A.~Slosar, ``{Non-Gaussianity and large-scale
  structure in a two-field inflationary model},'' 2010.

\bibitem{Shandera:2010ei}
S.~Shandera, N.~Dalal, and D.~Huterer, ``{A generalized local ansatz and its
  effect on halo bias},'' {\em JCAP}, vol.~1103, p.~017, 2011.

\bibitem{Smith:2010gx}
K.~M. Smith and M.~LoVerde, ``{Local stochastic non-Gaussianity and N-body
  simulations},'' {\em JCAP}, vol.~1111, p.~009, 2011.

\bibitem{Byrnes:2009pe}
C.~T. Byrnes, S.~Nurmi, G.~Tasinato, and D.~Wands, ``{Scale dependence of local
  fNL},'' {\em JCAP}, vol.~1002, p.~034, 2010.

\bibitem{Bramante:2013}
J.~Bramante, J.~Kumar, E.~Nelson, and S.~Shandera, ``{In preparation},''

\bibitem{Starobinsky:1986fx}
A.~A. Starobinsky, ``{STOCHASTIC DE SITTER (INFLATIONARY) STAGE IN THE EARLY
  UNIVERSE},'' 1986.

\bibitem{Salopek:1990re}
D.~Salopek and J.~Bond, ``{Stochastic inflation and nonlinear gravity},'' {\em
  Phys.Rev.}, vol.~D43, pp.~1005--1031, 1991.

\bibitem{Lyth:2007jh}
D.~H. Lyth, ``{The curvature perturbation in a box},'' {\em JCAP}, vol.~0712,
  p.~016, 2007.

\bibitem{Salem:2012ve}
M.~P. Salem, ``{The CMB and the measure of the multiverse},'' {\em JHEP},
  vol.~1206, p.~153, 2012.

\bibitem{Bull:2013fga}
P.~Bull and M.~Kamionkowski, ``{What if Planck's Universe isn't flat?},'' 2013.

\bibitem{Byrnes:2007tm}
C.~T. Byrnes, K.~Koyama, M.~Sasaki, and D.~Wands, ``{Diagrammatic approach to
  non-Gaussianity from inflation},'' {\em JCAP}, vol.~0711, p.~027, 2007.

\bibitem{Smith:2011if}
K.~M. Smith, M.~LoVerde, and M.~Zaldarriaga, ``{A universal bound on N-point
  correlations from inflation},'' {\em Phys.Rev.Lett.}, vol.~107, p.~191301,
  2011.

\end{thebibliography}

\end{document}